\documentclass[12pt,a4paper,onecolumn,titlepage,oneside,french]{report}
\usepackage{amsbsy}
\usepackage{amsfonts}
\usepackage{amsmath}
\usepackage{amssymb}
\usepackage{amstext}
\usepackage{fancyheadings}
\usepackage{array}
\usepackage{graphicx}
\usepackage{fancyhdr}
\usepackage{framed}
\usepackage{physics}
\usepackage{boxedminipage}
\usepackage[utf8]{inputenc}
\usepackage{curves}
\usepackage{fancyheadings}
\usepackage{float}
\usepackage{latexsym}
\usepackage[T1]{fontenc}
\usepackage[french]{babel}

\def\be{\begin{equation}}
\def\ee{\end{equation}}

\begin{document}

\newpage {\ }\thispagestyle{plain}

\begin{center}
{\Large INTRODUCTION AUX SYSTEMES }

\bigskip {\Large QUANTIQUES NON-HERMITIQUES:}

\bigskip

$\mathcal{PT-}${\Large sym\'{e}trie et pseudo Hermiticit\'{e}}

\bigskip

\bigskip

{\huge Maamache Mustapha}

\bigskip

Laboratoire de physique quantique et syst\`{e}mes dynamiques

D\'{e}partement de Physique

\ \ \ \ \ Universit\'{e} Ferhat Abbas S\'{e}tif 1, S\'{e}tif, Algeria.

\ \ \ \ \ \ \ \ \ \ \ \ \ 

\ \ \ \ \ \ \ \ \ \ \ \ E-mail: maamache@univ-setif.dz

\bigskip {\LARGE This course was delivered during the 4th Jijel School of
Theoretical Physics, held from September 25 to 29, 2016}

\bigskip

{\Large In memory of my mother Djabou Zoulikha }

\ \ \ \ \ \ \ \ \ {\Large who died on February 03, 2019}
\end{center}

{\huge Abstract }In recent decades, an important shift has taken place with
the growing role of non-Hermitian quantum mechanics. What makes this
framework remarkable is that the eigenvalues of the Hamiltonians involved
can still be real, just as in the Hermitian case.

Before turning to non-Hermitian quantum mechanics, we will first review some
basic ideas and mathematical tools from the standard Hermitian formulation.
This review is intended as a survey: most results will be stated without
proof and without going into detailed discussion.$\ \ $\ \ \ \ \ \ \ \ \ \ \
\ \ \ \ \ \ \ \ \ \ \ \ \ \ \ \ \ \ \ \ \ \ \ \ \ \ \ \ 

\thispagestyle{empty} 
\markboth{}{Introduction
} \addcontentsline{toc}{chapter}{Introduction}{\Huge Introduction}

La th\'{e}orie quantique est une th\'{e}orie\textbf{\ }qui permet de d\'{e}%
crire le comportement de la mati\`{e}re. Il s'agit d'un mod\`{e}le efficace,
mais totalement inutile pour d\'{e}crire par exemple la m\'{e}canique
d'objets de notre taille.

La pratique de la physique suppose l'intervention de deux entit\'{e}s, le
syst\`{e}me physique \'{e}tudi\'{e}, et l'observateur qui l'\'{e}tudie. La
mod\'{e}lisation d'une th\'{e}orie physique doit donc faire intervenir des
objets qui caract\'{e}risent ces deux entit\'{e}s. Le syst\`{e}me physique
va \^{e}tre caract\'{e}ris\'{e} par des \'{e}tats, c'est \`{a} dire des
quantit\'{e}s math\'{e}matiques d\'{e}crivant les propri\'{e}t\'{e}s intrins%
\`{e}ques du syst\`{e}me. L'intervention de l'observateur, qui effectue des
mesures sur le syst\`{e}me, va \^{e}tre caract\'{e}ris\'{e}e par des
observables, c'est \`{a} dire des quantit\'{e}s math\'{e}matiques qui vont d%
\'{e}crire les r\'{e}sultats des mesures effectu\'{e}es sur le syst\`{e}me
en fonction de l'\'{e}tat de celui-ci.

Les concepts fondamentaux, comme les amplitudes de probabilit\'{e}, les
superpositions lin\'{e}aires d'\'{e}tats, qui semblent si \'{e}tranges pour
notre intuition quand on les rencontre pour la premi\`{e}re fois, restent
toujours essentiels.

Une \'{e}volution importante s'est toutefois manifest\'{e}e au cours des
derni\`{e}res d\'{e}cennies. On assiste \'{e}galement \`{a} une invasion de
notre monde quotidien par les concepts de la m\'{e}canique quantique
non-Hermitique dont la particularit\'{e} est que les valeurs propres des
Hamiltoniens en jeu sont r\'{e}elles exactement que comme dans le cas
Hermitique. Il est donc clair qu'un enseignement moderne de la physique
quantique doit tenir compte de ces d\'{e}veloppements, pour donner \`{a} l'%
\'{e}tudiant ou au chercheur qui d\'{e}sire s'instruire une image plus pr%
\'{e}cise des progr\`{e}s r\'{e}alis\'{e}s et pour accro\^{\i}tre sa
motivation de mieux comprendre les ph\'{e}nom\`{e}nes physiques dont
l'importance conceptuelle et pratique est de plus en plus \'{e}vidente.
Avant d'aborder la m\'{e}canique quantique non-Hermitique, nous donnerons
succinctement quelques \ notions de base sur \ les concepts et les outils
math\'{e}matiques de la m\'{e}canique quantique Hermitique; il s'agira d'un
survol, et la grande majorit\'{e} des \'{e}nonc\'{e}s seront donn\'{e}s sans
d\'{e}monstration et sans discussion d\'{e}taill\'{e}e. Pour plus de d\'{e}%
tails, nous invitons le lecteur \`{a} consulter\textbf{\ \ }les ouvrages
suivants \ \cite{belac,Tanoudji, Messiah}.

A cet effet, on aborde les outils math\'{e}matiques et les postulats de la m%
\'{e}canique quantique afin de voir leurs changements dans le cas
non-Hermitique.

Le formalisme de la m\'{e}canique quantique est tr\`{e}s complexe et souvent
il est impossible de trouver une solution analytique \`{a} la dynamique des
particules. Dans le cas o\`{u} le syst\`{e}me quantique est ind\'{e}pendant
du temps, on peut obtenir une solution analytique. Par contre, lorsque le
syst\`{e}me quantique d\'{e}pend du temps, il n'y a que quelques cas qui
peuvent \^{e}tre solutionn\'{e}s analytiquement dans le cadre
d'approximations telle que: la th\'{e}orie des perturbations,
l'approximation soudaine, l'approximation adiabatique, ou la th\'{e}orie des
invariants.

Nous introduirons les concepts math\'{e}matiques et ensuite expliquer \`{a}
quoi \c{c}a sert, puis nous vous exposerons les principes de base et leurs
implications. Dans toute la suite, on appellera \guillemotleft\ syst\`{e}me
quantique \guillemotright\ un objet microscopique soumis aux lois de la m%
\'{e}canique quantique.

Un des axiomes principaux de la m\'{e}canique quantique impose aux
observables physiques d'\^{e}tre Hermitiques. Cela principalement dans le
but de leur assurer des valeurs propres r\'{e}elles. Mais en 1998, une large
classe d'Hamiltoniens s'est r\'{e}v\'{e}l\'{e}e poss\'{e}der des valeurs
propres r\'{e}elles en \'{e}tant non Hermitiques. Ces Hamiltoniens
respectent une condition plus faible pour l'obtention de valeurs propres r%
\'{e}elles : la sym\'{e}trie parit\'{e} temps ($\mathcal{PT}$).

L'objectif \ de ce cours est d'introduire la notion de $\mathcal{PT}$ -sym%
\'{e}trie puis celle plus g\'{e}n\'{e}rale la pseudo-Hermiticit\'{e} qui
permet passer d'un Hamiltonien non-hermitique \`{a} un Hamiltonien
hermitique \'{e}quivalent, puis leur g\'{e}n\'{e}ralisation aux syt\`{e}mes d%
\'{e}pendants du temps.

Le premier chapitre introduit un rappel g\'{e}n\'{e}ral de la m\'{e}canique
quantique Hermitique: les outils math\'{e}matiques n\'{e}cessaires et les
postulats.

On rappelle les sym\'{e}tries en physique dans le deuxi\`{e}me chapitre et
en particulier la sym\'{e}trie d'espace et du temps (la parit\'{e}, le
renversement du temps). Le premier et le deuxi\`{e}me chapitre sont
largement inspir\'{e}e de \cite{belac}.

Le troisi\`{e}me chapitre introduit bri\`{e}vement la m\'{e}canique
quantique des Hamiltoniens $\mathcal{PT-}$sym\'{e}triques et pseudo
Hermitiques.

\ Le chapitre quatre est d\'{e}di\'{e} aux syst\`{e}mes quantique Hermitique
d\'{e}pendants du temps, plusieurs m\'{e}thodes de r\'{e}solution de l'\'{e}%
quation de Schr\"{o}dinger associ\'{e}e seront abord\'{e}es.

Enfin, les syst\`{e}mes $\mathcal{PT-}$sym\'{e}triques et pseudo-hermitiques
d\'{e}pendant du temps sont d\'{e}taill\'{e}s au cinqui\`{e}me chapitre.\ 

\chapter{Rappel de la M\'{e}canique Quantique}

La naissance de la physique quantique date du dernier si\`{e}cle et cette
description des ph\'{e}nom\`{e}nes physiques, qui a transform\'{e} notre
vision du monde, n'est toujours pas remise en cause, ce qui est exceptionnel
pour une th\'{e}orie scientifique. Ses pr\'{e}dictions ont toujours \'{e}t%
\'{e} v\'{e}rifi\'{e}es par l'exp\'{e}rience avec une pr\'{e}cision
impressionnante. Les concepts fondamentaux, comme les amplitudes de
probabilit\'{e}, les superpositions lin\'{e}aires d'\'{e}tats, qui semblent
si \'{e}tranges pour notre intuition quand on les rencontre pour la premi%
\`{e}re fois, restent toujours essentiels. Le premier objectif de ce
chapitre introductif, avant d'aborder la m\'{e}canique quantique
non-Hermitique, est de donner succinctement quelques \ notions de base sur \
les concepts et les outils math\'{e}matiques de m\'{e}canique quantique
Hermitique.

Ce premier chapitre, largement inspir\'{e}e de \cite{belac}, d\'{e}crit les
concepts math\'{e}matiques g\'{e}n\'{e}raux utilis\'{e}es pour formaliser la
th\'{e}orie quantique. Comme toute th\'{e}orie physique, la th\'{e}orie
quantique poss\`{e}de plusieurs formalisations math\'{e}matiques, ayant des
niveaux de g\'{e}n\'{e}ralit\'{e} et d'abstraction vari\'{e}s.

\section{\protect\bigskip\ Premi\`{e}re approche de la physique quantique:
Dualit\'{e} onde--corpuscule}

la m\'{e}canique quantique est une th\'{e}orie tr\`{e}s ambitieuse: pr\'{e}%
dire (ou au moins expliquer) tous les comportements de la mati\`{e}re \`{a}
partir de ses constituants \'{e}l\'{e}mentaires (et par la m\^{e}me occasion
chercher ce que sont ces constituants). En fait, l'ambition moindre de ce
chapitre consiste, en partant des noyaux et des \'{e}lectrons, \`{a} trouver
des \'{e}quations qui permettent, au moins dans le principe, des pr\'{e}%
dictions sur tout le monde qui nous entoure (au niveau terrestre). Ces \'{e}%
quations permettent en fait avec quelques adaptations de d\'{e}crire aussi
la mati\`{e}re nucl\'{e}aire, ce qui permet d'arriver aux \'{e}toiles et
donc \`{a} tout le syst\`{e}me solaire. En fait, il faut aussi adapter la th%
\'{e}orie de Maxwell du champ e.m. \`{a} la th\'{e}orie quantique (ou
l'inverse), afin de d\'{e}crire la lumi\`{e}re \'{e}mise par les atomes. Il
se trouve qu'\`{a} petite \'{e}chelle, le monde n'est pas comme le n\^{o}%
tre. Les particules doivent \^{e}tre d\'{e}crites par des ondes et non des
points mat\'{e}riels. Toutefois, la description par des trajectoires est
valable \`{a} titre d'approximation quand la longueur d'onde devient tr\`{e}%
s petite devant les dimensions spatiales de ce qu'on cherche \`{a} mesurer.
C'est la m\^{e}me chose pour la description du champ e.m. en terme de rayons
(optique g\'{e}om\'{e}trique) : valable si $\lambda $ $\ll d$ o\`{u} $d$ est
la dimension des objets qui se trouvent sur le trajet du faisceau. Compte
tenu de ce que l'impulsion et le vecteur d'onde sont parall\`{e}les et de m%
\^{e}me sens, on aboutit \`{a} la relation vectorielle suivante entre
impulsion $\vec{p}$ et vecteur d'onde $\vec{k}$%
\begin{equation}
\vec{p}\mathbf{=\hbar \vec{k}}  \label{ch.1.7}
\end{equation}%
Cette \'{e}quation se traduit aussi par une relation (cette fois scalaire)
entre l'impulsion et la longueur d'onde $\lambda $, la longueur d'onde de de
Broglie

\begin{equation*}
p=\frac{h}{\lambda }.
\end{equation*}%
L'hypoth\`{e}se de Broglie est que ces deux derni\`{e}res relations sont
valables pour toutes les particules. Selon cette hypoth\`{e}se, une
particule d'impulsion $\vec{p}$ poss\`{e}de des propri\'{e}t\'{e}s
ondulatoires caract\'{e}ristiques d'une longueur d'onde $\lambda $ $=h/p$.
Si $\vec{v}\ll c$, on utilisera $\vec{p}=m$ $\vec{v}$, et sinon la formule g%
\'{e}n\'{e}rale relativiste d\'{e}crite plus haut, sauf bien s\^{u}r pour $%
m=0$, o\`{u} $p=E/c$. Si cette hypoth\`{e}se est correcte, on doit pouvoir
observer avec des particules des propri\'{e}t\'{e}s caract\'{e}ristiques des
ondes comme les interf\'{e}rences et la diffraction. C'est ce qu'on appelle
la dualit\'{e} onde--corpuscule.

\section{Math\'{e}matiques de la m\'{e}canique quantique}

Le principe de superposition est un principe fondateur de la m\'{e}canique
quantique, et nous pouvons s'appuyer sur ce principe pour rendre compte des
interf\'{e}rences. La m\'{e}canique quantique \'{e}tant une th\'{e}orie lin%
\'{e}aire et il est naturel que les espaces vectoriels y jouent un r\^{o}le
fondamental, ainsi le principe de superposition en est un principe
fondateur. Nous verrons qu'un \'{e}tat physique est repr\'{e}sent\'{e} math%
\'{e}matiquement par un vecteur dans un espace dont nous allons pr\'{e}ciser
les caract\'{e}ristiques, et qui sera appel\'{e} espace des \'{e}tats. Un
second principe fondateur, \'{e}galement d\'{e}duit des exp\'{e}riences
d'interf\'{e}rences, est l'existence d'amplitudes de probabilit\'{e}. Ces
amplitudes de probabilit\'{e} seront repr\'{e}sent\'{e}es math\'{e}%
matiquement par des produits scalaires d\'{e}finis sur l'espace des \'{e}%
tats. En m\'{e}canique quantique les amplitudes de probabilit\'{e} sont
fondamentalement des nombres complexes : le produit scalaire sera a priori
un nombre complexe. Les propri\'{e}t\'{e}s physiques: impulsion, position, 
\'{e}nergie . . . seront repr\'{e}sent\'{e}es par des op\'{e}rateurs
agissant dans l'espace des \'{e}tats. Avant, d'introduire les propri\'{e}t%
\'{e}s essentielles des espaces de Hilbert, c'est-\`{a}-dire les espaces
vectoriels munis d'un produit scalaire d\'{e}fini positif, en nous limitant
au cas de la dimension finie, nous survolerons les notions de Probabilit\'{e}
; Fonction d'onde et \'{e}quation de Schr\"{o}dinger.

\subsection{Probabilit\'{e} ; Fonction d'onde ; \'{e}quation de Schr\"{o}%
dinger}

\subsubsection{Description probabiliste et densit\'{e} de probabilit\'{e}}

En fait, toute la physique est une description probabiliste: un r\'{e}sultat
de mesure est donn\'{e} par un nombre et son incertitude, et on peut m\^{e}%
me dire que l'on peut souvent donner une loi de probabilit\'{e} de trouver
telle valeur de la mesure (on fait plusieurs fois la m\^{e}me exp\'{e}rience
et on fait un histogramme).

Pour fixer les id\'{e}es, nous rappelons la d\'{e}finition de la densit\'{e}
de probabilit\'{e} $p(\vec{r})$ o\`{u} $\vec{r}$ a pour coordonn\'{e}es $%
\left\{ x,y,z\right\} $. La probabilit\'{e} $dP(\vec{r})$ de trouver la
particule dans un petit volume $d\vec{r}$ est $dP(\vec{r})$) = $p(\vec{r})d%
\vec{r}$. Cela n'est pas limit\'{e} \`{a} la mesure de la position, mais
peut-\^{e}tre \'{e}tendu \`{a} la mesure de toute variable continue, par
exemple la vitesse, ou une seule coordonn\'{e}e $x$.

Dans le domaine classique, on pourrait \'{e}crire alors l'\'{e}volution de
ces densit\'{e}s de probabilit\'{e}s au cours du temps (en supposant qu'on
les conna\^{\i}t au temps $t=0$ et en r\'{e}solvant les \'{e}quations de
Newton). Il se trouve que cette m\'{e}thode ne donne pas les bons r\'{e}%
sultats d\`{e}s qu'on arrive dans le domaine microscopique (c.-\`{a}-d.
quand la longueur d'onde de de Broglie $\lambda $\ est de l'ordre de
grandeur des distances de variation de cette probabilit\'{e}). On doit alors
avoir recours \`{a} une autre description.

\subsubsection{Fonction d'onde}

Pour bien d\'{e}crire la physique d'une particule \`{a} des \'{e}chelles de
longueur de l'ordre de la longueur d'onde de Broglie $\lambda $, il fallait
introduire une nouvelle fonction de la position, appel\'{e}e fonction
d'onde, not\'{e}e en g\'{e}n\'{e}ral $\psi (\vec{r})$ dont les valeurs sont
complexes et qui a la propri\'{e}t\'{e} que $\left\vert \psi (%
\overrightarrow{r})\right\vert ^{2}$ = $p(\vec{r})$, la densit\'{e} de
probabilit\'{e}. En clair, on compl\`{e}te la description par $p(\vec{r})$
en ajoutant une phase $\psi (\vec{r})$ $=\sqrt{p(\vec{r})}\exp (i\varphi (%
\vec{r}))$.

\subsubsection{\'{E}quation de Schr\"{o}dinger}

Nous passons tr\`{e}s vite et invitons le lecteur \`{a} voir le premier
chapitre de \cite{Tanoudji} pour plus de d\'{e}tails.

\subsubsection{\'{E}quation de Schr\"{o}dinger d\'{e}pendante du temps}

Cette \'{e}quation d\'{e}crit comment la fonction d'onde se transforme au
cours du temps. Pour l'instant elle va vous para\^{\i}tre compliqu\'{e}e,
mais sa forme se rationalisera quand on sera un peu plus loin et qu'on aura
vu le Hamiltonien. On a donc une fonction $\psi (\vec{r})$ pour chaque
valeur du temps $t$. On d\'{e}finit donc une fonction $\psi (\vec{r},t)$.
Attention, cette fonction n'est en aucun cas reli\'{e}e \`{a} une densit\'{e}
de probabilit\'{e} de mesure d'un temps: $\left\vert \psi (\vec{r}%
,t)\right\vert ^{2}dt$ (noter l'\'{e}l\'{e}ment diff\'{e}rentiel $dt$) n'a
aucune signification physique! $t$ est ici un param\`{e}tre et non un r\'{e}%
sultat de mesure. L'\'{e}quation s'\'{e}crit 
\begin{equation}
i\hbar \frac{\partial }{\partial t}\Psi \left( \vec{r},t\right) =\left[ 
\frac{-\hbar ^{2}}{2m}\vec{\Delta}+V(\vec{r})\right] \Psi \left( \vec{r}%
,t\right) ,  \label{ch.1.8}
\end{equation}%
\qquad o\`{u} $\hbar =h/2\pi $, $\vec{\Delta}$ est le laplacien et $V(\vec{r}%
)$ est l'\'{e}nergie potentielle de la particule (qui pourrait d\'{e}pendre
aussi du temps d'ailleurs).

\subsubsection{Principe de superposition}

On peut faire une constatation tr\`{e}s importante en regardant l'\'{e}%
quation de Schr\"{o}dinger, c'est qu'elle est lin\'{e}aire. Cette
constatation anodine est \`{a} la base de m\'{e}thodes puissantes de r\'{e}%
solution de l'\'{e}quation. D'abord qu'est-ce que \c{c}a veut dire lin\'{e}%
aire ? Que si $\Psi _{1}\left( \vec{r},t\right) $ et \ $\Psi _{2}\left( \vec{%
r},t\right) $ sont deux solutions de l'\'{e}quation (\ref{ch.1.8}), alors $%
\alpha _{1}\Psi _{1}\left( \vec{r},t\right) +\alpha _{2}\Psi _{2}\left( \vec{%
r},t\right) $ est aussi solution de l'\'{e}quation ( $\alpha _{1}$ et $%
\alpha _{2}$ peuvent \^{e}tre deux constantes complexes quelconques). C'est
ce que l'on appelle le principe de superposition. Ceci permet d'essayer d'%
\'{e}crire la solution g\'{e}n\'{e}rale sous la forme d'une combinaison lin%
\'{e}aire de solutions ayant certaines propri\'{e}t\'{e}s qui simplifient l'%
\'{e}quation. En particulier, on peut \'{e}liminer le temps de l'\'{e}%
quation, et obtenir l'\'{e}quation de Schr\"{o}dinger ind\'{e}pendante du
temps (ou stationnaire) qui fait intervenir l'\'{e}nergie $E$ de la
particule :

\begin{equation}
\left[ \frac{-\hbar ^{2}}{2m}\vec{\Delta}+V(\vec{r})\right] \Psi \left( \vec{%
r},t\right) =E\Psi \left( \vec{r},t\right) .  \label{ch1.9}
\end{equation}

Il se trouve que, dans beaucoup de cas, cette \'{e}quation n'a de solution
physiquement acceptable (c.-\`{a}-d. normalisable) que pour un ensemble
discret de valeurs de $E$. Toutes les valeurs de $E$ ne sont pas n\'{e}%
cessairement autoris\'{e}es : c'est la quantification de l'\'{e}nergie.

\subsection{\protect\bigskip Outils math\'{e}matiques de la m\'{e}canique
quantique}

Les notions telle que: dualit\'{e} onde-corpuscule, fonction d'onde,
probabilit\'{e} de pr\'{e}sence d'une particule, incertitudes sur la mesure
des grandeurs physiques et la quantification des grandeurs physiques (l'\'{e}%
nergie); montrent l'importance jou\'{e}e par "la fonction d'onde" en
physique quantique et il est donc n\'{e}cessaire d'\'{e}tudier les \textbf{%
propri\'{e}t\'{e}s math\'{e}matiques }de l'espace des fonctions d'ondes et
des op\'{e}rateurs agissant sur ces fonctions \`{a} l'int\'{e}rieur de cet
espace. Dans cette partie, nous introduisons les propri\'{e}t\'{e}s
essentielles des espaces de Hilbert, c'est-\`{a}-dire les espaces vectoriels
munis d'un produit scalaire d\'{e}fini positif, en nous limitant au cas de
la dimension finie.

Il est impossible de pr\'{e}senter ici un formalisme complet et rigoureux,
mais donner des divers notions utiles en m\'{e}canique quantique telles que
la notion de repr\'{e}sentation, notion de Dirac, et l'alg\`{e}bre des op%
\'{e}rateurs \cite{belac,Tanoudji,Messiah,Jean,Zettili,Franz,Habib,land}.

\subsubsection{Espace des \'{e}tats et notations de Dirac}

\paragraph{ Espace des \'{e}tats}

On consid\`{e}re un syst\`{e}me physique dont l'ensemble des configurations $%
\mathcal{C}$ est d\'{e}crit par des variables $q_{i}$ ($i$ $=1...N$). La
fonction d'onde est une fonction $\Psi \left( q\right) $ qui associe une
valeur complexe \`{a} chaque configuration (on repr\'{e}sente une
configuration par la notation $q$, qui repr\'{e}sente en fait l'ensemble des
variables $q_{i}$). La fonction d'onde a la propri\'{e}t\'{e} que le carr%
\'{e} de son module est la densit\'{e} de probabilit\'{e} que le syst\`{e}me
soit dans la configuration $q$. En notant $dq$ un \'{e}l\'{e}ment de volume
dans l'espace de configuration (il est en g\'{e}n\'{e}ral de la forme $%
dq=A(q)\Pi _{i}$ $dq_{i}$), la densit\'{e} de probabilit\'{e} est donc $%
\left\vert \Psi (q)\right\vert ^{2}dq$. Ainsi, on doit avoir : 
\begin{equation*}
\int_{\mathcal{C}}\left\vert \Psi (q)\right\vert ^{2}dq=1.
\end{equation*}%
Math\'{e}matiquement, on dit que la fonction $\Psi (q)$ est de carr\'{e}
sommable. L'ensemble des fonctions de carr\'{e} sommable sur $\mathcal{C}$
est un espace vectoriel not\'{e} $L^{2}(\mathcal{C})$ en math\'{e}matiques.
C'est un espace de dimension infinie. Sa structure est celle d'un espace de
Hilbert $\mathcal{H}$. En fait, l'ensemble des fonctions de carr\'{e}
sommable est trop grand. Il contient en particulier des fonctions
discontinues, ce qui n'a aucun sens physiquement. On peut donc se
restreindre \`{a} des fonctions continues, voire d\'{e}rivables, voire ind%
\'{e}finiment d\'{e}rivable, voire analytiques. Nous dirons donc juste que
l'espace des fonctions d'onde de notre syst\`{e}me est un sous-espace de
l'espace de Hilbert $L^{2}(\mathcal{C})$), que nous appellerons espace des
fonctions d'onde et que nous noterons $\mathcal{F}$. Comme vu pr\'{e}c\'{e}%
demment, il est possible de repr\'{e}senter notre fonction d'onde dans diff%
\'{e}rentes bases, si bien que notre fonction d'onde aura diverses repr\'{e}%
sentations possibles, la forme $\Psi (q)$ n'\'{e}tant qu'une forme de repr%
\'{e}sentation parmi plein d'autres. On est donc conduit \`{a} envisager un
espace abstrait (notons le $\mathcal{E}$), sous-espace d'un espace de
Hilbert abstrait, dans lequel notre particule sera caract\'{e}ris\'{e}e par
un vecteur d'\'{e}tat, dont une des repr\'{e}sentations possibles sera d'%
\^{e}tre une fonction sur l'espace de configuration $\mathcal{C}$, ce qui
fait que l'espace $\mathcal{E}$ est isomorphe \`{a} l'espace $\mathcal{F}$.
L'avantage de proc\'{e}der ainsi est que tout ce que nous dirons par la
suite est applicable \`{a} n'importe quel espace abstrait, m\^{e}me si le
vecteur d'\'{e}tat ne peut pas \^{e}tre repr\'{e}sent\'{e} par une fonction,
par exemple si l'espace $\mathcal{E}$ est de dimension finie. Il se trouve
que des vecteurs d'\'{e}tats dans des espaces de dimension finie peuvent
exister en m\'{e}canique quantique, et sont m\^{e}me n\'{e}cessaires pour
expliquer des propri\'{e}t\'{e}s comme le spin (moment cin\'{e}tique intrins%
\`{e}que) des particules.

\subsubsection{Notations de Dirac}

Les notations de Dirac sont une mani\`{e}re de repr\'{e}senter les vecteurs
de l'espace $\mathcal{H}$, les formes lin\'{e}aires sur cet espace, ainsi
que le produit scalaire.

\textbf{Ket} : On d\'{e}signera par $\left\vert \psi \right\rangle $ un
vecteur de notre espace abstrait (on utilise la m\^{e}me lettre pour se
souvenir qu'il a comme repr\'{e}sentant $\psi (q)$ dans $\mathcal{F}$). On
appelle ce vecteur ket. Les combinaisons lin\'{e}aires de kets sont aussi
des kets. On note: $\alpha _{1}\left\vert \psi _{1}\right\rangle +\alpha
_{2}\left\vert \psi _{2}\right\rangle =\left\vert \alpha _{1}\psi
_{1}+\alpha _{2}\psi _{2}\right\rangle $ la combinaison lin\'{e}aire de deux
kets.

\textbf{Bra} : Un bra est une fonction lin\'{e}aire de l'espace $\mathcal{H}$
\`{a} valeurs dans $\mathcal{C}$, ce qu'on appelle aussi une forme lin\'{e}%
aire. On la note $\left\langle {}\right\vert $, par exemple $\left\langle
\chi \right\vert $. Pour l'instant l'image d'un ket

$\left\vert \psi \right\rangle $ par le bra $\left\langle \chi \right\vert $
est not\'{e}e $\left\langle \chi \right\vert (\left\vert \psi \right\rangle
)\equiv \left\langle \chi \right\vert \psi \rangle $. La lin\'{e}arit\'{e}
signifie que:

\begin{equation}
\left\langle \chi \right\vert (\alpha _{1}\left\vert \psi _{1}\right\rangle
+\alpha _{2}\left\vert \psi _{2}\right\rangle )=\alpha _{1}\left\langle \chi
\right\vert \psi _{1}\rangle +\alpha _{2}\left\langle \chi \right\vert \psi
_{2}\rangle .
\end{equation}

On peut bien s\^{u}r faire des combinaisons lin\'{e}aires de bras de type 
\begin{equation}
(\alpha _{1}\left\langle \chi _{1}\right\vert +\alpha _{2}\left\langle \chi
_{2}\right\vert )\left\vert \psi \right\rangle =\alpha _{1}\left\langle \chi
_{1}\right\vert \psi \rangle +\alpha _{2}\left\langle \chi _{2}\right\vert
\psi \rangle ,
\end{equation}%
pour tout ket $\left\vert \psi \right\rangle $ de l'espace $\mathcal{E}.$
L'ensemble des formes lin\'{e}aires (des bras) sur l'espace $\mathcal{E}$
est appel\'{e} l'espace dual de $\mathcal{E}$ et est not\'{e} $\mathcal{E}$ $%
^{\ast }$. Quand $\mathcal{E}$ est de dimension finie, $\mathcal{E}$ $^{\ast
}$ est de m\^{e}me dimension, et on peut donc trouver un isomorphisme entre
ces deux espaces. Ce n'est pas n\'{e}cessairement le cas en dimension
infinie. L'espace $\mathcal{E}$ est muni d'un produit scalaire d\'{e}fini
positif, ce qui est en fait un espace de Hilbert.

\textbf{Produit scalaire, norme}: Le produit scalaire de deux vecteurs $%
\left\vert \chi \right\rangle $ et $\left\vert \psi \right\rangle $ est not%
\'{e} $\left\langle \chi \right\vert \psi \rangle $. Il est lin\'{e}aire par
rapport \`{a} $\left\vert \psi \right\rangle $, c'est \`{a} dire que $%
\left\langle \chi \right\vert (\alpha _{1}\left\vert \psi _{1}\right\rangle
+\alpha _{2}\left\vert \psi _{2}\right\rangle )=\alpha _{1}\left\langle \chi
\right\vert \psi _{1}\rangle +\alpha _{2}\left\langle \chi \right\vert \psi
_{2}\rangle $ et v\'{e}rifie la propri\'{e}t\'{e} de conjugaison complexe 
\begin{equation}
\left\langle \chi \right\vert \psi \rangle =\left\langle \psi \right\vert
\chi \rangle ^{\ast },
\end{equation}%
ce qui implique que $\left\langle \psi \right\vert \psi \rangle $ est un
nombre r\'{e}el. Il est anti-lin\'{e}aire par rapport \`{a} $\left\vert \chi
\right\rangle $, c'est \`{a} dire que $(\alpha _{1}\left\langle \chi
_{1}\right\vert +\alpha _{2}\left\langle \chi _{2}\right\vert )\left\vert
\psi \right\rangle =\alpha _{1}^{\ast }\left\langle \chi _{1}\right\vert
\psi \rangle +\alpha _{2}^{\ast }\left\langle \chi _{2}\right\vert \psi
\rangle $ . Enfin, le produit scalaire est d\'{e}fini positif : $%
\left\langle \chi \right\vert \psi \rangle \geqslant 0$ , $\left\langle \psi
\right\vert \psi \rangle =0\Longleftrightarrow \left\vert \psi \right\rangle
=0.$

Il sera commode de choisir dans $\mathcal{E}$ une base orthonorm\'{e}e de $N$
vecteurs \{$\left\vert 1\right\rangle $, $\left\vert 2\right\rangle $ , . .
. , $\left\vert n\right\rangle $ , . . . , $\left\vert N\right\rangle $\} 
\begin{equation}
\left\langle n\right\vert m\rangle =\delta _{nm}.
\end{equation}%
Tout vecteur $\left\vert \psi \right\rangle $ peut se d\'{e}composer sur
cette base avec des coefficients $c_{n}$ qui sont les composantes de $%
\left\vert \psi \right\rangle $ dans cette base

\begin{equation}
\left| \psi \right\rangle = \sum_{n=1}^{N} c_{n} \left| n \right\rangle .
\end{equation}
Prenant le produit scalaire de cette derni\`{e}re \'{e}quation avec le
vecteur de base $\left\vert m\right\rangle $ , on trouve pour $c_{m}$

\begin{equation}
c_{m}=\left\langle m\right\vert \psi \rangle .
\end{equation}%
Si un vecteur $\left| \chi \right\rangle$ se d\'{e}compose sur cette m\^{e}me base 
suivant $\left| \chi \right\rangle = \sum_{n} d_{n}\left| n \right\rangle$, 
alors le produit scalaire $\left\langle \chi \middle| \psi \right\rangle$ s'\'{e}crit,
\begin{equation}
\left\langle \chi \right\vert \psi \rangle
=\sum\limits_{n=1}^{N}d_{n}^{\ast }c_{n}.
\end{equation}%
La norme de $\left\vert \psi \right\rangle $, not\'{e}e $\left\Vert \psi
\right\Vert $, est d\'{e}finie \`{a} partir du produit scalaire

\begin{equation}
\left\Vert \psi \right\Vert ^{2}=\left\langle \psi \right\vert \psi \rangle
=\sum\limits_{n=1}^{N}\left\vert c_{n}\right\vert ^{2}\geqslant 0.
\end{equation}%
Une propri\'{e}t\'{e} importante du produit scalaire est l'in\'{e}galit\'{e}
de Schwarz

\begin{equation}
\left\vert \left\langle \chi \right\vert \psi \rangle \right\vert
^{2}\leqslant \left\langle \chi \right\vert \chi \rangle \left\langle \psi
\right\vert \psi \rangle =\left\Vert \chi \right\Vert ^{2}\left\Vert \psi
\right\Vert ^{2}.
\end{equation}%
L'\'{e}galit\'{e} est vraie si et seulement si $\left\vert \psi
\right\rangle $ et $\left\vert \chi \right\rangle $ sont proportionnels: $%
\left\vert \chi \right\rangle =$ $\lambda \left\vert \psi \right\rangle $.

\subsubsection{Op\'{e}rateurs lin\'{e}aires sur $\mathcal{E}$}

\paragraph{ Op\'{e}rateurs lin\'{e}aires, Hermitiques, unitaires}

Un op\'{e}rateur lin\'{e}aire $A$ fait correspondre au vecteur $\left\vert
\psi \right\rangle $ un vecteur $\left\vert A\psi \right\rangle $ v\'{e}%
rifiant la propri\'{e}t\'{e} de lin\'{e}arit\'{e} 
\begin{equation}
\left\vert A(\alpha \psi +\lambda \chi )\right\rangle =\alpha \left\vert
A\psi \right\rangle +\lambda \left\vert A\chi \right\rangle .
\end{equation}%
Dans une base d\'{e}termin\'{e}e, cet op\'{e}rateur est repr\'{e}sent\'{e}
par une matrice d'\'{e}l\'{e}ments $A_{mn}$. En effet gr\^{a}ce \`{a} la lin%
\'{e}arit\'{e} et en utilisant la d\'{e}composition

\begin{equation}
\left\vert A\psi \right\rangle =\sum\limits_{n=1}^{N}c_{n}\left\vert
An\right\rangle ,
\end{equation}%
on obtient les composantes $d_{m}$ de $\left\vert A\psi \right\rangle $ $%
=\sum_{m}$ $d_{m}\left\vert m\right\rangle $

\begin{equation}
d_{m}=\left\langle m\right\vert A\psi \rangle
=\sum\limits_{n=1}^{N}c_{n}\left\langle m\right\vert An\rangle
=\sum\limits_{n=1}^{N}c_{n}A_{mn}.
\end{equation}%
L'\'{e}l\'{e}ment de matrice $A_{mn}$ est donc

\begin{equation*}
A_{mn}=\left\langle m\right\vert An\rangle .
\end{equation*}%
L'op\'{e}rateur conjugu\'{e} Hermitique (ou adjoint) $A^{+}$ de $A$ est d%
\'{e}fini par

\begin{equation}
\left\langle \chi \right\vert A\psi \rangle =\left\langle A\chi \right\vert
\psi \rangle =\left\langle \psi \right\vert A\chi \rangle ^{\ast },
\end{equation}%
pour tout couple de vecteurs $\left\vert \psi \right\rangle $, $\left\vert
\chi \right\rangle $. On montre facilement que $A^{+}$ est bien un op\'{e}%
rateur lin\'{e}aire. Ses \'{e}l\'{e}ments de matrice dans la base \{$%
\left\vert 1\right\rangle $, $\left\vert 2\right\rangle $, . . . , $%
\left\vert n\right\rangle $, . . . , $\left\vert N\right\rangle $ \} sont
obtenus en prenant pour $\left\vert \psi \right\rangle $ et $\left\vert \chi
\right\rangle $ les vecteurs de base et $(A^{+})_{mn}$ v\'{e}rifie%
\begin{equation}
(A^{+})_{mn}=A^{\ast }{}_{nm}.
\end{equation}%
Le conjugu\'{e} Hermitique du produit $AB$ de deux op\'{e}rateurs est $%
B^{+}A^{+}$ ; en effet%
\begin{equation}
\left\langle \chi \right\vert \left( AB\right) ^{+}\psi \rangle
=\left\langle AB\chi \right\vert \psi \rangle =\left\langle B\chi
\right\vert A^{+}\psi \rangle =\left\langle \chi \right\vert B^{+}A^{+}\psi
\rangle .
\end{equation}%
Un op\'{e}rateur v\'{e}rifiant $A=A^{+}$ est appel\'{e} Hermitique, ou
auto-adjoint. Les deux termes sont \'{e}quivalents pour les espaces de
dimension finie, mais non dans le cas de la dimension infinie.

Un op\'{e}rateur tel que $UU^{+}=U^{+}U=I$, ou de fa\c{c}on \'{e}quivalente $%
U^{-1}$ $=U^{+}$, est appel\'{e} op\'{e}rateur unitaire, $I$ d\'{e}signe l'op%
\'{e}rateur identit\'{e} de l'espace de Hilbert. Dans un espace de dimension
finie, une condition n\'{e}cessaire et suffisante pour qu'un op\'{e}rateur $%
U $ soit unitaire est qu'il conserve la norme 
\begin{equation*}
\left\Vert U\psi \right\Vert ^{2}=\left\Vert \psi \right\Vert ^{2}\text{ \
ou \ }\left\langle U\psi \right\vert U\psi \rangle =\left\langle \psi
\right\vert \psi \rangle \text{ \ }\forall \text{ }\left\vert \psi
\right\rangle \in \mathcal{E}
\end{equation*}%
Les op\'{e}rateurs unitaires effectuent les changements de base orthonorm%
\'{e}e dans $\mathcal{E}$. Soit $\left\vert n^{\prime }\right\rangle
=\left\vert Un\right\rangle $ , alors

\begin{equation*}
\left\langle m^{\prime }\right\vert n^{\prime }\rangle =\left\langle
Um\right\vert Un\rangle =\left\langle m\right\vert n\rangle =\delta
_{mn}=\delta _{m^{\prime }n^{\prime }}
\end{equation*}%
et l'ensemble des vecteurs \{$\left\vert n^{\prime }\right\rangle $\} forme
une base orthonorm\'{e}e.

\paragraph{Relation de fermeture}

Pour une base orthonorm\'{e}e, on a la relation suivante pour tout ket $%
\left\vert \psi \right\rangle $:

\begin{equation}
\left\vert \psi \right\rangle =\sum\limits_{n=1}\left\vert n\right\rangle
\left\langle n\right\vert \psi \rangle ,
\end{equation}%
le symbole $\left\vert n\right\rangle \left\langle n\right\vert $ repr\'{e}%
sente ce que l'on appelle un projecteur $P_{n}.$ Dans un espace vectoriel de
dimension finie, il transforme un vecteur en sa projection orthogonale sur
le vecteur $\left\vert n\right\rangle $. C'est une application lin\'{e}aire
(on dit un op\'{e}rateur lin\'{e}aire dans le cas d'un espace de dimension
infinie) de $\mathcal{E}$ dans lui-m\^{e}me. Ce que dit \ cette derni\`{e}re 
\'{e}quation c'est que la somme de tous les projecteurs sur les vecteurs de
la base est l'op\'{e}rateur identit\'{e}. On \'{e}crit cela sous la forme : 
\begin{equation}
\sum\limits_{n}\left\vert n\right\rangle \left\langle n\right\vert
=\sum\limits_{n}P_{n}=I,
\end{equation}%
cette relation est appel\'{e}e relation de fermeture. Elle est souvent tr%
\`{e}s commode dans les calculs. Par exemple, elle redonne simplement la loi
de multiplication des matrices.

\paragraph{Op\'{e}rateurs unitaires et op\'{e}rateurs Hermitiques}

Les propri\'{e}t\'{e}s des op\'{e}rateurs unitaires $U^{+}=U^{-1}$ sont
intimement li\'{e}es \`{a} celles des op\'{e}rateurs Hermitiques, et en
particulier ils peuvent toujours \^{e}tre diagonalis\'{e}s.

Le th\'{e}or\`{e}me de base pour les op\'{e}rateurs unitaires s'\'{e}nonce
comme suit.

\textbf{Th\'{e}or\`{e}me}: a. Les valeurs propres $a_{n}$ d'un op\'{e}rateur
unitaire sont de module unit\'{e}: $a_{n}=\exp (i\alpha _{n}),\alpha _{n}$ r%
\'{e}el.

\ \ \ \ \ \ \ \ \ \ \ \ \ \ \ \ \ \ \ b. Les vecteurs propres correspondant 
\`{a} deux valeurs propres diff\'{e}rentes sont orthogonaux.

\ \ \ \ \ \ \ \ \ \ \ \ \ \ \ \ \ \ \ c. La d\'{e}composition spectrale d'un
op\'{e}rateur unitaire s'\'{e}crit en fonction de projecteurs $P_{n}$ sous
la forme%
\begin{equation}
U=\sum\limits_{n}a_{n}P_{n}=\sum\limits_{n}e^{i\alpha _{n}}P_{n}.
\end{equation}

\section{\protect\bigskip Postulats de la m\'{e}canique quantique}

Nous allons \'{e}noncer dans ce chapitre les postulats de base de la
physique quantique. Les postulats tels qu'ils sont \'{e}nonc\'{e}s dans ce
chapitre fixent le cadre conceptuel g\'{e}n\'{e}ral de la m\'{e}canique
quantique, et ne donnent pas directement les outils n\'{e}cessaires pour r%
\'{e}soudre des probl\`{e}mes sp\'{e}cifiques. Il est possible d'utiliser
d'autres ensembles de postulats : par exemple une autre approche de la m\'{e}%
canique quantique consiste \`{a} \'{e}noncer des postulats sur les int\'{e}%
grales de chemin.

\subsection{Les postulats de la m\'{e}canique quantique}

\subsubsection{Premier postulat I: la description de l'\'{e}tat quantique
d'un syst\`{e}me}

\textit{Un syst\`{e}me quantique est compl\`{e}tement d\'{e}crit \`{a}
chaque instant par un vecteur d'\'{e}tat norm\'{e}, not\'{e} }$\left\vert 
\text{ket}\right\rangle $, \textit{appartenant \`{a} un espace de Hilbert}.

Cela veut dire qu'un syst\`{e}me quantique se d\'{e}crit dans un espace math%
\'{e}matique particulier, complexe. Il existe un vecteur de cet espace qui
contient toute l'information sur le syst\`{e}me.

Le fait qu'un \'{e}tat physique soit repr\'{e}sent\'{e} par un vecteur
implique sous certaines conditions le principe de superposition, caract\'{e}%
ristique de la lin\'{e}arit\'{e} de la th\'{e}orie : si nous avons deux
vecteurs appartenant \`{a} l'espace de Hilbert, alors leur somme appartient
aussi \`{a} l'espace de Hilbert (comme la somme de deux vecteurs de l'espace
donne toujours un vecteur de l'espace).

\subsubsection{Deuxi\`{e}me postulat II: l'observable}

\textit{Toute grandeur physique mesurable} $\mathcal{A}$ \textit{est associ%
\'{e}e \`{a} un op\'{e}rateur lin\'{e}aire Hermitique }$A$ \textit{appel\'{e}
\guillemotleft\ observable \guillemotright }.

Une grandeur physique correspond \`{a} une quantit\'{e} que l'on peut
mesurer. La position, la vitesse, l'\'{e}nergie par exemple. Comme le dit le
principe, chaque grandeur que l'on peut mesurer est associ\'{e}e \`{a} un op%
\'{e}rateur qui agit sur le vecteur d'\'{e}tat (un op\'{e}rateur c'est comme
une fonction). On note souvent une fonction r\'{e}elle $f(x)$ o\`{u} $x$ est
la variable r\'{e}elle du probl\`{e}me. La fonction associe \`{a} chaque
nombre x un autre nombre $f(x)$. Ici il s'agit d'un raisonnement analogue :
chaque observable transforme un vecteur d'\'{e}tat en un autre vecteur d'%
\'{e}tat de l'espace hilbertien (lin\'{e}arit\'{e}).

En r\'{e}sum\'{e} et \`{a} ce stade, un syst\`{e}me est enti\`{e}rement d%
\'{e}crit par un vecteur d'\'{e}tat, et \`{a} chaque grandeur physique on
associe un op\'{e}rateur qui agit sur le vecteur d'\'{e}tat.

\subsubsection{Troisi\`{e}me postulat III: le r\'{e}sultat d'une mesure}

\textit{Les r\'{e}sultats possibles de la mesure d'une grandeur physique } $%
\mathcal{A}$ \textit{sont les valeurs propres de l'observable associ\'{e}e.}

Mais quel est la diff\'{e}rence avec ce qui pr\'{e}c\`{e}de? En fait, ce
qu'on vient de d\'{e}terminer pr\'{e}c\'{e}demment, c'est comment on associe
une grandeur physique \`{a} la description d'un syst\`{e}me. On applique au
vecteur d'\'{e}tat un op\'{e}rateur, ce qui nous donne un nouveau vecteur d'%
\'{e}tat. Mais d\'{e}sormais, ce principe nous donne ce qui se passe quand
on fait exp\'{e}rimentalement la mesure de cette grandeur physique. Ce qui
est dit ici, c'est qu'alors plusieurs r\'{e}sultats sont possibles! C'est l%
\`{a} qu'intervient le probabilisme de la m\'{e}canique quantique.

Ces valeurs propres correspondent aux r\'{e}sultats possibles de la mesure
physique (ce sont toujours des valeurs r\'{e}elles et non pas complexes, du
fait des propri\'{e}t\'{e}s Hermitiennes de l'observable). Et -- pr\'{e}%
cision cruciale -- dans la mesure o\`{u} il y a un nombre fini (ou plut\^{o}%
t d\'{e}nombrable) de valeurs propres, cela signifie que les r\'{e}sultats
de la mesure sont quantifi\'{e}s. S'il y a un nombre fini de valeurs propres
alors il y a un nombre fini de r\'{e}sultats de mesures possibles! C'est
notamment le cas pour l'\'{e}nergie tr\`{e}s souvent (lorsque le syst\`{e}me
est soumis \`{a} des conditions aux limites par exemple), si bien qu'on voit
alors \'{e}merger le concept de quantification de l'\'{e}nergie.

Ainsi, pour chaque grandeur physique correspondent des valeurs propres qui
sont les r\'{e}sultats possibles de la mesure. Chaque valeur propre a une
probabilit\'{e} particuli\`{e}re d'\^{e}tre mesur\'{e}e. C'est tr\`{e}s
puissant, parce que cela signifie que si on conna\^{\i}t la mani\`{e}re dont
s'\'{e}crit l'observable, on peut gr\^{a}ce \`{a} une simple op\'{e}ration
math\'{e}matique d\'{e}terminer les r\'{e}sultats possibles de la mesure de
la grandeur physique associ\'{e}e.

\subsubsection{Quatri\`{e}me postulat IV: la projection de mesure}

\textit{Apr\`{e}s une mesure, le syst\`{e}me se trouve projet\'{e} dans l'%
\'{e}tat propre correspondant au r\'{e}sultat de la mesure.}

Cela veut dire qu'\`{a} chaque valeur propre est associ\'{e} ce qu'on
appelle un \'{e}tat propre, qui correspond ici au vecteur d'\'{e}tat dans
lequel a \'{e}t\'{e} \guillemotleft\ projet\'{e} \guillemotright\ le vecteur
d'\'{e}tat initial lors de la mesure. On peut dire qu'en quelque sorte dans
la base de ces \'{e}tats propres (qui forment bien une base de l'espace
hilbertien) on n'a gard\'{e} que la composante du vecteur d'\'{e}tat suivant
l'\'{e}tat propre associ\'{e} \`{a} la valeur propre mesur\'{e}e. En clair, 
\`{a} chaque r\'{e}sultat possible de la mesure, le vecteur d'\'{e}tat apr%
\`{e}s la mesure est un \'{e}tat bien d\'{e}termin\'{e}.

Pour d\'{e}terminer la probabilit\'{e} avec laquelle un r\'{e}sultat donn%
\'{e} va \^{e}tre mesur\'{e}, il faut conna\^{\i}tre le vecteur d'\'{e}tat
initial, le vecteur d'\'{e}tat projet\'{e} associ\'{e} \`{a} la mesure, en
faire le produit scalaire (on projette l'\'{e}tat initial sur l'\'{e}tat
mesur\'{e}) et en prendre le module, c'est-\`{a}-dire le mettre au carr\'{e}
: $\mathcal{P}_{\text{mesurer }a}$=$\left\vert \left\langle \varphi
\right\vert \psi _{a}\rangle \right\vert ^{2}$, o\`{u} $\left\vert \varphi
\right\rangle $ est l'\'{e}tat initial et $\left\vert \psi _{a}\right\rangle 
$ le vecteur propre associ\'{e} \`{a} la valeur propre $a$. On conna\^{\i}t
ainsi la probabilit\'{e} de mesurer chaque r\'{e}sultat possible. On peut
alors calculer des probabilit\'{e}s de mesure \`{a} partir du vecteur d'\'{e}%
tat, et faire des v\'{e}rifications exp\'{e}rimentales des r\'{e}sultats pr%
\'{e}dits par la th\'{e}orie. Et \c{c}a marche redoutablement bien!

Nous voici arriv\'{e}s \`{a} l'un des aspects les plus probl\'{e}matiques de
la th\'{e}orie quantique. Il nous faut maintenant expliquer ce qu'il advient
\ du vecteur d'\'{e}tat quand on effectue une mesure. Ce r\'{e}sultat est pr%
\'{e}sent\'{e} dans la plupart des manuels comme un postulat suppl\'{e}%
mentaire de la m\'{e}canique quantique, le \textquotedblleft postulat de r%
\'{e}duction du paquet d'ondes\textquotedblright\ (RPO),

\subsubsection{Cinqui\`{e}me Postulat RPO V: r\'{e}duction du paquet d'ondes}

Si le syst\`{e}me \'{e}tait initialement dans l'\'{e}tat $\left\vert \varphi
\right\rangle $, et si le r\'{e}sultat de la mesure de $\mathcal{A}$ est $%
a_{n}$, alors imm\'{e}diatement apr\`{e}s la mesure, le syst\`{e}me se
trouve dans l'\'{e}tat projet\'{e} sur le sous-espace de la valeur propre $%
a_{n}$

\begin{equation}
\left\vert \varphi \right\rangle \rightarrow \left\vert \psi \right\rangle =%
\frac{P_{n}\left\vert \varphi \right\rangle }{\left\langle \varphi
\left\vert P_{n}\right\vert \varphi \right\rangle ^{1/2}}.
\end{equation}%
Le vecteur $\left\vert \psi \right\rangle $ est bien normalis\'{e} \`{a}
l'unit\'{e} car $\left\Vert P_{n}\left\vert \varphi \right\rangle
\right\Vert ^{2}=\left\langle \varphi \left\vert P_{n}^{+}P_{n}\right\vert
\varphi \right\rangle =\left\langle \varphi \left\vert P_{n}\right\vert
\varphi \right\rangle $ compte tenu des propri\'{e}t\'{e}s des projecteurs $%
P_{n}$. L'\'{e}nonc\'{e} de ce \textquotedblleft pseudo\textquotedblright\
postulat appelle quelques remarques. Il ne faut surtout pas imaginer que
cette\ derni\`{e}re transformation correspond \`{a} un processus physique r%
\'{e}el. Elle n'a de sens que si l'on s'int\'{e}resse \`{a} une \'{e}%
volution effective, qui fait abstraction de l'appareil de mesure pour se
focaliser uniquement sur le syst\`{e}me. En fait, c'est une simple commodit%
\'{e} d'\'{e}criture dans l'espace de Hilbert des \'{e}tats du syst\`{e}me,
qui isole artificiellement le syst\`{e}me de l'appareil de mesure et de son
environnement.

Jusqu'\`{a} pr\'{e}sent, nous avons consid\'{e}r\'{e} le syst\`{e}me
physique \`{a} un instant donn\'{e}, ou pendant l'intervalle de temps suppos%
\'{e} infiniment court d'une mesure. Nous allons maintenant prendre en consid%
\'{e}ration l'\'{e}volution temporelle du vecteur d'\'{e}tat, auquel nous
donnerons une d\'{e}pendance explicite $\left\vert \varphi (t)\right\rangle $
par rapport au temps $t$.

\subsubsection{Sixi\`{e}me postulat VI: l'\'{e}volution temporelle d'un syst%
\`{e}me quantique}

\textit{L'\'{e}volution temporelle du syst\`{e}me quantique est donn\'{e}e
par l'\'{e}quation de Schr\"{o}dinger :}

\begin{equation}
i\hbar \frac{\partial }{\partial t}\left\vert \Psi (t)\right\rangle
=H\left\vert \Psi (t)\right\rangle .  \label{ch2.1}
\end{equation}

On a jusqu'ici d\'{e}crit l'effet d'observables physiques sur un vecteur d'%
\'{e}tat, mais on s'int\'{e}resse maintenant \`{a} son \'{e}volution dans le
temps. Il s'agit en fait ici d'une \'{e}quation analogue au principe
fondamental de la dynamique (la troisi\`{e}me loi de Newton), mais en
beaucoup plus g\'{e}n\'{e}rale. $\left\vert \Psi (t)\right\rangle $ est le
vecteur d'\'{e}tat, qui comprend toutes les informations sur le syst\`{e}me. 
$H$ est une observable qui comprend l'\'{e}nergie du syst\`{e}me. On
l'appelle le Hamiltonien. Cette observable prend comme entr\'{e}e le vecteur
d'\'{e}tat. Ainsi on retrouve une relation entre la d\'{e}riv\'{e}e
temporelle du vecteur d'\'{e}tat, et le vecteur d'\'{e}tat lui-m\^{e}me : il
s'agit d'une \'{e}quation diff\'{e}rentielle, qui permet de d\'{e}terminer l'%
\'{e}volution du vecteur d'\'{e}tat au court du temps. Ce r\'{e}sultat d\'{e}%
pend des diff\'{e}rentes \'{e}nergies pr\'{e}sentes dans le syst\`{e}me.

Notons que $\hbar $ est une constante appel\'{e}e la constante de Planck r%
\'{e}duite, et que $i$ est le nombre complexe. Il est tout \`{a} faire
remarquable que cette \'{e}quation, cens\'{e}e d\'{e}crire le comportement
de syst\`{e}mes physiques, soit complexe. Mais comme on l'a vu, le vecteur d'%
\'{e}tat est complexe, et ce n'est qu'en repassant aux probabilit\'{e}s
qu'on retrouve des r\'{e}sultats physiques, et donc r\'{e}els.

Bon alors \c{c}a para\^{\i}t ensuite plut\^{o}t facile, il suffit de r\'{e}%
soudre cette \'{e}quation et on peut conna\^{\i}tre le comportement de tout
syst\`{e}me quantique ! Oui sauf que non, car tr\`{e}s souvent le
Hamiltonien est tr\`{e}s difficile \`{a} \'{e}crire ce qui rend l'\'{e}%
quation parfois impossible \`{a} r\'{e}soudre analytiquement (notamment
quand il y a beaucoup de particules). Il faut alors faire des simulations num%
\'{e}riques et/ou des approximations.

La (n\'{e}cessaire) conservation de la norme du vecteur d'\'{e}tat est assur%
\'{e}e par l'Hermiticit\'{e} de $H$. En effet,

\begin{equation}
\frac{d}{dt}\left\Vert \left\vert \Psi (t)\right\rangle \right\Vert ^{2}=%
\frac{d}{dt}\left\langle \Psi (t)\right\vert \Psi (t)\rangle =\frac{i}{\hbar 
}\left\langle \Psi (t)\right\vert H-H^{+}\left\vert \Psi (t)\right\rangle =0
\end{equation}%
car $H=H^{+}$. La conservation de la norme de $\left\vert \Psi
(t)\right\rangle $ implique, dans un espace de Hilbert de dimension finie,
que l'\'{e}volution $\left\vert \Psi (0)\right\rangle \rightarrow $ $%
\left\vert \Psi (t)\right\rangle $est unitaire : on parle alors d'\'{e}%
volution Hamiltonienne, ou d'\'{e}volution unitaire. Nous avons donn\'{e} en
(\ref{ch2.1}) l'\'{e}quation d'\'{e}volution sous forme diff\'{e}rentielle.
Il existe une formulation int\'{e}grale de cette \'{e}quation qui fait
intervenir l'op\'{e}rateur d'\'{e}volution $U(t,t_{0})$. Dans cette
formulation, le postulat VI devient :

\subsubsection{Sixi\`{e}me postulat VI': Op\'{e}rateur d'\'{e}volution}

Le vecteur d'\'{e}tat $\left\vert \Psi (t)\right\rangle $ au temps $t$ se d%
\'{e}duit du vecteur d'\'{e}tat $\left\vert \Psi (t_{0})\right\rangle $ au
temps $t_{0}$ par application d'un op\'{e}rateur unitaire $U(t,t_{0})$, appel%
\'{e} op\'{e}rateur d'\'{e}volution%
\begin{equation}
\left\vert \Psi (t)\right\rangle =U(t,t_{0})\left\vert \Psi
(t_{0})\right\rangle .  \label{ch2.2}
\end{equation}%
L'unitarit\'{e} de $U$: $U^{+}U$ $=UU^{+}$ $=I,$ assure la conservation de
la norme $\langle \Psi (t)\left\vert \Psi (t)\right\rangle =\left\langle
\Psi (t_{0})\right\vert U^{+}(t,t_{0})U(t,t_{0})\left\vert \Psi
(t_{0})\right\rangle =\langle \Psi (t_{0})\left\vert \Psi
(t_{0})\right\rangle =I.$ Inversement on aurait pu partir de la conservation
de la norme pour montrer que $U^{+}U$ $=I$. L'op\'{e}rateur d'\'{e}volution
ob\'{e}it aussi \`{a} la propri\'{e}t\'{e} de groupe

\begin{equation}
U(t,t_{1})U(t_{1},t_{0})=U(t,t_{0}).\text{ \ \ \ \ \ \ \ }t_{0}\leq
t_{1}\leq t
\end{equation}%
En effet, il est \'{e}quivalent d'aller directement de $t_{0}$ \`{a} $t$, ou
d'aller d'abord de $t_{0}$ \`{a} $t_{1}$ et ensuite de $t_{1}$ \`{a} $t.$
Les postulats d'\'{e}volution temporelle IV et IV' ne sont bien s\^{u}r pas
ind\'{e}pendants. En effet, il est facile \`{a} partir de (\ref{ch2.1}) d'%
\'{e}crire une \'{e}quation diff\'{e}rentielle pour $U(t,t_{0})$. En diff%
\'{e}rentiant (\ref{ch2.2}) par rapport au temps et en comparant avec (\ref%
{ch2.1}), on en d\'{e}duit une \'{e}quation diff\'{e}rentielle pour $%
U(t,t_{0})$

\begin{equation}
\frac{\partial }{\partial t}U(t,t_{0})=-\frac{i}{\hbar }H(t)U(t,t_{0}).
\label{ch2.3}
\end{equation}%
Il est donc ais\'{e} de passer de la formulation int\'{e}grale (\ref{ch2.2}) 
\`{a} la formulation diff\'{e}rentielle (\ref{ch2.1}). Le passage inverse
est plus compliqu\'{e}: en effet, si $H(t)$ \'{e}tait un nombre, l'\'{e}%
quation (\ref{ch2.3}) s'int\`{e}grerait imm\'{e}diatement. Mais $H(t)$ est
un op\'{e}rateur et en g\'{e}n\'{e}ral

\begin{equation}
U(t,t_{0}) \neq \exp \left( -\frac{i}{\hbar} \int_{t_{0}}^{t} H(t^{\prime}) \, dt^{\prime} \right)
\end{equation}
parce qu'il n'y a aucune raison pour que $[H(t\prime ),H(t\prime \prime )]=0$%
. Cependant il existe une formule g\'{e}n\'{e}rale \cite{Messiah}, pour
calculer $U(t,t_{0})$ \`{a} partir de $H(t)$, et les postulats IV et IV'
sont strictement \'{e}quivalents.

Naturellement, il peut parfaitement arriver que le Hamiltonien soit ind\'{e}%
pendant du temps, m\^{e}me pour un syst\`{e}me non isol\'{e}. Lorsque le
Hamiltonien est ind\'{e}pendant du temps, l'\'{e}quation diff\'{e}rentielle (%
\ref{ch2.3}) s'int\`{e}gre sans probl\`{e}me et 
\begin{equation}
U(t,t_{0})=\exp \left( -\frac{i}{\hbar }H(t-t_{0})\right)
\end{equation}%
qui ne d\'{e}pend que de $(t-t_{0})$. L'op\'{e}rateur $U(t-t_{0})$ est
obtenu par exponentiation de l'op\'{e}rateur Hermitique $H$ ; $U(t-t_{0})$
effectue une translation de temps de $(t-t_{0})$ sur le vecteur d'\'{e}tat,
et si $(t-t_{0})$ devient infinit\'{e}simal

\begin{equation*}
U(t-t_{0})\simeq I-\frac{i(t-t_{0})}{\hbar }H
\end{equation*}%
Cette \'{e}quation s'interpr\`{e}te ainsi : $H$ est le g\'{e}n\'{e}rateur
infinit\'{e}simal des translations de temps, et, pour un syst\`{e}me isol%
\'{e}, la d\'{e}finition la plus g\'{e}n\'{e}rale du Hamiltonien est d'\^{e}%
tre pr\'{e}cis\'{e}ment ce g\'{e}n\'{e}rateur infinit\'{e}simal.

Maintenant, si l'Hamiltonien $H$ ne d\'{e}pend pas du temps, l'int\'{e}%
gration de (\ref{ch2.3}) (et on tenant compte la condition $U(t_{0},t_{0})=I$%
) donne

\begin{equation}
U(t,t_{0})=e^{-\frac{i}{\hbar }(t-t_{0})H},
\end{equation}

et%
\begin{equation*}
\left\vert \Psi (t)\right\rangle =e^{-\frac{i}{\hbar }(t-t_{0})H}\left\vert
\Psi (t_{0})\right\rangle .
\end{equation*}

En plus, il faut aussi que l'\'{e}volution des \'{e}tats propres dans le
temps d'un Hamiltonien doit \^{e}tre unitaire

\begin{eqnarray}
U(t,t_{0})U^{+}(t,t_{0}) &=&U(t,t_{0})U^{-1}(t,t_{0})  \notag \\
&=&e^{-\frac{i}{\hbar }(t-t_{0})H}e^{\frac{i}{\hbar }(t-t_{0})H}=I.
\end{eqnarray}

Cette condition garantie justement l'ind\'{e}pendance du probabilit\'{e} par
rapport au temps.

\begin{equation*}
\langle \Psi (t)\left\vert \Psi (t)\right\rangle =\left\langle \Psi
(t_{0})\right\vert U^{+}(t,t_{0})U(t,t_{0})\left\vert \Psi
(t_{0})\right\rangle =\langle \Psi (t_{0})\left\vert \Psi
(t_{0})\right\rangle .
\end{equation*}

Lorsque l'on examine l'\'{e}volution d'un syst\`{e}me ouvert, on rencontre
automatiquement une \'{e}volution non unitaire qui peut parfois \^{e}tre repr%
\'{e}sent\'{e}e par un \textbf{Hamiltonien non-Hermitique }(qu'on verra au
chapitre 3).

Par d\'{e}finition, un op\'{e}rateur antiunitaire est un op\'{e}rateur
anti-lin\'{e}aire inversible \ dont l'adjoint est \'{e}gal \`{a} l'inverse:%
\begin{equation*}
T\text{ est antiunitaire}\Leftrightarrow T\text{ est antilin\'{e}aire et }%
T^{+}=T^{-1}
\end{equation*}

Un op\'{e}rateur antiunitaire a la propri\'{e}t\'{e} de transformer le
produit scalaire en son complexe conjugu\'{e}:

\begin{equation*}
T\text{ \ antiunitaire}\Leftrightarrow \left( T\left\vert \psi \right\rangle
,T\left\vert \phi \right\rangle \right) =\left( \left\vert \psi
\right\rangle ,\left\vert \phi \right\rangle \right) ^{\ast },\text{ \ \ \ \
\ \ \ \ \ }\forall \text{\ \ }\left\vert \psi \right\rangle ,\text{\ }%
\left\vert \phi \right\rangle \text{\ }
\end{equation*}

\ On rencontrera un op\'{e}rateur anti-unitaire lors de l'\'{e}tude de
l'invariance par renversement du temps.

Voil\`{a}, on a donn\'{e} un aper\c{c}u sur les bases de la m\'{e}canique
quantique. Comme vous le voyez, c'est tr\`{e}s math\'{e}matis\'{e} et tout y
est parfaitement \'{e}tabli \`{a} travers la notion de vecteur d'\'{e}tat
agissant dans un espace complexe.

\chapter{\protect\bigskip Les sym\'{e}tries en Physique}

\section{Sym\'{e}tries en m\'{e}canique quantique}

Les sym\'{e}tries jouent un r\^{o}le extr\^{e}mement important en physique th%
\'{e}orique. Non seulement facilitent-elles beaucoup la solution de nombreux
probl\`{e}mes, mais elles sont aussi \`{a} la base des th\'{e}ories des
interactions fondamentales.

En physique on appelle sym\'{e}tries les transformations qui laissent
invariant un objet (sens g\'{e}om\'{e}trique du terme), mais aussi une loi (%
\'{e}quation de Newton par renversement du temps si la force ne d\'{e}pend
que de la position, par exemple) ou un observable en m\'{e}canique quantique
(op\'{e}rateur Hamiltonien, par exemple). Les sym\'{e}tries forment ce que
les math\'{e}maticiens appellent des groupes. C'est pour cela que l'\'{e}%
tude des sym\'{e}tries en physique est souvent appel\'{e}e th\'{e}orie des
groupes. Il y a deux grandes classes de sym\'{e}tries : les transformations
continues et les transformations discr\`{e}tes.

En m\'{e}canique classique les sym\'{e}tries sont associ\'{e}es \`{a} des
lois de conservation (l'invariance par translation donne la conservation de
l'impulsion totale, l'invariance par rotation la conservation du moment cin%
\'{e}tique, etc...).

Les propri\'{e}t\'{e}s de sym\'{e}trie jouent un r\^{o}le encore plus
important en m\'{e}canique quantique. Elles permettent d'obtenir des r\'{e}%
sultats tr\`{e}s g\'{e}n\'{e}raux, qui sont ind\'{e}pendants des
approximations faites par exemple pour le Hamiltonien (bien s\^{u}r si ces
approximations respectent les sym\'{e}tries du probl\`{e}me !).

On souhaite transcrire les id\'{e}es d\'{e}velopp\'{e}es dans le cadre de la
m\'{e}canique analytique \`{a} la m\'{e}canique quantique. On s'int\'{e}%
resse donc \`{a} l'effet induit par une transformation (translation,
rotation, ...) sur les grandeurs math\'{e}matiques qui mod\'{e}lisent le syst%
\`{e}me physique. Une diff\'{e}rence importante entre la m\'{e}canique
classique et la m\'{e}canique quantique porte d'embl\'{e}e sur cette mod\'{e}%
lisation math\'{e}matique. Dans tous les cas (classique et quantique) le syst%
\`{e}me physique est un objet (une particule ou un ensemble de particules,
...) \'{e}voluant dans l'espace ordinaire. La m\'{e}canique classique d\'{e}%
crit le syst\`{e}me par la simple donn\'{e}e d'une collection de points rep%
\'{e}rant les coordonn\'{e}es du syst\`{e}me dans $\mathbf{R}^{3}$. C'est
donc naturellement sur ces variables de l'espace ordinaire que les
transformations agissent en m\'{e}canique classique. Au contraire, en m\'{e}%
canique quantique, la mod\'{e}lisation math\'{e}matique du m\^{e}me syst\`{e}%
me fait intervenir un ket d'\'{e}tat qui appartient \`{a} un espace abstrait
qui n'a rien \`{a} voir avec l'espace ordinaire. On consid\`{e}re aussi des
observables agissant sur ce ket et qui mod\'{e}lisent des grandeurs
physiques. Il s'agit donc dans un premier temps de d\'{e}finir la repr\'{e}%
sentation des transformations dans l'espace des \'{e}tats ainsi que la mani%
\`{e}re dont ces transformations agissent sur les \'{e}tats et les
observables.

\subsection{Cons\'{e}quences de la sym\'{e}trie en m\'{e}canique quantique}

Les op\'{e}rations de sym\'{e}trie les plus utiles en physique atomique et
mol\'{e}culaire sont : les translations dans l'espace, les rotations dans
l'espace, L'inversion par rapport \`{a} l'origine de l'espace ou parit\'{e}
etc..., d'autres sym\'{e}tries sont importantes dans d'autres domaines par
exemple : les transformations relativistes. les translations dans le temps,
la conjugaison de charge, le renversement du temps, les transformations de
jauge locales ect...

\subsection{Propri\'{e}t\'{e}s g\'{e}n\'{e}rales des transformations de sym%
\'{e}trie}

Par op\'{e}ration de sym\'{e}trie on entend toute transformation d'une
quantit\'{e} qui ne change pas certaines de ses propri\'{e}t\'{e}s. Par
exemple, la rotation simultan\'{e}e d'un ensemble de particules autour d'un
axe n'affecte pas leur \'{e}nergie d'interaction, tout comme la translation
en bloc des m\^{e}mes particules. En m\'{e}canique classique, une op\'{e}%
ration de sym\'{e}trie se traduit par une application de l'espace des
configurations sur lui-m\^{e}me. Par exemple, une translation de la coordonn%
\'{e}e $x$ peut s'\'{e}crire $x$ $\rightarrow $ $x+a$, o\`{u} $a$ est une
constante. En m\'{e}canique quantique, la m\^{e}me transformation peut \^{e}%
tre appliqu\'{e}e aux op\'{e}rateurs observables : $\hat{x}\rightarrow $ $%
\hat{x}+a$ que l'on notera $x$ $\rightarrow $ $x+a.$

En m\'{e}canique quantique une transformation du syst\`{e}me physique est
repr\'{e}sent\'{e}e par un op\'{e}rateur tel que, quand il s'applique au
vecteur d'\'{e}tat \ $\left\vert \Psi (t)\right\rangle $ d\'{e}crivant le
syst\`{e}me, donne le vecteur d'\'{e}tat transform\'{e}:

\begin{equation}
\left\vert \tilde{\Psi}(t)\right\rangle =T\left\vert \Psi (t)\right\rangle .
\end{equation}%
Si l'op\'{e}rateur $T$ est ind\'{e}pendante du temps et lin\'{e}aire ($%
T\alpha $ $\left\vert \phi \right\rangle =\alpha T$ $\left\vert \phi
\right\rangle $) et si $T$ a un op\'{e}rateur inverse $T^{-1}$, alors l'\'{e}%
volution dans le temps du ket $\left\vert \tilde{\Psi}(t)\right\rangle $ est
donn\'{e} par: 
\begin{equation}
i\hbar \frac{\partial }{\partial t}\left\vert \tilde{\Psi}(t)\right\rangle =%
\tilde{H}\left\vert \tilde{\Psi}(t)\right\rangle ,
\end{equation}%
o\`{u} on a d\'{e}fini: $\tilde{H}=THT^{-1}$ l'op\'{e}rateur transform\'{e}
par $T$.

\subsection{ Op\'{e}ration de sym\'{e}trie}

Une transformation $T$ est dite de sym\'{e}trie si $\tilde{H}=H$, c'est-\`{a}%
-dire que l'\'{e}volution du syst\`{e}me transform\'{e} est la m\^{e}me que
celle sans transformation. La condition $\tilde{H}=H$ implique que l'op\'{e}%
rateur $T$ commute avec $H:$ $\left[ H,T\right] =0.$

\begin{quote}
Remarques:
\end{quote}

$\blacktriangleright $ On remarque qu'une transformation de sym\'{e}trie
laisse l'Hamiltonien invariant, mais pas n\'{e}cessairement le vecteur d'%
\'{e}tat $\left\vert \Psi (t)\right\rangle $.

$\blacktriangleright $ L'op\'{e}rateur $T$ n'est pas n\'{e}cessairement
Hermitique (il n'est donc pas, en g\'{e}n\'{e}ral, un observable). En fait,
si l'op\'{e}rateur est lin\'{e}aire, il sera n\'{e}cessairement unitaire (la
signification physique du vecteur d'\'{e}tat impose la condition $\langle 
\tilde{\Psi}\left\vert \tilde{\Psi}\right\rangle =$ $\langle \Psi \left\vert
\Psi \right\rangle $, laquelle implique $TT^{+}=1$).

$\blacktriangleright $ Il existe de transformations plus g\'{e}n\'{e}rales
que celles consid\'{e}r\'{e}es jusqu'ici. La transformation renversement du
temps, par exemple, est antilin\'{e}aire ($T\alpha $ $\left\vert \phi
\right\rangle =\alpha ^{\ast }T$ $\left\vert \phi \right\rangle $). On peut 
\'{e}galement, avoir des transformations d\'{e}pendantes du temps
(transformation de jauge locale, par exemple).

$\blacktriangleright $Transformations des observables: Consid\'{e}rons un op%
\'{e}rateur $A$ Hermitique et sa valeur moyenne $\left\langle \phi
\left\vert A\right\vert \phi \right\rangle $ qui est un nombre r\'{e}el et
qui correspond \`{a} une quantit\'{e} mesurable (observable) si $\left\vert
\phi \right\rangle $ est un \'{e}tat possible du syst\`{e}me. Si on consid%
\`{e}re une transformation $T$ des kets telle que: $\left\vert \tilde{\phi}%
\right\rangle =T$ $\left\vert \phi \right\rangle $ on peut se demander
qu'elle la nouvelle forme $\tilde{A}$ de l'op\'{e}rateur qui donne la m\^{e}%
me valeur moyenne dans l'\'{e}tat transform\'{e}, on obtient: $A=T^{+}\tilde{%
A}T$ et si la transformation $T$ est unitaire ($TT^{+}=1$): $\tilde{A}%
=TAT^{+}.$

$\blacktriangleright $ La condition g\'{e}n\'{e}rale pour qu'une
transformation soit de sym\'{e}trie est: 
\begin{equation}
U^{+}(t,t_{0})T(t_{0})=T(t)U(t,t_{0}),
\end{equation}

o\`{u} $U(t,t_{0})$ est l'op\'{e}rateur d'\'{e}volution.

\section{Les sym\'{e}tries fondamentales}

Dans cette section, nous allons passer en revue les transformations d'espace
et de temps fondamentales en m\'{e}canique quantique non relativiste, et
nous allons \'{e}tablir

la forme de l'op\'{e}rateur qui leur est associ\'{e}.

Une sym\'{e}trie est une transformation qui change un syst\`{e}me physique
en un autre syst\`{e}me, lui aussi physique. Les lois physiques concern\'{e}%
es sont invariantes sous cette transformation.\textbf{\ }Les sym\'{e}tries
sont des transformations qui agissent sur un objet g\'{e}om\'{e}trique ou un
syst\`{e}me physique en pr\'{e}servant :

En physique, la notion de sym\'{e}trie est intimement associ\'{e}e \`{a} la
notion d'invariance : elle renvoie \`{a} la possibilit\'{e} de consid\'{e}%
rer un m\^{e}me syst\`{e}me physique selon plusieurs points de vue distincts
en termes de description, mais \'{e}quivalents quant aux pr\'{e}dictions
effectu\'{e}es sur son \'{e}volution.

On parle donc du groupe de sym\'{e}trie (ou d'invariance) de l'objet consid%
\'{e}r\'{e} : la composition de deux sym\'{e}tries est encore une sym\'{e}%
trie. La correspondance entre ces deux notions - sym\'{e}trie et loi
d'invariance - est attribu\'{e}e \`{a} Emmy Noether (1882-1935) en 1917 : 
\`{a} toute loi de conservation correspond une sym\'{e}trie et \`{a} toute
sym\'{e}trie correspond une loi de conservation\textbf{\ }\cite{Yvette}%
\textbf{.} Cette correspondance est d\'{e}montr\'{e}e pour :

$\rhd $la translation:

\ \ $\bullet $ dans le temps: conservation de l'\'{e}nergie (r\'{e}sultat
d'une exp\'{e}rience de physique est ind\'{e}pendante du choix de l'\'{e}%
poque ou de la date \`{a} laquelle cette exp\'{e}rience a ou eu lieu);

\ \ $\bullet $ dans l'espace: conservation de l'impulsion (r\'{e}sultat
d'une exp\'{e}rience de physique est ind\'{e}pendante du lieu de r\'{e}%
alisation);

$\rhd $la rotation: conservation du moment angulaire ou cin\'{e}tique (r\'{e}%
sultat d'une exp\'{e}rience de physique est ind\'{e}pendante du choix de
l'orientation des axes du rep\`{e}re servant \`{a} la d\'{e}crire).

\subsection{Sym\'{e}trie spatiale continue: translations et rotations}

Une sym\'{e}trie spatiale continue se caract\'{e}rise par le fait que l'on
peut lui associer des transformations infinit\'{e}simales, arbitrairement
proches de la transformation identit\'{e}: on peut translater un point sur
une longueur aussi petite que l'on veut, ou le faire tourner \ d'un angle
arbitrairement petit. S'en tenant strictement au plan g\'{e}om\'{e}trique,
deux types de transformations jouent un r\^{o}le de tout premier plan en M%
\'{e}canique (classique ou quantique), les translations et les rotations. 
\`{A} ces sym\'{e}tries continues s'opposent les sym\'{e}tries discr\`{e}tes
(la r\'{e}flexion, la parit\'{e}, le renversement du temps, . . .), pour
lesquelles on ne peut d\'{e}finir d'op\'{e}rations infinit\'{e}simales. Les
translations et les rotations appartiennent \`{a} la classe des d\'{e}%
placements, op\'{e}rations g\'{e}om\'{e}triques qui ne changent ni les
longueurs, ni les angles, ni la chiralit\'{e}. Il existe aussi, dans le plan 
$\mathbf{R}^{2}$, des transformations conformes qui modifient les longueurs
mais pr\'{e}servent les angles.

\paragraph{Les translations d'espace}

Soit $\psi (x)$ la fonction d'onde qui d\'{e}crit un \'{e}tat d'une
particule localis\'{e}e au voisinage du point $x_{0}$. La particule subit
une translation de a selon ox, donc au point $(x_{0}+a)$, elle sera d\'{e}%
crite par $\psi _{a}=\psi (x-a).$ La transformation $T_{a}\psi (x)=\psi
_{a}(x)=\psi (x-a)$ repr\'{e}sente la translation de a dans l'espace des
fonctions d'ondes.

Ecrivons le d\'{e}veloppement de Taylor de $\psi _{a}(x)$

\begin{equation}
\psi _{a}(x)=\psi (x-a)\approx \psi (x)-a\frac{\partial }{\partial x}\psi
(x)+...+(-)^{n}\frac{a^{n}}{n!}\frac{\partial ^{n}}{\partial x^{n}}\psi (x).
\end{equation}

en utilisant $p_{x}=-\frac{i}{\hbar }\frac{\partial }{\partial x},$ on
obtient

\begin{eqnarray}
\psi _{a}(x) &=&\psi (x-a)=\psi (x)-\frac{ia}{\hbar }p_{x}\psi
(x)+...+(-)^{n}\left( \frac{ia}{\hbar }\right) ^{n}\frac{p_{x}^{n}}{n!}\psi
(x)  \notag \\
&=&e^{-\frac{ia}{\hbar }p_{x}}\psi (x).
\end{eqnarray}

On voit que la translation de $a$ le long de l'axe ox est repr\'{e}sent\'{e}%
e par l'op\'{e}rateur $T_{a}=e^{-\frac{ia}{\hbar }p_{x}}.$

La g\'{e}n\'{e}ralisation \`{a} trois dimensions est imm\'{e}diate: $%
T_{a}=e^{-\frac{i}{\hbar }\vec{a}.\vec{p}}.$

\begin{equation}
T_{a}\left\vert \psi \right\rangle =e^{-\frac{i}{\hbar }\vec{a}.\vec{p}%
}\left\vert \psi \right\rangle \text{.}
\end{equation}

o\`{u} $T_{a}$ est un op\'{e}rateur unitaire.

Consid\'{e}rons une translation de la quantit\'{e} infinit\'{e}simale $%
\delta a$. Elle est repr\'{e}sent\'{e}e par un op\'{e}rateur infiniment
voisin de l'identit\'{e} que l'on peut \'{e}crire%
\begin{equation}
T_{\delta a}\approx 1-\frac{i\delta a}{\hbar }p.
\end{equation}

On en conclut que si les \'{e}quations de mouvement sont laiss\'{e}es
invariantes suite \`{a} une translation cela implique que l'impulsion est
conserv\'{e}e, c'est-\`{a}-dire que $\left[ T_{\delta a},H\right] =T_{\delta
a}H-HT_{\delta a}=0\Rightarrow \vec{p}$ (impulsion) totale est conserv\'{e}e.

Comme on \'{e}tudie les translations dans l'espace, on peut aussi \'{e}%
tudier les translations dans le temps: $t\longmapsto t+a$. Cependant, le
temps $t$ n'est pas une variable dynamique en m\'{e}canique quantique: il
n'y a pas d'"op\'{e}rateur du temps". On peut toutefois d\'{e}finir un op%
\'{e}rateur unitaire $U(t)$, appel\'{e} op\'{e}rateur d'\'{e}volution, qui
effectue l'\'{e}volution temporelle du syst\`{e}me sur un temps $t$. Par
analogie avec les translations spatiales, l'\'{e}tat $\left\vert \psi
(t)\right\rangle $ est alors obtenu de l'\'{e}tat $\left\vert \psi
(t)\right\rangle $ par la relation $\left\vert \psi (t)\right\rangle $ $=$ $%
U(t)\left\vert \psi (t)\right\rangle $. Ceci s'applique \'{e}videmment dans
le point de vue de Schr\"{o}dinger.

\paragraph{Les translations dans le temps}

On sait qu'a tout intervalle de temps $\Delta t$ est associ\'{e} un op\'{e}%
rateur unitaire $U(\Delta t),$ et la multiplication de ces op\'{e}rateurs
correspond \`{a} l'addition des intervalles, \ c'est-\`{a}-dire l'op\'{e}%
ration m\^{e}me qui d\'{e}finit le groupe des translations de temps, ou
"groupe d'\'{e}volution". La transformation $U$ repr\'{e}sente la
translation dans le temps des fonctions d'ondes

\begin{equation}
U\psi (t)=\psi (t^{\prime })=\psi (t+\delta t).
\end{equation}

Ecrivons le d\'{e}veloppement de Taylor de $\psi (t^{\prime })$%
\begin{equation}
\psi (t^{\prime })=\psi (t+\delta t)=\psi (t)+\delta t\frac{\partial }{%
\partial t}\psi (t)+...\simeq (1+\delta t\frac{\partial }{\partial t})\psi
(t),
\end{equation}

en rempla\c{c}ant l'op\'{e}rateur $\frac{\partial }{\partial t}$ par $-\frac{%
i}{\hbar }H$ , l'op\'{e}rateur $U$ est alors obtenu par exponentiation de
l'op\'{e}rateur Hermitique $H$;

\begin{equation}
U\simeq (1+\delta t\frac{\partial }{\partial t})\simeq (1-\delta t\frac{i}{%
\hbar }H)=\exp (-\frac{i}{\hbar }H\delta t).
\end{equation}

On voit que la translation temporelle est repr\'{e}sent\'{e}e par l'op\'{e}%
rateur $\hat{U}=(1-\delta t\frac{i}{\hbar }\hat{H})$ qui s'interpr\`{e}te
ainsi: $H$ est le g\'{e}n\'{e}rateur infinit\'{e}simal des translations de
temps, et, pour un syst\`{e}me isol\'{e} (\'{e}nergie totale conserv\'{e}e),
la d\'{e}finition la plus g\'{e}n\'{e}rale du Hamiltonien est d'\^{e}tre pr%
\'{e}cis\'{e}ment ce g\'{e}n\'{e}rateur infinit\'{e}simal. Donc une
observable est invariante par translation de temps si elle commute l'\'{e}%
nergie totale.

\paragraph{Les Rotations}

Les arguments d\'{e}velopp\'{e}s \`{a} propos des translations dans l'espace
peuvent \^{e}tre repris pour \ les rotations. L'invariance par translation
implique l'homog\'{e}n\'{e}it\'{e} pr\'{e}suppos\'{e}e de l'espace (suppos%
\'{e} illimit\'{e}), telle que l'affirme le principe euclidien (pas
d'origine privil\'{e}gi\'{e}e) ; en ce qui concerne les rotations, c'est
l'isotropie de l'espace qui est \`{a} l'\oe uvre (pas de direction privil%
\'{e}gi\'{e}e). Les rotations sont \'{e}galement des d\'{e}placements et,
comme il existe des transformations infinit\'{e}simales, arbitrairement
voisines de la transformation identit\'{e}, les op\'{e}rateurs associ\'{e}s
aux rotations et agissant dans l'espace des \'{e}tats seront aussi unitaires.

La conservation du moment cin\'{e}tique d\'{e}coule de l'invariance par
rotation tout comme la conservation de l'impulsion est la cons\'{e}quence de
l'invariance par translation. L'invariance par rotation repose ici sur la
constatation que toutes les directions dans l'espace sont physiquement \'{e}%
quivalentes. Autrement dit, les propri\'{e}t\'{e}s d'un syst\`{e}me ferm\'{e}
restent inchang\'{e}es suite \`{a} une rotation autour de son centre de masse

L'effet de la rotation sur une fonction d'onde passe par un changement
infinit\'{e}simal de la variable angulaire $\varphi \rightarrow \ \varphi
+\delta \varphi $ (en coordonn\'{e}es sph\'{e}riques), soit $R_{z}$ la
transformation qui repr\'{e}sente une rotation infinit\'{e}simale $d\varphi $
autour de l'axe $z$

\begin{equation}
R_{z}\psi (\varphi )=\psi (\varphi +\delta \varphi ).
\end{equation}

\begin{figure}[htbp]
    \centering
    \includegraphics[width=0.7\textwidth]{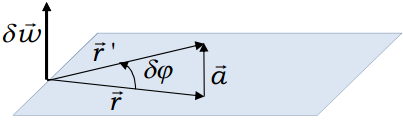}
    \caption{This is my figure caption.}
    \label{fig:myfigure}
\end{figure}

La transformation $R_{z}$ peut \^{e}tre repr\'{e}sent\'{e}e par la
translation $\vec{a}\mathbf{=}\delta \vec{\omega}\times \vec{r}.$

\begin{eqnarray}
\psi (\varphi ) &\rightarrow &\psi ^{\prime }(\varphi )=\psi (\varphi
+\delta \varphi )=\psi (\varphi )+\delta \varphi \frac{\partial \psi
(\varphi )}{\partial \varphi }+\mathcal{O}(\left( \delta \varphi \right)
^{2})  \notag \\
&=&(1+\delta \varphi \frac{\partial }{\partial \varphi })\psi (\varphi )+%
\mathcal{O}(\left( \delta \varphi \right) ^{2})\equiv R_{z}\psi (\varphi )
\end{eqnarray}

o\`{u} $R_{z}$ est l'op\'{e}rateur infinit\'{e}simal de la rotation 
\begin{equation}
R_{z}=(1+\delta \varphi \frac{\partial }{\partial \varphi })=(1+\frac{i}{%
\hbar }\delta \varphi L_{z})
\end{equation}

et $L_{z}=i\hbar \frac{\partial }{\partial \varphi }=i\hbar (x\frac{\partial 
}{\partial y}-y\frac{\partial }{\partial x})$ l'op\'{e}rateur de moment cin%
\'{e}tique dans la direction des $z$. Visiblement, $p$ et $L_{z}$ jouent des
r\^{o}les analogues pour les translations et les rotations respectivement: $%
L_{z}$ est ainsi appel\'{e} le g\'{e}n\'{e}rateur des rotations .

Une rotation finie dans l'espace par $\Delta \varphi $ s'obtient par une
action r\'{e}p\'{e}t\'{e}e de la rotation infinit\'{e}simale, soit 
\begin{equation}
R_{z}=\exp (\frac{i}{\hbar }L_{z}\Delta \varphi ).
\end{equation}%
Pour obtenir l'op\'{e}rateur associ\'{e} \`{a} une rotation d'angle fini $%
\varphi $ autour d'un axe \ donn\'{e}, il sufit d'appliquer $N$ fois la
rotation $\ R$ et de faire tendre $N$ $\rightarrow \infty $, on obtient
ainsi 
\begin{equation}
R_{u,\varphi }=\exp (-\frac{i}{\hbar }\varphi \text{ }\vec{u}.\vec{L}).
\end{equation}%
Cet op\'{e}rateur est visiblement unitaire puisque le moment cin\'{e}tique $%
\vec{L}$ est Hermitique.

En d\'{e}signant d'une fa\c{c}on g\'{e}n\'{e}rale par $\vec{J}$ \ le moment
cin\'{e}tique, l'op\'{e}rateur de rotation est%
\begin{equation}
R_{u,\varphi }=\exp (-\frac{i}{\hbar }\varphi \text{ }\vec{u}.\vec{J}).
\end{equation}%
Il est \`{a} noter que $\vec{J}$ est l'op\'{e}rateur de moment angulaire
total. Lorsque le spin d'une particule est non nul

\begin{equation*}
\vec{J}\mathbf{=}\vec{L}\mathbf{+}\vec{S},
\end{equation*}%
o\`{u} $\vec{L}$ est le moment angulaire orbital et $\vec{S}$ est celui de
spin. L'invariance par rotation implique la conservation de $\vec{J}$ mais
ne signifie pas n\'{e}cessairement que $\vec{L}$ et $\vec{S}$ sont conserv%
\'{e}s s\'{e}par\'{e}ment. De plus, en m\'{e}canique quantique, toutes les
composantes du moment angulaire ne commutent pas entre-elles et donc
seulement $\left\vert \vec{J}\right\vert ^{2}$et $L_{z}$ sont observables
simultan\'{e}ment. En r\'{e}sum\'{e}, pour les transformations continues on
a les situations suivantes

\begin{figure}[htbp]
    \centering
    \includegraphics[width=0.7\textwidth]{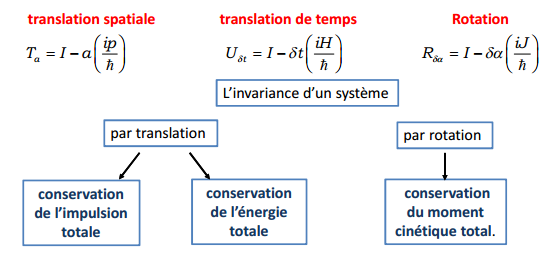}
    \caption{This is my figure caption.}
    \label{fig:myfigure}
\end{figure}
\subsection{Sym\'{e}tries discr\`{e}tes}

Toutes les sym\'{e}tries rencontr\'{e}es jusqu'\`{a} pr\'{e}sent sont associ%
\'{e}es \`{a} des transformations continues, au sens o\`{u} celles-ci d\'{e}%
pendent contin\^{u}ment d'un param\`{e}tre (angle de rotation, amplitude de
la translation, charge \'{e}lectrique pour la transformation de jauge,
etc...). En cons\'{e}quence, il existe des transformations infinit\'{e}%
simales, aussi proches que l'on veut de la transformation identit\'{e} ;
ceci assure que les op\'{e}rateurs associ\'{e}s ne sauraient \^{e}tre
antiunitaires : par le th\'{e}or\`{e}me de Wigner, ils sont donc forc\'{e}%
ment unitaires.

Il existe aussi des op\'{e}rations de sym\'{e}trie discr\`{e}tes; par
exemple l'inversion d'espace, traditionnellement appel\'{e}e parit\'{e} en M%
\'{e}canique quantique.

\subsubsection{L'op\'{e}ration parit\'{e}: sym\'{e}trie d'espace par rapport 
\`{a} l'origine}

L'op\'{e}ration parit\'{e} consiste \`{a} inverser le signe des coordonn\'{e}%
es: $\vec{r}\rightarrow -\vec{r}$.%
\begin{equation}
\begin{pmatrix}
x \\ 
y \\ 
z%
\end{pmatrix}%
\text{ \ }\longrightarrow \text{ \ \ }%
\begin{pmatrix}
-x \\ 
-y \\ 
-z%
\end{pmatrix}%
\end{equation}%

   \begin{figure}[htbp]
    \centering
    \includegraphics[width=0.7\textwidth]{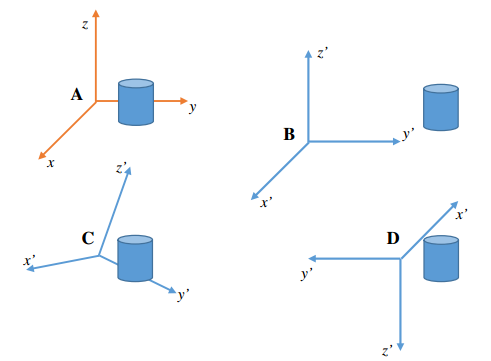}
    \caption{Transformation de la sym\'{e}trie spatiale:translation de l'origine des coordonn,Rotation de l'origine des coordonn\'{e}es, Rotation de l'origine des coordonn\'{e}es, Renversement de l'origine des coordonn\'{e}s.}
    \label{fig:myfigure}
\end{figure}

Comme l'invariance par rotation est en g\'{e}n\'{e}ral valable, on exprime
de fa\c{c}on imag\'{e}e l'invariance par parit\'{e} en disant que l'image
dans un miroir d'une exp\'{e}rience de physique doit appara\^{\i}tre comme
physiquement possible. L'op\'{e}ration parit\'{e} agit diff\'{e}remment sur
les vecteurs proprement dits, ou vecteurs polaires comme la position $\vec{r}
$, l'impulsion $\vec{p}$ ou le champ \'{e}lectrique $\vec{E}$

\begin{equation}
\vec{r}\rightarrow -\vec{r}\mathbf{,}\text{ \ \ \ \ }\vec{p}\rightarrow -%
\vec{p}\mathbf{,}\text{ \ \ \ \ \ }\vec{E}\rightarrow -\vec{E}\mathbf{,}
\end{equation}%
et sur les pseudo vecteurs, ou vecteurs axiaux, comme le moment angulaire $%
\vec{j}=\vec{r}\times \vec{p}$ (produit vectoriel de deux vecteurs polaires)
ou le champ magn\'{e}tique $\vec{B}$ , qui sont associ\'{e}s \`{a} un sens
de rotation autour d'un axe, et non \`{a} une direction

\begin{equation}
\vec{j}\rightarrow \vec{j}\mathbf{,}\text{\ \ \ \ \ \ \ \ \ }\vec{B}%
\rightarrow \vec{B}\mathbf{,}
\end{equation}
\ La transformation correspondant \`{a} la r\'{e}flexion dans l'espace d\'{e}%
finit l'op\'{e}rateur de parit\'{e} sur la fonction d'onde $\Psi (\vec{r}%
,t)\longrightarrow \Psi ^{\prime }(\vec{r},t)=\mathcal{P}\Psi (\vec{r}%
,t)=\Psi (-\vec{r},t).$

Les interactions faibles ne respectent pas l'invariance par parit\'{e} :
ceci a \'{e}t\'{e} montr\'{e} pour la premi\`{e}re fois par C.S. Wu \cite%
{cwu}, en utilisant la d\'{e}sint\'{e}gration $\beta $ de noyaux de $^{60}Co$
polaris\'{e}s en un \'{e}tat excit\'{e} du $^{60}Ni$

\begin{equation}
^{60}Co\longrightarrow ^{60}Ni^{\ast }+e^{-}+\bar{\nu}\longrightarrow
^{60}Ni+e^{-}+\bar{\nu}+2\gamma
\end{equation}

La valeur moyenne du moment angulaire $\left\langle \vec{J}\right\rangle $
du $^{60}Co$ a une orientation fix\'{e}e (voir Figure II). On constate que
les \'{e}lectrons de la d\'{e}sint\'{e}gration sont \'{e}mis de fa\c{c}on pr%
\'{e}f\'{e}rentielle dans la direction oppos\'{e}e \`{a} celle du moment
angulaire : si $\vec{P}$ est l'impulsion des \'{e}lectrons, $\left\langle 
\vec{J}\mathbf{.}\vec{P}\right\rangle <0$. Mais $\left\langle \vec{J}\mathbf{%
.}\vec{P}\right\rangle $, valeur moyenne du produit scalaire d'un vecteur
polaire et d'un vecteur axial, est un pseudo-scalaire, qui change de signe
dans une op\'{e}ration parit\'{e}.

L'exp\'{e}rience vue dans le miroir est diff\'{e}rente de l'exp\'{e}rience r%
\'{e}alis\'{e}e avec un appareil identique \`{a} celui vu dans le miroir.
L'interaction faible viole la parit\'{e}. En d'autres termes, on peut
distinguer un ph\'{e}nom\`{e}ne physique de son image dans un miroir.
L'image de l'exp\'{e}rience dans un miroir (Figure II) n'appara\^{\i}t pas
comme physiquement possible: dans le miroir les sens de rotation sont invers%
\'{e}s, et les \'{e}lectrons partent pr\'{e}f\'{e}rentiellement dans la
direction de $\vec{J}$. Exp\'{e}rimentalement, Il suffit de voir dans quelle
direction l'\'{e}lectron est \'{e}mis pour d\'{e}terminer si l'on voit le
vrai ph\'{e}nom\`{e}ne ou son image dans un miroir:

\textbullet\ Si les \'{e}lectrons \'{e}taient toujours \'{e}mis dans la m%
\^{e}me direction et dans la m\^{e}me proportion que les rayons gamma, la
conservation de la parit\'{e} serait vraie.

\textbullet\ Si la distribution des \'{e}lectrons ne suivait pas la
distribution des rayons gamma, alors la violation de la parit\'{e} serait 
\'{e}tablie.

C.S. Wu \cite{cwu} a observ\'{e} que les \'{e}lectrons \'{e}taient \'{e}mis
dans une direction pr\'{e}f\'{e}rentiellement oppos\'{e}e \`{a} celle des
rayons gamma.
\begin{figure}[htbp]
    \centering
    \includegraphics[width=0.7\textwidth]{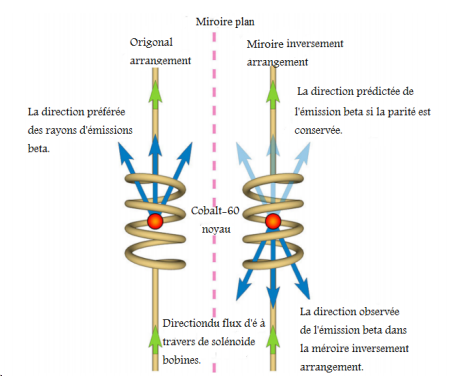}
    \caption{L'exp\'{e}rience de d\'{e}sint\'{e}gration du
cobalt polaris\'{e} et son image dans un miroir.}
    \label{fig:myfigure}
\end{figure}
Le groupe $\mathcal{G}$ correspondant \`{a} l'op\'{e}ration parit\'{e} est
le groupe multiplicatif \`{a} deux \'{e}l\'{e}ments $\left\{ +1,-1\right\} $%
. Comme on ne peut pas relier contin\^{u}ment $-1$ \`{a} l'identit\'{e}, il
nous faut trouver un argument pour d\'{e}cider si l'op\'{e}rateur $\mathcal{P%
}$, qui repr\'{e}sente l'op\'{e}ration parit\'{e} dans l'espace des \'{e}%
tats, est unitaire ou antiunitaire. Soit $\left\vert \chi \right\rangle $ et 
$\left\vert \phi \right\rangle $ deux vecteurs arbitraires et $\left(
\left\vert \chi \right\rangle ,\left\vert \phi \right\rangle \right) $ leur
produit scalaire . Si la parit\'{e} est une sym\'{e}trie,

\begin{equation}
\left\vert \left( \mathcal{P}\left\vert \chi \right\rangle ,\mathcal{P}%
\left\vert \phi \right\rangle \right) \right\vert =\left\vert \left(
\left\vert \chi \right\rangle ,\left\vert \phi \right\rangle \right)
\right\vert .
\end{equation}%
Comme dans l'op\'{e}ration parit\'{e}, les op\'{e}rateurs position et
impulsion doivent se transformer tous deux comme des vecteurs:

\begin{eqnarray}
\vec{r} &\mathbf{\rightarrow }&\mathcal{P}^{-1}\vec{r}\mathcal{P=}-\vec{r} \\
\vec{p} &\rightarrow &\mathcal{P}^{-1}\vec{p}\mathcal{P=-}\vec{p}\mathbf{,}
\end{eqnarray}%
leur commutateur est inchang\'{e}%
\begin{equation}
\mathcal{P}^{-1}\left[ x_{i},p_{j}\right] \mathcal{P=}i\hbar \delta _{ij}I.
\end{equation}

Examinons l'\'{e}l\'{e}ment de matrice

\begin{eqnarray}
\left( \mathcal{P}\left\vert \chi \right\rangle ,\mathcal{P}\left[
x_{i},p_{j}\right] \left\vert \phi \right\rangle \right) &=&\left( \mathcal{P%
}\left\vert \chi \right\rangle ,\mathcal{P}\text{ }\left[ x_{i},p_{j}\right] 
\mathcal{P}^{-1}\mathcal{P}\left\vert \phi \right\rangle \right)  \notag \\
&=&\left( \mathcal{P}\left\vert \chi \right\rangle ,i\hbar \delta _{ij}%
\mathcal{P}\left\vert \phi \right\rangle \right) =i\hbar \delta _{ij}\left( 
\mathcal{P}\left\vert \chi \right\rangle ,\mathcal{P}\left\vert \phi
\right\rangle \right) .  \label{ch2.4}
\end{eqnarray}%
Mais, on a \'{e}galement%
\begin{eqnarray}
\left( \mathcal{P}\left\vert \chi \right\rangle ,\mathcal{P}\left[
x_{i},p_{j}\right] \left\vert \phi \right\rangle \right) &=&\left( \mathcal{P%
}\left\vert \chi \right\rangle ,\mathcal{P}i\hbar \delta _{ij}\left\vert
\phi \right\rangle \right)  \notag \\
&=&i\hbar \delta _{ij}\left( \mathcal{P}\left\vert \chi \right\rangle ,%
\mathcal{P}\left\vert \phi \right\rangle \right) ,  \label{ch2.5}
\end{eqnarray}%
si l'on suppose que $\mathcal{P}$ est unitaire. En effet, pour un op\'{e}%
rateur unitaire

\begin{equation}
\left( U\left\vert \chi \right\rangle ,Ui\phi \right) =\left( \left\vert
\chi \right\rangle ,i\left\vert \phi \right\rangle \right) =i\left(
\left\vert \chi \right\rangle ,\left\vert \phi \right\rangle \right) ,
\end{equation}%
tandis que pour un op\'{e}rateur antiunitaire

\begin{equation}
\left( U\left\vert \chi \right\rangle ,Ui\left\vert \phi \right\rangle
\right) =\left( i\left\vert \phi \right\rangle ,\left\vert \chi
\right\rangle \right) =i\left( \left\vert \phi \right\rangle ,\left\vert
\chi \right\rangle \right) ,
\end{equation}

Les \'{e}quations (\ref{ch2.4}) et (\ref{ch2.5}) sont compatibles uniquement
si $\mathcal{P}$ est unitaire. En revanche, si au lieu de la parit\'{e} $%
\mathcal{P}$ on consid\`{e}re le renversement du sens du temps $\mathcal{T}$
: $\vec{r}\mathbf{\rightarrow }\vec{r}$ \ et $\vec{p}\mathbf{\rightarrow -}%
\vec{p}$, alors

\begin{equation}
\mathcal{T\ }\left[ x_{i},p_{j}\right] \mathcal{T}^{-1}\mathcal{=-}\left[
x_{i},p_{j}\right] =-i\hbar \delta _{ij},
\end{equation}%
et ce changement de signe entra\^{\i}ne que $\mathcal{T}$ est antiunitaire.

\subsubsection{Renversement du sens du temps}

\paragraph{Le renversement du temps en physique classique}

En m\'{e}canique classique, l'\'{e}quation de Newton%
\begin{equation}
m\frac{d^{2}\overrightarrow{r}(t)}{dt^{2}}=\overrightarrow{F}\mathbf{(}%
\overrightarrow{r}\mathbf{(}t\mathbf{)),}
\end{equation}%
est invariante par renversement du sens du temps $t\rightarrow -t$. Posons
en effet $\overrightarrow{r}^{\prime }(t)=\overrightarrow{r}(-t)$

\begin{equation}
m\frac{d^{2}\overrightarrow{r}(t)}{dt^{2}}=m\frac{d^{2}\overrightarrow{r}(-t)%
}{dt^{2}}=\overrightarrow{F}\mathbf{(}\overrightarrow{r}\mathbf{(-}t\mathbf{%
))=}\overrightarrow{F}\mathbf{\mathbf{(}}\overrightarrow{r}^{\prime }\mathbf{%
\mathbf{(}}t\mathbf{\mathbf{))},}
\end{equation}%
on constate que $\overrightarrow{r}^{\prime }\mathbf{\mathbf{(}}t\mathbf{%
\mathbf{)}}$ ob\'{e}it bien aux \'{e}quations de Newton. La raison en est 
\'{e}videmment que ces \'{e}quations ne d\'{e}pendent que de la d\'{e}riv%
\'{e}e seconde par rapport au temps de $r$ et pas de la d\'{e}riv\'{e}e premi%
\`{e}re.

Une image intuitive du renversement du temps est la suivante : imaginons que
nous suivions la trajectoire d'une particule de $t=-\infty $ \`{a} $t=0$ et
qu'\`{a} $t=0$, nous renversions brutalement le sens de l'impulsion (ou de
la vitesse) : \ $p(0)\rightarrow -p(0)$. Dans ces conditions, la particule
va \textquotedblleft remonter sa trajectoire\textquotedblright , elle
repassera au temps $t$ par la position qu'elle avait au temps $-t$ avec une
impulsion oppos\'{e}e (Figure III)
\begin{figure}[htbp]
    \centering
    \includegraphics[width=0.7\textwidth]{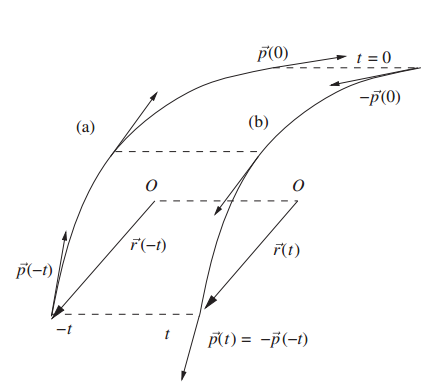}
    \caption{Renversement du temps sur une trajectoire
classique.}
    \label{fig:myfigure}
\end{figure}

\begin{equation}
\overrightarrow{r}^{\prime }\mathbf{\mathbf{(}}t\mathbf{\mathbf{)=}}\vec{r}%
\mathbf{\ (-}t\mathbf{)}\text{ \ \ }\overrightarrow{p}^{\prime
}(-t)\rightarrow -\overrightarrow{p}(t)
\end{equation}

Le vecteur position\ $\overrightarrow{r}$ est pair par renversement du
temps, et $\overrightarrow{p}$ est impair dans cette m\^{e}me op\'{e}ration.
L'invariance par renversement du temps est appel\'{e}e micror\'{e}versibilit%
\'{e}. Si l'on filme le mouvement de particules et que l'on projette la
projection appara\^{\i}t physiquement possible. On sait que tel n'est pas le
cas dans la vie courante, qui est fondamentalement irr\'{e}versible, et il
n'est pas \'{e}vident de comprendre comment une dynamique r\'{e}versible 
\`{a} l'\'{e}chelle microscopique peut conduire \`{a} des ph\'{e}nom\`{e}nes
irr\'{e}versibles \`{a} l'\'{e}chelle macroscopique.

\paragraph{Le renversement du temps en m\'{e}canique quantique}

Revenons \`{a} la m\'{e}canique quantique, en appelant $\mathcal{\tilde{T}}$
\ l'op\'{e}rateur qui r\'{e}alise le renversement du temps dans $\mathcal{E}$%
. Cet op\'{e}rateur est l'analogue temporel de la r\'{e}flexion de l'espace
doit transformer $\vec{r}$ et $\vec{p}$ suivant%
\begin{equation}
\left\{ 
\begin{array}{c}
\mathcal{\tilde{T}}\vec{r}\mathcal{\tilde{T}}^{-1}=\vec{r}, \\ 
\mathcal{\tilde{T}}\vec{p}\mathcal{\tilde{T}}^{-1}=-\vec{p} \\ 
\mathcal{\tilde{T}}\vec{j}\mathcal{\tilde{T}}^{-1}=-\vec{j}\mathbf{.}%
\end{array}%
\right.  \label{ch2.6}
\end{equation}%
En effet, $\vec{j}$ doit se transformer comme $\vec{r}\mathbf{\times }\vec{p}
$, qui est impair par renversement du temps : le moment angulaire d\'{e}%
finit un sens de rotation qui est invers\'{e} par renversement du temps.

On d\'{e}finit l'op\'{e}rateur de renversement du temps, $\mathcal{\tilde{T}}
$, sur une fonction d'onde par

\begin{equation}
\psi (\vec{r},t)\longrightarrow \psi ^{\prime }(\vec{r},t)=\mathcal{\tilde{T}%
}\psi (\vec{r},t)=\psi (\vec{r},-t).
\end{equation}%
\`{A} titre d'exemple, mentionnons quelques quantit\'{e}s ou op\'{e}rateurs
qui se transforment par renversement du temps selon les r\`{e}gles
suivantes: 
\begin{equation}
\left\{ 
\begin{array}{c}
t \\ 
\vec{r} \\ 
\vec{p} \\ 
\vec{\sigma},\vec{j}\mathbf{,}\vec{L} \\ 
\vec{E} \\ 
\vec{B}%
\end{array}%
\right. \overset{\mathcal{\tilde{T}}\ }{\longrightarrow }\left\{ 
\begin{array}{c}
-t \\ 
\vec{r} \\ 
-\vec{p} \\ 
-\vec{\sigma},-\vec{j}\mathbf{,-}\vec{L} \\ 
\vec{E} \\ 
-\vec{B}%
\end{array}%
\right. .
\end{equation}%
Malgr\'{e} l'analogie entre $\mathcal{\tilde{T}}\ $et $\mathcal{P}$, il est
facile de d\'{e}montrer que $\mathcal{\tilde{T}}\ \ $ne peut \^{e}tre un op%
\'{e}rateur unitaire. En effet, une relation importante de la m\'{e}canique
quantique, la relation de commutation $\left[ x_{i},p_{j}\right] =i\hbar
\delta _{ij}$ devient $\left[ x_{i},p_{j}\right] =-i\hbar \delta _{ij}$ apr%
\`{e}s renversement du temps et n'est donc pas pr\'{e}serv\'{e}e. $\mathcal{%
\tilde{T}}$ ne peut donc pas \^{e}tre unitaire et par cons\'{e}quent, ne poss%
\`{e}de pas de valeurs propres et on ne peut y associer des observables.

Nous voulons construire une transformation qui renverse le temps tout en pr%
\'{e}servant $\left[ x_{i},p_{j}\right] =i\hbar \delta _{ij}.$Pour y
arriver, il suffit de combiner $\mathcal{\tilde{T}}\ \ $\`{a} la
transformation dite anti-unitaire $\mathcal{K}$ qui transforme un nombre
complexe en son complexe conjugu\'{e}. D\'{e}finissons maintenant $\mathcal{%
T=}$ $\mathcal{K\tilde{T}}\ $tel que%
\begin{equation}
\mathcal{T=K\tilde{T}}\ \text{\ : \ \ \ \ \ \ \ \ }\left\{ 
\begin{array}{c}
x_{i} \\ 
p_{i} \\ 
i%
\end{array}%
\right. \overset{\mathcal{T}\ }{\longrightarrow }\left\{ 
\begin{array}{c}
x_{i}\longrightarrow \mathcal{T}x_{i}\mathcal{T}^{-1}=x_{i} \\ 
p_{i}\longrightarrow \mathcal{T}p_{i}\mathcal{T}^{-1}=-p_{i} \\ 
-i%
\end{array}%
\right. .
\end{equation}

L'examen de la transformation par $\mathcal{T}$ des relations de commutation
canoniques montre que $\mathcal{T}$ doit \^{e}tre antiunitaire. Calculons de
deux fa\c{c}ons diff\'{e}rentes un \'{e}l\'{e}ment de matrice du commutateur 
$\left[ x_{i},p_{j}\right] =i\hbar \delta _{ij}$

\begin{eqnarray}
\left( \mathcal{T}\left\vert \varphi \right\rangle ,\mathcal{T}\left[
x_{i},p_{j}\right] \left\vert \psi \right\rangle \right) &=&\left( \mathcal{T%
}\left\vert \varphi \right\rangle ,\mathcal{T}i\hbar \delta _{ij}I\left\vert
\psi \right\rangle \right) =\hbar \delta _{ij}\left( \left\vert \varphi
\right\rangle ,i\left\vert \psi \right\rangle \right) ^{\ast }=-i\hbar
\delta _{ij}\left( \left\vert \varphi \right\rangle ,\left\vert \psi
\right\rangle \right) ^{\ast }  \notag \\
&=&\left( \mathcal{T}\left\vert \varphi \right\rangle ,\mathcal{T}\left[
x_{i},p_{j}\right] \mathcal{T}^{-1}\mathcal{T}\left\vert \psi \right\rangle
\right)  \notag \\
&=&\left( \mathcal{T}\left\vert \varphi \right\rangle ,-i\hbar \delta _{ij}I%
\mathcal{T}\left\vert \psi \right\rangle \right) =-i\hbar \delta _{ij}\left(
\left\vert \varphi \right\rangle ,\left\vert \psi \right\rangle \right)
^{\ast }
\end{eqnarray}%
o\`{u} nous avons utilis\'{e} dans la seconde ligne les lois de
transformation (\ref{ch2.6}) de $x_{i}$ et $p_{j}$

\begin{equation}
\mathcal{T}\left[ x_{i},p_{j}\right] \mathcal{T}^{-1}=-\left[ x_{i},p_{j}%
\right]
\end{equation}%
Les deux lignes de l'\'{e}quation pr\'{e}c\'{e}dente sont compatibles, ce
qui ne serait pas le cas si la transformation $\mathcal{T}$ \'{e}tait
unitaire.

Il existe un autre argument tr\`{e}s instructif prouvant le caract\`{e}re
antiunitaire de $\mathcal{T}$. Soit $\varphi (t)$, le vecteur d'\'{e}tat
d'un syst\`{e}me quantique au temps $t$, $\left\vert \varphi \right\rangle
=\left\vert \varphi (t=0)\right\rangle $son \'{e}tat au temps $t=0$

\begin{equation}
\left\vert \varphi (t)\right\rangle =e^{-iHt}\left\vert \varphi
\right\rangle .
\end{equation}

L'invariance par rapport au renversement du temps implique que l'\'{e}tat
transform\'{e} de $\left\vert \varphi (-t)\right\rangle $ par renversement
du temps, $\mathcal{T}\left\vert \varphi (-t)\right\rangle $, co\"{\i}ncide
avec l'\'{e}tat obtenu par \'{e}volution temporelle de $\mathcal{T}%
\left\vert \varphi (t=0)\right\rangle $

\begin{equation}
\mathcal{T}\left\vert \varphi (-t)\right\rangle =e^{-iHt}\mathcal{T}%
\left\vert \varphi \right\rangle .
\end{equation}%
et comme les \'{e}quations sont valables pour tout $\left\vert \varphi
\right\rangle $

\begin{equation}
\mathcal{T}e^{iHt}=e^{-iHt}\mathcal{T}.  \label{ch2.7}
\end{equation}%
Si $\mathcal{T}$ \'{e}tait unitaire, cela impliquerait que%
\begin{equation*}
\mathcal{T}H=-H\mathcal{T}
\end{equation*}%
et \`{a} tout vecteur propre $\left\vert \varphi _{E}\right\rangle $ de $H$
d'\'{e}nergie $E$ correspondrait un vecteur propre $\mathcal{T}\left\vert
\varphi _{E}\right\rangle $ avec une \'{e}nergie $-E$. Dans ces conditions,
l'\'{e}nergie ne serait pas born\'{e}e inf\'{e}rieurement et il existerait
une instabilit\'{e} fondamentale. Si au contraire $\mathcal{T}$ est
antiunitaire puisque dans ce cas $\mathcal{T}i=-i\mathcal{T}$, gr\^{a}ce 
\`{a} $\mathcal{T}iH=-iH\mathcal{T}$ \ l'\'{e}quation (\ref{ch2.7}) implique

\begin{equation*}
\mathcal{T}H=H\mathcal{T}\text{ \ \ \ \ \ \ ou }\mathcal{T}H\mathcal{T}%
^{-1}=H\text{\ }
\end{equation*}%
Cette derni\`{e}re \'{e}quation traduit l'invariance de $H$ par renversement
du sens du temps et implique donc que $\mathcal{T}$ et $H\mathcal{\ }$
doivent commuter.

Pour une particule sans spin, l'op\'{e}ration de renversement du temps est
simplement la conjugaison complexe. En effet, si $\psi (\mathbf{r},t)$ v\'{e}%
rifie l'\'{e}quation de Schr\"{o}dinger

\begin{equation}
i\hbar \frac{\partial }{\partial t}\psi (\vec{r},t)=\left( \frac{p^{2}}{2m}%
+V(\vec{r})\right) \psi (\vec{r},t),
\end{equation}%
en changeant $t$ en $-t$ dans cette \'{e}quation on voit que $\psi (\mathbf{r%
},-t)$ n'est pas solution de la m\^{e}me \'{e}quation. En effet, si $\psi (%
\mathbf{r},t)$ est solution de l'\'{e}quation de Schr\"{o}dinger c'est la
fonction $\ \psi _{renv}(\mathbf{r},t)$ d\'{e}finie par:%
\begin{equation}
\psi _{renv}(\overrightarrow{r},t)=\psi ^{\ast }(\vec{r},-t)  \label{ch2.8}
\end{equation}%
qui est solution de la m\^{e}me \'{e}quation

\begin{equation}
i\hbar \frac{\partial }{\partial t}\psi ^{\ast }(\vec{r},-t)=\left( \frac{%
p^{2}}{2m}+V(\vec{r})\right) \psi ^{\ast }(\vec{r},-t),
\end{equation}%
pourvu que le potentiel $V(\vec{r})$ soit r\'{e}el. Les fonctions $\psi (%
\vec{r},-t)$ et $\psi ^{\ast }(\vec{r},-t)$ solutions diff\'{e}rentes de la m%
\^{e}me \'{e}quation de Schr\"{o}dinger se correspondent par renversement du
temps sont en g\'{e}n\'{e}ral deux solutions diff\'{e}rentes de la m\^{e}me 
\'{e}quation de Schr\"{o}dinger. L'invariance de l'\'{e}quation de Schr\"{o}%
dinger par renversement du temps porte le nom de principe de micro-r\'{e}%
versibilit\'{e}, voir \cite{Messiah}. Tout comme dans le cas classique cette
sym\'{e}trie est bris\'{e}e lorsque la particule est soumise \`{a} un champ
magn\'{e}tique (champ semi-classique donc n\'{e}cessairement ext\'{e}rieur
au syst\`{e}me quantique).

D'apr\`{e}s l'Eq. (\ref{ch2.8}), le renversement du temps en m\'{e}canique
quantique est intimement li\'{e} \`{a} la conjugaison complexe. L'action de
l'op\'{e}rateur $\mathcal{K}$ dans l'espace des \'{e}tats \ est d\'{e}fini
par :

\begin{equation}
\psi _{renv}(\vec{r},t)=\mathcal{K}\psi (\vec{r},-t)
\end{equation}%
Notons d'embl\'{e}e que, puisque $t$ est un param\`{e}tre en m\'{e}canique
quantique, l'op\'{e}rateur $\mathcal{K}$ ne peut agir sur $t$. Comme en m%
\'{e}canique classique, le changement $t$ en $-t$ doit se faire
\textquotedblleft \`{a} la main\textquotedblright\ en m\'{e}canique
quantique. L'op\'{e}rateur $\mathcal{K}$, qui porte toutefois le nom d'op%
\'{e}rateur de renversement du temps, est charg\'{e} des op\'{e}rations qui
n'ont pas d'\'{e}quivalent classique. En repr\'{e}sentation-position et pour
une particule sans spin, $\mathcal{K}$ est simplement l'op\'{e}rateur de
conjugaison complexe :

\begin{equation}
\psi _{renv}(\vec{r},t)=\mathcal{K}\psi (\vec{r},t)=\psi ^{\ast }(\vec{r},t)%
\text{ ,\ \ \ \ \ \ \ \ \ \ \ \ (particule sans spin)}
\end{equation}

En r\'{e}sum\'{e}

\textbullet\ Un op\'{e}rateur de translation, rotation... agissant sur une
variable classique est repr\'{e}sent\'{e}, en m\'{e}canique quantique, par
un op\'{e}rateur de translation, rotation... agissant dans l'espace des \'{e}%
tats. Les op\'{e}rateurs, classiques et quantiques, ont les m\^{e}mes propri%
\'{e}t\'{e}s de groupe. En m\'{e}canique quantique, la repr\'{e}sentation
est d\'{e}finie \`{a} une phase pr\`{e}s.

\textbullet\ En m\'{e}canique quantique, le principe de relativit\'{e} de
Galil\'{e}e implique l'\'{e}galit\'{e} en module du produit scalaire. Ceci
impose une contrainte aux op\'{e}rateurs mod\'{e}lisant une transformation
en m\'{e}canique quantique : ils doivent \^{e}tre unitaires ou
anti-unitaires. De tels op\'{e}rateurs sont parfois qualifi\'{e}s d'op\'{e}%
rateurs de sym\'{e}trie (mais cela n'a rien \`{a} voir avec la sym\'{e}trie
propre au syst\`{e}me).

\textbullet\ Les op\'{e}rateurs de sym\'{e}trie les plus courants sont des op%
\'{e}rateurs unitaires. C'est le cas des op\'{e}rateurs de translations et
rotation.

\textbullet\ A un op\'{e}rateur de sym\'{e}trie (translation, rotation) est
associ\'{e} un g\'{e}n\'{e}rateur (impulsion, moment cin\'{e}tique). Il
s'agit du g\'{e}n\'{e}rateur de la transformation repr\'{e}sent\'{e}e par
l'op\'{e}rateur de sym\'{e}trie.

\textbullet\ Un syst\`{e}me est invariant sous l'action d'un op\'{e}rateur
de sym\'{e}trie lorsque son hamiltonien commute avec cet op\'{e}rateur ou
son g\'{e}n\'{e}rateur. A l'invariance (sym\'{e}trie propre au syst\`{e}me)
est associ\'{e}e une loi de conservation (le g\'{e}n\'{e}rateur est une
constante du mouvement). Il faut savoir mettre en \'{e}vidence les sym\'{e}%
tries et d\'{e}terminer les constantes du mouvement associ\'{e}es en m\'{e}%
canique quantique. Pour le faire, le point de vue de Heisenberg est particuli%
\`{e}rement bien adapt\'{e}.

\textbullet\ Les transformations de jauge sont des transformations locales
du type. On rencontre ce type de transformation lors de l'\'{e}tude de la
dynamique (classique ou quantique) d'une particule soumise \`{a} un champ 
\'{e}lectromagn\'{e}tique.

\textbullet\ En m\'{e}canique classique comme en m\'{e}canique quantique la
physique est invariante de jauge, c'est-\`{a}-dire invariante sous l'action
d'une transformation de jauge. En m\'{e}canique quantique, ceci se traduit
par le fait que l'\'{e}quation de Schr\"{o}dinger garde la m\^{e}me forme
lors d'une transformation de jauge.

\textbullet\ La fonction d'onde transform\'{e}e de jauge est, \`{a} une
phase locale pr\`{e}s, \'{e}gale \`{a} la fonction non transform\'{e}e.

\textbullet\ En pratique, les calculs sont men\'{e}s pour un choix de jauge
donn\'{e} (jauge de Coulomb par exemple).

\textbullet\ Les sym\'{e}tries discr\`{e}tes les plus importantes sont la
parit\'{e} et le renversement du temps.

\textbullet\ Selon leur transformation sous l'action de la parit\'{e} les
grandeurs physiques peuvent \^{e}tre des scalaires ou des pseudo-scalaires,
des vecteurs ou des pseudo-vecteurs.

\textbullet\ Les sym\'{e}tries discr\`{e}tes ne sont pas forc\'{e}ment repr%
\'{e}sent\'{e}es par des op\'{e}rateurs unitaires. Le renversement du temps
(qui correspond \`{a} l'op\'{e}ration de conjugaison complexe) est repr\'{e}%
sent\'{e} par un op\'{e}rateur anti-unitaire, c'est-\`{a}-dire un op\'{e}%
rateur anti-lin\'{e}aire et unitaire.

La Figure IV sch\'{e}matise, les sym\'{e}tries spatiales et discr\`{e}tes
qui jouent un r\^{o}le tr\`{e}s sp\'{e}cial en physique fondamentale: 

\begin{figure}[htbp]
    \centering
    \includegraphics[width=0.7\textwidth]{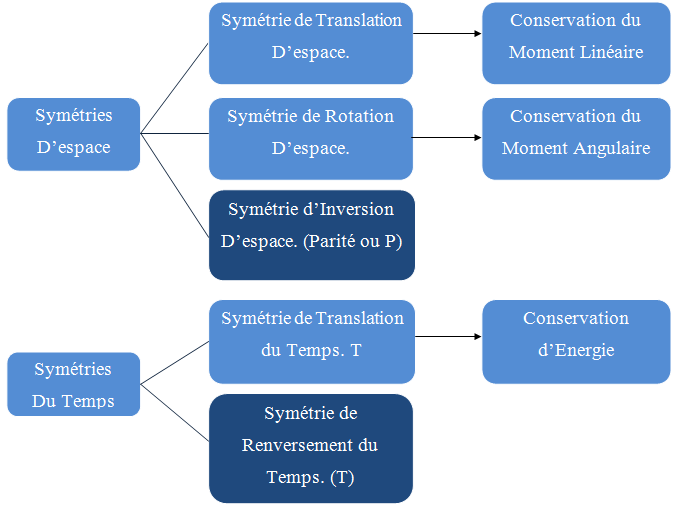}
    \caption{Symétries d’espace et du temps.}
    \label{fig:myfigure}
\end{figure}

\chapter{$\mathcal{PT}$ sym\'{e}trie et pseudo-Hermiticit\'{e}}

Un des axiomes principaux de la m\'{e}canique quantique impose aux
observables physiques d'\^{e}tre Hermitiques. Cela principalement dans le
but de leur assurer des valeurs propres \ r\'{e}elles. En m\'{e}canique
quantique, l'\'{e}tat d'un syst\`{e}me, ses niveaux d'\'{e}nergie et son 
\'{e}volution dans le temps sont d\'{e}termin\'{e}s par un op\'{e}rateur $H$
appel\'{e} Hamiltonien. Cette th\'{e}orie est b\^{a}tie sur un certain
nombre d'axiomes fondamentaux, dont la plupart sont impos\'{e}s par des ph%
\'{e}nom\`{e}nes physiques. Par exemple, le fait que le spectre de $H$, qui
repr\'{e}sente les niveaux d'%
\'{}%
energie du syst\`{e}me, doit \^{e}tre r\'{e}el, est une hypoth\`{e}se
naturelle du point de vue physique. L'\'{e}volution de l'\'{e}tat d'un syst%
\`{e}me quantique \'{e}tant d\'{e}crite par les solutions $\Psi (x,t)=e^{-%
\frac{it}{\hbar }H}\Psi _{0}(x)$ de l'%
\'{}%
\'{e}quation de Schr\"{o}dinger associ\'{e}e au Hamiltonien $H$, il est \'{e}%
galement naturel d'imposer \`{a} l'op\'{e}rateur d'\'{e}volution $U(t)=e^{-%
\frac{it}{\hbar }H}$ d'\^{e}tre unitaire, c'est-\`{a}-dire $\left\vert e^{-%
\frac{it}{\hbar }H}\right\vert ^{2}=1$ pour tout $t$ $\in $ $%
\mathbb{R}
$. En effet, les solutions $\Psi (x,t)$ repr\'{e}sentant la densit\'{e} de
probabilit\'{e} de pr\'{e}sence d'une particule quantique au temps $t$, il
est indispensable que leur norme soit pr\'{e}serv\'{e}e dans le temps. Pour
ces raisons, les physiciens supposent assez syst\'{e}matiquement que l'op%
\'{e}rateur $H$ est auto-adjoint, ce qui garantit ces propri\'{e}t\'{e}s.

Mais en 1998, une large classe d'Hamiltoniens s'est r\'{e}v\'{e}l\'{e}e poss%
\'{e}der des valeurs propres r\'{e}elles en \'{e}tant non Hermitiques. Ces
Hamiltoniens respectent une condition plus faible pour l'obtention de
valeurs propres r\'{e}elles: la sym\'{e}trie parit\'{e} temps ($\mathcal{PT}$%
). Selon certains physiciens, il serait donc possible de donner un sens,
dans le cadre de la m\'{e}canique quantique, \`{a} des th\'{e}ories faisant
intervenir des op\'{e}rateurs non-auto-adjoints, pour peu que leur spectre
soit r\'{e}el et leur \'{e}volution dans le temps unitaire. Ainsi, depuis
quelques ann\'{e}es, physiciens et math\'{e}maticiens ont cherch\'{e} \`{a}
remplacer la sym\'{e}trie propre aux op\'{e}rateurs auto-adjoints par un
autre type de sym\'{e}trie dans l'espace-temps, plus faible, appel\'{e}e $%
\mathcal{PT}$-sym\'{e}trie.

Ce chapitre aborde des questions ayant donn\'{e} lieu \`{a} une intense
activit\'{e} ces derni\`{e}res ann\'{e}es, \`{a} la suite des travaux
pionniers de Bender et Boettcher sur les Hamiltoniens non Hermitiques \cite%
{Bender,bender2,bender3,Ben}. L'un des "dogmes" de la th\'{e}orie quantique
est l'association d'un op\'{e}rateur Hermitique \`{a} toute grandeur
physique, une propri\'{e}t\'{e} qui garantit la r\'{e}alit\'{e} des valeurs
propres. En r\'{e}alit\'{e}, la condition d' Hermiticit\'{e} est une
condition suffisante, nullement n\'{e}cessaire, \ puisqu'il existe des op%
\'{e}rateurs non Hermitiques dont le spectre est r\'{e}el. L'id\'{e}e
centrale est de remplacer la condition d'Hermiticit\'{e} par une condition
plus faible assurant \'{e}videmment elle aussi la r\'{e}alit\'{e} des
valeurs propres. Ceci propose une question importante: est-ce que la th\'{e}%
orie quantique conventionnelle est applicable pour des Hamiltoniens
non-Hermitiques? Ou bien il faut construire une nouvelle th\'{e}orie?

\subsection{$\mathcal{PT}$-sym\'{e}trie}

L'un des premiers r\'{e}sultats importants fut obtenu dans l'\'{e}tude d'une
famille de Hamiltoniens de la forme:

\begin{equation}
H=p^{2}-\lambda (ix)^{N}  \label{ch4.1.1}
\end{equation}%
o\`{u} $\lambda $ et$\ N$ sont des param\`{e}tres positifs. . Un tel
Hamiltonien est visiblement non-Hermitique, puisque $H$ $\neq H^{+}$ si $N$
n'est pas un entier pair. Par ailleurs, $H$ \ n'est pas (en g\'{e}n\'{e}ral)
invariant par la parit\'{e} $\mathcal{P}$, qui change $p$ en $-p$ et $x$ en $%
-x$ - sauf \ si $N=4k+2$, $k\in N$ , ni par renversement du temps $\mathcal{T%
}$ qui change $p$ en $-p$, laisse $\ x$ inchang\'{e} mais complexe conjugue.
En revanche, comme Bender et Boettcher \cite{Bender} l'ont r\'{e}alis\'{e},
tous les Hamiltoniens (\ref{ch4.1.1}), sont invariants si on effectue les
deux op\'{e}rations $\mathcal{P}$ et $\mathcal{T}$ , d'o\`{u} la
terminologie invariance $\mathcal{PT}$. C'est ce qui les a conduits \`{a}
proposer de remplacer la condition d'Hermiticit\'{e} par celle de
l'invariance $\mathcal{PT}$, bien que, pour les Hamiltoniens de la forme (%
\ref{ch4.1.1}), le spectre n'est r\'{e}el que si $N$ est sup\'{e}rieur \`{a}
une certaine valeur $N_{c}$ $\simeq 1,42207$. La non-suffisance de \
l'invariance $\mathcal{PT}$ pour la r\'{e}alit\'{e} du spectre force \`{a}
distinguer les r\'{e}gions de l'espace des param\`{e}tres (pour (\ref%
{ch4.1.1}), le seul param\`{e}tre est $N$) selon la r\'{e}alit\'{e} ou non
du spectre, et r\'{e}cup\`{e}re la notion de sym\'{e}trie bris\'{e}e, cette
fois \`{a} propos de la sym\'{e}trie $\mathcal{PT}$. \ Par la suite,
d'autres Hamiltoniens ont \'{e}t\'{e} \'{e}tudi\'{e}s et fournissent des
exemples o\`{u} les r\'{e}gions de sym\'{e}trie non-bris\'{e}e sont aussi
celles o\`{u} le spectre est r\'{e}el.

Un Hamiltonien $H$ est dit $\mathcal{PT}$-Sym\'{e}trique s'il est invariant
par la transformation $\mathcal{PT}$, c'est-\`{a}-dire

\begin{equation}
H=\mathcal{PT}H\mathcal{PT},  \label{ch4.1.2}
\end{equation}%
o\`{u} l'op\'{e}rateur de parit\'{e} $\ \mathcal{P}$ et l'op\'{e}rateur
renversement du temps $\mathcal{T}$ \ commutent entre eux 
\begin{equation}
\lbrack \mathcal{P},\mathcal{T}]=0,  \label{ch4.1.3}
\end{equation}%
et tel que leurs carr\'{e}s donnent l'op\'{e}rateur unit\'{e}

\begin{equation}
\mathcal{P}^{2}=\mathcal{T}^{2}=I.  \label{ch4.1.4}
\end{equation}%
Ainsi, un Hamiltonien $H$ est dit $\mathcal{PT}$-sym\'{e}trique s'il commute
avec l'op\'{e}rateur $\mathcal{PT}$

\begin{equation}
\lbrack H;\mathcal{PT}]=0,
\end{equation}%
La $\mathcal{PT}$-sym\'{e}trie est dite non bris\'{e}e si les fonctions
propres de l'Hamiltonien $\mathcal{PT}$-sym\'{e}trique sont aussi des
fonctions propres de l'op\'{e}rateur $\mathcal{PT}$. Dans le cas contraire,
si les fonctions propres de l'Hamiltonien $\mathcal{PT}$-sym\'{e}trique ne
sont pas des fonctions propres de l'op\'{e}rateur $\mathcal{PT}$ , elle est
dite bris\'{e}e.

\subsection{Valeurs propres des Hamiltoniens $\mathcal{PT}$-sym\'{e}triques}

Ainsi, pour construire une th\'{e}orie quantique \`{a} partir des
Hamiltoniens $\mathcal{PT}$-sym\'{e}triques, nous exigeons de plus que la sym%
\'{e}trie ne soit pas bris\'{e}e. Il faut noter cependant que cette
condition n'est pas triviale car il n'existe aucun moyen pour affirmer \`{a}
priori qu'une telle sym\'{e}trie d'un Hamiltonien $\mathcal{PT}$-sym\'{e}%
trique est bris\'{e}e ou pas. Il faut tout d'abord d\'{e}terminer les
fonctions propres pour en tirer une conclusion. Avec cette condition suppl%
\'{e}mentaire, on peut d\'{e}montrer la r\'{e}alit\'{e} des valeurs propres
d'un Hamiltonien $\mathcal{PT}$-sym\'{e}trique. En effet, soit $\left\{ \psi
_{n}\left( x\right) ,n=1,2...\right\} $, l'ensemble des fonctions propres
communes \`{a} $H$ et $\mathcal{PT}$ ,

\begin{equation}
H\psi _{n}=E_{n}\psi _{n},  \label{ch4.1.5}
\end{equation}%
et\qquad \qquad \qquad \qquad \qquad \qquad \qquad \qquad

\begin{equation}
\mathcal{PT}\psi _{n}=\theta \psi _{n},  \label{ch4.1.6}
\end{equation}%
avec $E_{n}$ et $\theta $ les valeurs propres correspondantes, qui sont a
priori complexes. Comme

\begin{equation}
(\mathcal{PT})^{2}=1,  \label{ch4.1.7}
\end{equation}%
il en r\'{e}sulte que

\begin{equation}
\left\vert \theta \right\vert ^{2}=1.  \label{ch4.1.8}
\end{equation}%
La relation (\ref{ch4.1.2}) permet d'\'{e}crire

\begin{eqnarray}
\mathcal{PT}H\mathcal{PT}\psi _{n} &=&E_{n}^{\ast }\theta ^{\ast }\mathcal{PT%
}\psi _{n}  \notag \\
&=&|\theta |^{2}E_{n}^{\ast }\psi _{n}=E_{n}^{\ast }\psi _{n}=E_{n}\psi _{n},
\label{ch4.1.9}
\end{eqnarray}%
d'o\`{u}

\begin{equation}
E_{n}=E_{n}^{\ast }.  \label{ch4.1.10}
\end{equation}

\subsection{$\mathcal{PT-}$ produit scalaire}

La question qui se pose maintenant est de savoir si les Hamiltoniens poss%
\'{e}dant une $\mathcal{PT}$- sym\'{e}trie non bris\'{e}e peuvent d\'{e}%
crire la dynamique de syst\`{e}mes physiques r\'{e}els. En d'autres termes
il faut v\'{e}rifier que la norme d'un vecteur \ propre de $H$ dans l'espace
de Hilbert doit \^{e}tre positive et que l'\'{e}volution au cours du temps
des \'{e}tats propres demeure unitaire. Bien entendu, ces deux exigences
sont satisfaites avec des Hamiltoniens Hermitiques. La premi\`{e}re permet
d'interpr\'{e}ter la norme d'un \'{e}tat comme une probabilit\'{e}, qui doit 
\^{e}tre d\'{e}fini positive, alors que la deuxi\`{e}me condition garantie
justement l'ind\'{e}pendance de cette probabilit\'{e} par rapport au temps.

A cet effet, Bender et Boettcher \cite{Bender} ont d\'{e}finit un $\mathcal{%
PT}$-produit scalaire

\begin{equation}
(f,g)=\int_{C}dx\left[ \mathcal{PT}f(x)\right] g(x),  \label{ch4.1.11}
\end{equation}%
associ\'{e} aux Hamiltoniens $\mathcal{PT}$-sym\'{e}triques et o\`{u}

\begin{equation}
\mathcal{PT}f(x)=f^{\ast }(-x).  \label{ch4.1.12}
\end{equation}%
L'avantage de cette d\'{e}finition du produit scalaire est que, comme en m%
\'{e}canique quantique ordinaire, la norme de toute fonction d'onde est une
quantit\'{e} ind\'{e}pendante de sa phase globale et de plus elle est conserv%
\'{e}e dans le temps. Cependant, cette d\'{e}finition a un inconv\'{e}nient
majeur qui r\'{e}side dans le fait que les normes de certains \'{e}tats
propres d'Hamiltoniens $\mathcal{PT-}$sym\'{e}triques sont n\'{e}gatives.

En effet, d\'{e}signons par $\psi _{n}$ et $\psi _{m}$ les fonctions propres
de $H$ orthogonales pour $n\neq m$, leur $\mathcal{PT}$-produit scalaire s'%
\'{e}crit

\begin{eqnarray}
\left\langle \psi _{m},\psi _{n}\right\rangle _{\mathcal{PT}} &=&\int dx[%
\mathcal{PT}\psi _{m}(x)]\psi _{n}(x)  \notag \\
&=&\int dx\psi _{m}^{\ast }(-x)\psi _{n}(x)=(-1)^{n}\delta mn.
\label{ch4.1.13}
\end{eqnarray}%
Si $n=m$,

\begin{equation}
\left\langle \psi _{m},\psi _{n}\right\rangle _{\mathcal{PT}}=(-1)^{m},
\label{ch4.1.14}
\end{equation}%
il est clair que la norme n'est pas toujours positive, contrairement \ \`{a}
ce qui est admis en m\'{e}canique quantique, usuelle. La relation de
fermeture s'\'{e}crit comme

\begin{equation}
\overset{\infty }{\sum_{n=\grave{a}}}(-1)^{n}\psi _{n}(x)\psi _{n}(y)=1,
\label{ch4.1.15}
\end{equation}
Par cons\'{e}quent ces normes ne peuvent pas \^{e}tre interpr\'{e}t\'{e}s
comme des probabilit\'{e}s. Ce qui plus tard, \ a incit\'{e} Bender et al 
\cite{bender2,bender3} \`{a} montrer que tous les Hamiltoniens $\mathcal{PT}$%
-sym\'{e}triques dont la sym\'{e}trie n'est pas bris\'{e}e poss\`{e}dent une
autre sym\'{e}trie cach\'{e}e, engendr\'{e}e par un nouveau operateur d\'{e}%
not\'{e} $\mathcal{C}$, appel\'{e} op\'{e}rateur de conjugaison de charge.\
\ Les propri\'{e}t\'{e}s de l'op\'{e}rateur $\mathcal{C}$ sont presque
identiques \`{a} ceux de l'op\'{e}rateur de la conjugaison de charge mais
son action est diff\'{e}rente de celle de l'op\'{e}rateur $\mathcal{C}$ de
la m\'{e}canique quantique\textbf{\ }qui transforme tous les nombres
quantiques additifs en leurs oppos\'{e}s. Ces nombres comprennent la charge 
\'{e}lectrique, le nombre leptonique (\'{e}lectronique, muonique,
tauonique), l'isospin, l'hypercharge, l'\'{e}tranget\'{e}, la couleur, le
nombre baryonique.

\subsection{L'op\'{e}rateur $\mathcal{C}$ et le $\mathcal{CPT}$ produit
scalaire}

Pour r\'{e}soudre le probl\`{e}me de la norme n\'{e}gative, Bender et al 
\cite{bender2,bender3} ont montr\'{e} que tous les Hamiltoniens $\mathcal{PT}
$-sym\'{e}trique dont la sym\'{e}trie n'est pas bris\'{e}e poss\`{e}dent une
autre sym\'{e}trie engendr\'{e}e par un nouvel op\'{e}rateur lin\'{e}aire not%
\'{e} $\mathcal{C}$ \ qui commute avec $H$ et $\mathcal{PT}$.

\begin{equation}
\lbrack \mathcal{C},H]=[\mathcal{C},PT]=0,  \label{ch4.1.16}
\end{equation}%
et par cons\'{e}quent $H$ commute avec le produit $\mathcal{CPT}$, $\ [H,%
\mathcal{C}PT]=0.$ On montre alors que les fonctions propres communes \`{a} $%
H$ et $CPT$ sont toutes de norme \ d\'{e}finie positive. Une fois
l'operateur $\mathcal{C}$ d\'{e}termin\'{e}, il sera possible donc de
construire une nouvelle th\'{e}orie quantique qui satisfait \`{a} toutes les
contraintes requises. Nous pouvons contruire, dans l'espace des coordonn\'{e}%
es, l'op\'{e}rateur $\mathcal{C}$ explicitement en fonction des \'{e}tats
propres de l'Hamitonien $H$

\begin{equation}
\mathcal{C}(x,y)=\overset{\infty }{\underset{n=0}{\sum }}\psi _{n}(x)\psi
_{n}(y),  \label{ch4.1.17}
\end{equation}%
de sorte que le carr\'{e} est \'{e}gal \`{a} l'unit\'{e} \cite%
{bender2,bender3}:

\begin{equation}
\int dx\mathcal{C}(x,y)\mathcal{C}(y,z)=\delta (x-z),  \label{ch4.1.18}
\end{equation}%
et par cons\'{e}quent,

\begin{equation}
\mathcal{C}^{2}=1.  \label{ch4.1.19}
\end{equation}%
d'o\`{u} les valeurs propres de l'op\'{e}rateur $\mathcal{C}$ sont $\pm 1$.
L'action de $\mathcal{C}$ sur les fonctions propres $\psi _{n}(x)$ de $H$
est donn\'{e}e par

\begin{eqnarray}
\mathcal{C}\psi _{n}(x) &=&\int dyC(x,y)\phi _{n}(y)  \notag \\
&=&\overset{\infty }{\underset{m=0}{\sum }}\psi _{m}(x)\int dy\psi
_{m}(y)\psi _{n}(y)=(-1)^{n}\psi _{n}(x).  \label{ch4.1.20}
\end{eqnarray}%
\ Nous pouvons \'{e}galement construire l'op\'{e}rateur de parit\'{e} $%
\mathcal{P}$ en termes fonctions propres de $H$

\begin{equation}
\mathcal{P}(x,y)=\sum_{n}(-1)^{n}\psi _{n}(x)\psi _{n}(-y)=\delta (x+y).
\label{ch4.1.21}
\end{equation}%
Comme l'op\'{e}rateur $\mathcal{C}$, le carr\'{e} de l'op\'{e}rateur de parit%
\'{e} est \'{e}galement \'{e}gal \`{a} l'unit\'{e}. Malgr\'{e} que, $%
\mathcal{P}^{2}=1$ et $\mathcal{C}^{2}=1$, les deux op\'{e}rateurs ne sont
pas identiques; $\mathcal{P}$ est un op\'{e}rateur r\'{e}el mais $\mathcal{C}
$ est un op\'{e}rateur complexe. De plus, ces deux op\'{e}rateurs ne
commutent pas 
\begin{equation}
\lbrack \mathcal{C},\mathcal{P}]\neq 0;[\mathcal{C},\mathcal{T}]\neq 0,
\label{ch4.1.22}
\end{equation}%
\'{e}videmment, $\mathcal{C}$ commute avec le produit $\mathcal{PT}$. \
Enfin, ayant obtenu l'op\'{e}rateur $\mathcal{C}$ nous d\'{e}finissons un
nouveau produit scalaire appel\'{e} $\mathcal{CPT}$-produit scalaire

\begin{equation}
\left\langle \phi _{m},\phi _{n}\right\rangle _{\mathcal{CPT}}=\int dx\left[ 
\mathcal{CPT}\phi _{m}(x)\right] \phi _{n}(x).  \label{ch4.1.23}
\end{equation}%
o\`{u}

\begin{equation}
\mathcal{CPT}\phi _{m}(x)=\int dy\mathcal{CPT}\phi _{m}^{\ast }(-y),
\label{ch4.1.24}
\end{equation}%
donc

\begin{equation}
\left\langle \phi _{m},\phi _{n}\right\rangle _{\mathcal{CPT}}=\int dx[%
\mathcal{CPT}\phi _{m}(x)]\phi _{n}(x)=\delta _{mn}.  \label{ch4.1.25}
\end{equation}

Le $\mathcal{CPT}$-produit scalaire est d\'{e}finit positif, et les
fonctions propres de $H$ sont orthogonales. Etant donn\'{e} que $\mathcal{C}$
est une fonction de $H$, il est un op\'{e}rateur sp\'{e}cifique qui d\'{e}%
pend particuli\`{e}rement du syst\`{e}me \'{e}tudi\'{e}, contrairement, aux
op\'{e}rateurs $\mathcal{P}$ et $\mathcal{T}$.

\subsection{Application:}

\subparagraph{Etude du syst\`{e}me \`{a} deux niveaux \guillemotleft\ %
Brachistochrone \guillemotright\ : cas $\mathcal{PT}$-sym\'{e}trique}

Le brachistochrone est l'un des probl\`{e}mes les plus adapt\'{e} \`{a} l'%
\'{e}tude des syst\`{e}mes \ $\mathcal{PT}$-sym\'{e}trique et est d\'{e}crit
par l'Hamiltonien

\begin{equation}
H=%
\begin{pmatrix}
re^{i\theta } & s \\ 
s & re^{-i\theta }%
\end{pmatrix}%
,  \label{ch4.1.26}
\end{equation}%
(les param\`{e}tres $r,s$ et $\theta $ sont r\'{e}els). $\mathcal{P}=%
\begin{pmatrix}
0 & 1 \\ 
1 & 0%
\end{pmatrix}%
$, $\mathcal{T}$ complexe et conjugue. En effectuant le produit des
matrices, on trouve que: $H=(\mathcal{PT})H(\mathcal{PT})^{-1}$. Les valeurs
propres $\varepsilon _{\pm }$ de $H$ sont :

\begin{equation}
\varepsilon _{\pm }=r\cos \theta \pm \sqrt{s%
{{}^2}%
-r^{2}\sin ^{2}\theta }.  \label{ch4.1.27}
\end{equation}%
\ La sym\'{e}trie $\mathcal{PT}$ est\emph{\ bris\'{e}e} quand les valeurs
propres ne sont pas r\'{e}elles: soit dans la r\'{e}gion o\`{u} $s%
{{}^2}%
\mathcal{<}r^{2}\sin ^{2}\theta $ . Dans la r\'{e}gion o\`{u} $s%
{{}^2}%
\mathcal{>}r^{2}\sin ^{2}\theta $ les valeurs propres sont r\'{e}elles donc
la $\mathcal{PT}$-Sym\'{e}trie est \emph{non bris\'{e}e}. Si on pose $\sin
\alpha =\frac{r}{\sin \theta }$, les valeurs propres s'\'{e}crivent: $%
\varepsilon _{\pm }=r\cos \theta \pm s\cos \alpha ,$ et les vecteurs propres
de $H$ s'obtiennent facilement; on trouve :%
\begin{eqnarray}
\left\vert +\right\rangle &=&C_{+}%
\begin{pmatrix}
e^{\frac{i\alpha }{2}} \\ 
e^{-\frac{i\alpha }{2}}%
\end{pmatrix}%
, \\
\left\vert -\right\rangle &=&C_{-}%
\begin{pmatrix}
e^{-\frac{i\alpha }{2}} \\ 
-e^{\frac{i\alpha }{2}}%
\end{pmatrix}%
\end{eqnarray}%
o\`{u} les constantes $C_{\pm }$ \ sont pour l'instant quelconques. On a : 
\begin{equation}
\mathcal{PT}\left\vert +\right\rangle =C_{+}^{\ast }%
\begin{pmatrix}
e^{\frac{i\alpha }{2}} \\ 
e^{-\frac{i\alpha }{2}}%
\end{pmatrix}%
,
\end{equation}%
$\left\vert +\right\rangle $ est donc vecteur propre de $\mathcal{PT}$ avec
la valeur propre $+1$ si et seulement si $C_{+}^{\ast }=$ $+C_{+}$. Le m\^{e}%
me calcul avec $\left\vert -\right\rangle $ donne $C_{-}^{\ast }=$ $-C_{-}$
pour avoir $\mathcal{PT}\left\vert -\right\rangle =\left\vert -\right\rangle
.$

Avec le produit scalaire habituel, on a :%
\begin{equation}
\langle +\left\vert -\right\rangle =-2iC_{+}^{\ast }C_{-}\sin \alpha \neq 0,
\label{ch4.1.28}
\end{equation}%
et%
\begin{equation}
\langle +\left\vert +\right\rangle =\frac{1}{\cos \alpha }\neq 1,
\label{ch4.1.29}
\end{equation}%
Quand on a choisi $C_{+}\in \mathcal{R}$ , puisque $\mathcal{PT}\left\vert
+\right\rangle =\left\vert +\right\rangle $ le produit scalaire d\'{e}fini
dans le texte ou $\mathcal{PT}$-produit scalaire donne :

\begin{equation*}
\langle +\left\vert -\right\rangle _{\mathcal{PT}}=0.
\end{equation*}%
Le carr\'{e} de la norme au sens de ce $\mathcal{PT}$ produit scalaire est
ainsi :%
\begin{equation}
\langle +\left\vert +\right\rangle _{\mathcal{PT}}=2C_{+}^{2}\cos \alpha ,
\end{equation}%
pour pr\'{e}server $\mathcal{PT}\left\vert +\right\rangle =\left\vert
+\right\rangle $, il faut $C_{+}\in \mathcal{R}$, d'o\`{u} $C_{+}=\left(
2\cos \alpha \right) ^{-1/2}$, \`{a} un signe pr\`{e}s inessentiel. De m\^{e}%
me $\langle -\left\vert -\right\rangle _{\mathcal{PT}}=$ $2C_{-}^{2}\cos
\alpha $ ; pour avoir $C_{-}^{\ast }$ $=-C_{-}$, il faut prendre $C_{-}$ $=$ 
$i\left( 2\cos \alpha \right) ^{-1/2}$. D'o\`{u} les vecteurs propres
normalis\'{e}s de $H$ qui sont \'{e}galement propres de $\mathcal{PT}$ avec
la valeur propre $+1$ (donc $\mathcal{PT}$ invariants) : 
\begin{eqnarray}
\left\vert +\right\rangle &=&\frac{1}{\sqrt{2\cos \alpha }}%
\begin{pmatrix}
e^{\frac{i\alpha }{2}} \\ 
e^{-\frac{i\alpha }{2}}%
\end{pmatrix}%
,  \label{ch4.1.30} \\
\left\vert -\right\rangle &=&\frac{i}{\sqrt{2\cos \alpha }}%
\begin{pmatrix}
e^{-\frac{i\alpha }{2}} \\ 
-e^{\frac{i\alpha }{2}}%
\end{pmatrix}
\label{ch4.1.31}
\end{eqnarray}%
Noter que le carr\'{e} de la norme de $\left\vert -\right\rangle $ vaut $-1$
: \ le produit scalaire ainsi d\'{e}finit n'est pas positif. Donc, il faut
red\'{e}finir le produit scalaire.

Soit l'op\'{e}rateur (\textquotedblleft de charge") $\mathcal{C}$ :

\begin{equation}
\mathcal{C}=\frac{1}{\cos \alpha }%
\begin{pmatrix}
i\sin \alpha & 1 \\ 
1 & -i\sin \alpha%
\end{pmatrix}%
,
\end{equation}%
On montre que :

\begin{equation}
\lbrack H,\mathcal{C}]=0\text{ \ \ et \ \ }\mathcal{C}^{2}=1.
\label{ch4.1.32}
\end{equation}%
Comme $\mathcal{PT}\left\vert \pm \right\rangle =\pm \left\vert \pm
\right\rangle $, \ on a donc maintenant $\mathcal{CPT}\left\vert \pm
\right\rangle =\left\vert \pm \right\rangle $, par cons\'{e}quent :

\begin{equation}
\langle \pm \left\vert \pm \right\rangle _{\mathcal{CPT}}=1\text{ \ et \ }%
\langle \pm \left\vert \mp \right\rangle _{\mathcal{CPT}}=0.
\label{ch4.1.33}
\end{equation}

\section{Pseudo-Hermiticit\'{e}}

Les situations o\`{u} les Hamiltoniens ne sont pas Hermitiques mais ne d\'{e}%
crivent pas non plus des syst\`{e}mes dissipatifs peuvent \^{e}tre formul%
\'{e}s de mani\`{e}re pr\'{e}cise et coh\'{e}rente afin de permettre une 
\'{e}volution temporelle unitaire. La mani\`{e}re la plus efficace et la
plus g\'{e}n\'{e}rale est de faire correspondre des Hamiltoniens non
Hermitiques \`{a} leurs homologues Hermitiques par l'action d'une
transformation de similarit\'{e}.

Ainsi, l'objectif de remplacer la condition math\'{e}matique de l'Hermiticit%
\'{e} par la condition plus physique de la $\mathcal{PT}$ sym\'{e}trie peut 
\^{e}tre plac\'{e} dans un contexte math\'{e}matique plus g\'{e}n\'{e}ral
connu sous le nom de pseudo-Hermiticit\'{e}. Un op\'{e}rateur lin\'{e}aire $%
A $ est pseudo Hermitique s'il existe un op\'{e}rateur Hermitique $\eta $
tel que

\begin{equation}
A^{+}=\eta A\eta ^{-1}.  \label{ch4.2.1}
\end{equation}%
L'op\'{e}rateur $\eta $ est souvent appel\'{e} op\'{e}rateur m\'{e}trique.
La condition dans (\ref{ch4.2.1}) se r\'{e}duit \`{a} l'Hermiticit\'{e}
ordinaire quand l'op\'{e}rateur m\'{e}trique $\eta $ est \'{e}gal \`{a}
l'identit\'{e} $I$ et \`{a} la $\mathcal{PT}$-sym\'{e}trie quand $\eta $ $=$ 
$\mathcal{PT}$. Le concept de pseudo-Hermiticit\'{e} a \'{e}t\'{e} introduit
dans les ann\'{e}es 1940 par Dirac et Pauli \cite{pauli,gupta,gupta2}, et
discut\'{e} plus tard par Lee, Wick et Sudarshan \cite{ed,td}, qui
essayaient de r\'{e}soudre les probl\`{e}mes qui se posent dans la
quantification de l'\'{e}lectrodynamique et dans d'autres th\'{e}ories
quantiques des champs o\`{u} les \'{e}tats de norme n\'{e}gative
apparaissent comme une cons\'{e}quence de la renormalisation.

La notion de quasi-Hermiticit\'{e} a \'{e}t\'{e} discut\'{e}e en d\'{e}tail
en 1992 par Scholtz et al \cite{scho}. Ce dernier article est pertinent pour
la $\mathcal{PT}$ sym\'{e}trie car il a \'{e}t\'{e} le premier \`{a} montrer
comment construire une transformation de similarit\'{e} qui met en
correspondance les op\'{e}rateurs Hermitiques et les op\'{e}rateurs
quasi-Hermitiques correspondants et aussi les premiers \`{a} consid\'{e}rer
les transformations correspondantes des produits scalaires de l'espace de
Hilbert.

En 2002, Ali Mostafazadah \cite{mostafa,mos1,mos2} a soulign\'{e} qu'il
existe des Hamiltoniens qui ne sont ni Hermitiques ni $\mathcal{PT}$ sym\'{e}%
triques mais poss\`{e}dent des spectres r\'{e}els et positifs. Ce qui
l'amener \`{a} conclure que la $\mathcal{PT}$-sym\'{e}trie n'est pas
suffisante ou n\'{e}cessaire pour garantir la r\'{e}alit\'{e} du spectre;
mais que cel\`{a} rel\`{e}ve d'une th\'{e}orie plus g\'{e}n\'{e}rale dans
laquelle les Hamiltoniens sont pseudo-Hermitiques.

La pseudo-Hermiticit\'{e} permet de faire passer d'un Hamiltonien Hermitique 
\`{a} un Hamiltonien pseudo-Hermitique \'{e}quivalent, autrement dit, tout
Hamiltonien pseudo-Hermitique poss\`{e}de un Hamiltonien Hermitique \'{e}%
quivalent, les deux Hamiltoniens sont reli\'{e}s par la relation

\begin{equation}
h=\rho H\rho ^{-1},  \label{ch4.2.2}
\end{equation}%
\ o\`{u} $\rho $ est un op\'{e}rateur lin\'{e}aire, inversible et born\'{e}
connu sous le nom d'op\'{e}rateur de transformation de Dyson,$\ h$ et $H$ \
sont respectivement des Hamiltoniens Hermitiques et pseudo-Hermitiques.

Consid\'{e}rons les \'{e}quations aux valeurs propres des deux Hamiltoniens

\begin{equation}
h\left\vert \psi _{n}\right\rangle =E_{n}\left\vert \psi _{n}\right\rangle ,
\label{ch4.2.3}
\end{equation}%
et\qquad \qquad \qquad \qquad \qquad \qquad \qquad \qquad

\begin{equation}
H\left\vert \phi _{n}\right\rangle =E_{n}\left\vert \phi _{n}\right\rangle .
\label{ch4.2.4}
\end{equation}%
La transformation $\rho $ permet donc de relier les vecteurs propres de l'op%
\'{e}rateur non-Hermitique $H$ \`{a} ceux de l'op\'{e}rateur Hermitique $h$

\begin{equation}
\left\vert \phi _{n}\right\rangle =\rho ^{-1}\left\vert \psi
_{n}\right\rangle .  \label{ch4.2.5}
\end{equation}%
Sachant que tout Hamiltonien Hermitique poss\`{e}de un spectre r\'{e}el,
donc \`{a}\ partir de l'\'{e}quation (\ref{ch4.2.4}) on d\'{e}duit que le
spectre des Hamiltoniens pseudo-Hermitiques est r\'{e}el.

Les \'{e}quations de Schr\"{o}dinger correspondantes sont:

\begin{equation}
i\hbar \frac{\partial }{\partial t}\left\vert \Psi _{n}\right\rangle
=h\left\vert \Psi _{n}\right\rangle ,  \label{ch4.2.6}
\end{equation}%
et\qquad \qquad \qquad \qquad \qquad \qquad \qquad \qquad

\begin{equation}
i\hbar \frac{\partial }{\partial t}\left\vert \Phi _{n}\right\rangle
=H\left\vert \Phi _{n}\right\rangle ,  \label{ch4.2.7}
\end{equation}%
comme $h=h^{+}$ on a

\begin{equation}
h=\rho H\rho ^{-1}\rightarrow h^{+}=(\rho ^{-1})^{+}H^{+}\rho ^{+},
\label{ch4.2.8}
\end{equation}%
donc

\begin{equation}
\rho H\rho ^{-1}=(\rho ^{-1})^{+}H^{+}\rho ^{+},  \label{ch4.2.9}
\end{equation}

\begin{equation}
H^{+}=\eta H\eta ^{-1},  \label{ch4.2.10}
\end{equation}%
avec\qquad \qquad \qquad \qquad \qquad \qquad \qquad \qquad \qquad

\begin{equation}
\eta =\rho ^{+}\rho ,\text{ \ \ \ \ \ }\eta ^{-1}=\left( \rho ^{+}\rho
\right) ^{-1},  \label{ch4.2.11}
\end{equation}%
o\`{u} $\eta $ est un op\'{e}rateur lin\'{e}aire, Hermitique et inversible.
On dit qu'un Hamiltonien est pseudo-Hermitique ou quasi-Hermitique, s'il
satisfait la relation (\ref{ch4.2.10}).

\subsection{Le pseudo produit scalaire\qquad}

L'Hamiltonien Hermitique $h$ pr\'{e}serve le produit scalaire usuel c'est-%
\`{a}-dire

\begin{equation}
\langle \psi _{m}\left\vert \psi _{n}\right\rangle =\delta _{nm},
\label{ch4.2.12}
\end{equation}%
La transformation (\ref{ch4.2.5} ) d'exprimer ce produit scalaire en
fonction des \'{e}tats propres de l'Hamiltonien pseudo-Hermitique $H,$ en
effet

\begin{equation}
\langle \psi _{m}\left\vert \psi _{n}\right\rangle ==\left\langle \phi
_{m}\right\vert \rho ^{+}\rho \left\vert \phi _{n}\right\rangle
=\left\langle \phi _{m}\right\vert \eta \left\vert \phi _{n}\right\rangle
\equiv \langle \phi _{m}\left\vert \phi _{n}\right\rangle _{\eta }=\delta
_{nm},  \label{ch4.2.13}
\end{equation}%
on constate que le produit scalaire standard n'est pas pr\'{e}serv\'{e}. Le
pseudo produit scalaire (\ref{ch4.2.13}) est d\'{e}finit, positif et
conserve la norme, c'est \`{a} dire pr\'{e}serve l'unitarit\'{e} de l'\'{e}%
volution. En effet, les \'{e}quations (\ref{ch4.2.7}) et (\ref{ch4.2.13})
impliquent que

\begin{equation}
i\hbar \frac{\partial }{\partial t}\langle \Phi _{m}\left\vert \Phi
_{n}\right\rangle _{\eta }=\left\langle \Phi _{m}\right\vert \eta
H-H^{+}\eta \left\vert \Phi _{n}\right\rangle =0,  \label{ch4.2.14}
\end{equation}%
de ce fait les postulats \'{e}nonc\'{e}s au Chapitre 1 sont toujours
d'actualit\'{e} moyennant la notion du pseudo produit scalaire (\ref%
{ch4.2.13}) et (\ref{ch4.2.14}).

\subsection{Hamiltoniens pseudo-Hermitiques ayant une base bi-orthonorm\'{e}%
e compl\`{e}te}

Les Hamiltonien pseudo Hermitiques poss\`{e}dent une base bi-orthonorm\'{e}e%
\begin{equation}
\langle \chi _{m}\left\vert \phi _{n}\right\rangle =\delta _{nm},
\label{ch4.2.15}
\end{equation}

o\`{u} $\left\vert \phi _{n}\right\rangle $ et $\left\vert \chi
_{n}\right\rangle $ sont respectivement les fonctions propres de $H$ et de $%
H^{+}$ v\'{e}rifiant les \'{e}quations aux valeurs propres 
\begin{equation}
H\left\vert \phi _{n}\right\rangle =E_{n}\left\vert \phi _{n}\right\rangle ,
\label{ch4.2.16}
\end{equation}%
\begin{equation}
H^{+}\left\vert \chi _{n}\right\rangle =E_{n}\left\vert \chi
_{n}\right\rangle ,  \label{ch4.2.17}
\end{equation}%
et la relation de fermeture est donn\'{e}e par :

\begin{equation}
\sum_{n}\left\vert \chi _{n}\right\rangle \left\langle \phi _{n}\right\vert
=\sum_{n}\left\vert \phi _{n}\right\rangle \left\langle \chi _{n}\right\vert
=1.  \label{ch4.2.18}
\end{equation}%
Dans cette base $H$ et $H^{+}$ sont donn\'{e}s respectivement par:%
\begin{equation}
H=\underset{n}{\sum }\left\vert \psi _{n}\right\rangle E_{n}\left\langle
\phi _{n}\right\vert ,\text{ \ \ \ \ \ \ \ }H^{+}=\underset{n}{\sum }%
\left\vert \phi _{n}\right\rangle E_{n}\left\langle \psi _{n}\right\vert .
\label{ch4.2.19}
\end{equation}

En utilisant (\ref{ch4.2.15}) et (\ref{ch4.2.18}) et dans le cas des valeurs
propres non d\'{e}g\'{e}n\'{e}r\'{e}es, on peut construire la m\'{e}trique 
\`{a} partir des vecteurs propres de l'op\'{e}rateur non-Hermitique $H$ et
son conjugu\'{e} Hermitique $H^{+}$, on a 
\begin{equation}
\eta =\sum_{n}\left\vert \chi _{n}\right\rangle \left\langle \chi
_{n}\right\vert ,  \label{ch4.2.20}
\end{equation}%
et son inverse est

\begin{equation}
\eta ^{-1}=\sum_{n}\left\vert \phi _{n}\right\rangle \left\langle \phi
_{n}\right\vert .  \label{ch4.2.21}
\end{equation}%
par contre l'op\'{e}rateur de transformation de Dyson $\rho $ et son inverse 
$\rho ^{-1}$ sont donn\'{e}s par

\begin{equation}
\rho =\sum_{n}\left\vert \psi _{n}\right\rangle \left\langle \chi
_{n}\right\vert ,\text{ \ \ \ \ \ \ }\rho ^{-1}=\sum_{n}\left\vert \phi
_{n}\right\rangle \left\langle \psi _{n}\right\vert .
\end{equation}%
Notons que dans le cas des Hamiltoniens $\mathcal{PT}$-sym\'{e}triques, le r%
\^{o}le de $\eta $ est jou\'{e} par $\mathcal{PC}$\ \cite{jones}:

\begin{equation}
\eta =\mathcal{PC}
\end{equation}%
\qquad \qquad \qquad \qquad \qquad \qquad \qquad \qquad

\subsection{Application:}

\subsubsection{Etude du syst\`{e}me \`{a} deux niveaux \guillemotleft\ %
Brachistochrone \guillemotright :\ cas pseudo-Hermitique}

On reprend l'exemple pr\'{e}c\'{e}dent (\ref{ch4.1.26}) dont les valeurs
propres sont donn\'{e}es par(\ref{ch4.1.27}). L'op\'{e}rateur m\'{e}trique $%
\eta =\rho ^{+}\rho $ qui par son action transforme $H$ en $H^{+}=\eta H\eta
^{-1}$ est donn\'{e} par :

\begin{equation}
\eta =\frac{1}{\cos \alpha }%
\begin{pmatrix}
1 & -i\sin \alpha \\ 
i\sin \alpha & 1%
\end{pmatrix}%
,  \label{ch4.2.22}
\end{equation}%
En calculant $\eta ^{1/2}$, nous obtenons\ $\rho $ \ sous la forme suivante

\begin{equation}
\rho =\frac{1}{\sqrt{\cos \alpha }}%
\begin{pmatrix}
\sin \frac{\alpha }{2} & -i\cos \frac{\alpha }{2} \\ 
i\cos \frac{\alpha }{2} & \sin \frac{\alpha }{2}%
\end{pmatrix}%
,  \label{ch4.2.23}
\end{equation}%
ainsi nous d\'{e}duisons facilement \`{a} partir (\ref{ch4.2.2})
l'Hamiltonien Hermitique $h$ ,

\begin{equation}
h=\rho H\rho ^{-1}=%
\begin{pmatrix}
r\cos \theta & -\frac{\omega }{2} \\ 
-\frac{\omega }{2} & r\cos \theta%
\end{pmatrix}%
,  \label{ch4.2.24}
\end{equation}%
o\`{u}

\begin{equation}
\omega =2\sqrt{s%
{{}^2}%
-r%
{{}^2}%
\sin ^{2}\theta }.  \label{ch4.2.25}
\end{equation}%
\ Les valeurs propres de $h$ sont donn\'{e}e par : $\varepsilon _{\pm
}=r\cos \alpha \pm s\cos \alpha .$ Les \'{e}tats propres $\left\vert \phi
_{\pm }\right\rangle $ de $H$ sont reli\'{e}s \`{a} ceux de $h$ par:

\begin{equation}
\left\vert \phi _{\pm }\right\rangle =\rho ^{-1}\left\vert \psi _{\pm
}\right\rangle .  \label{ch4.2.26}
\end{equation}%
d'o\`{u}

\begin{eqnarray}
\left\vert \phi _{+}\right\rangle &=&\frac{1}{\sqrt{2\cos \alpha }}%
\begin{pmatrix}
e^{i\frac{\alpha }{2}} \\ 
e^{-i\frac{\alpha }{2}}%
\end{pmatrix}
\label{ch4.2.27} \\
\text{et }\left\vert \phi _{-}\right\rangle &=&\frac{i}{\sqrt{2\cos \alpha }}%
\begin{pmatrix}
e^{-i\frac{\alpha }{2}} \\ 
-e^{i\frac{\alpha }{2}}%
\end{pmatrix}%
.
\end{eqnarray}%
On peut facilement v\'{e}rifier que :

\begin{equation}
\left\langle \phi _{\pm }\right\vert \eta \left\vert \phi _{\pm
}\right\rangle =1.  \label{ch4.2.28}
\end{equation}

\qquad \qquad \qquad \qquad \qquad \qquad \qquad \qquad

\chapter{Syst\`{e}mes quantiques d\'{e}pendants du temps}

Il est g\'{e}n\'{e}ralement impossible de r\'{e}soudre l'\'{e}quation de Schr%
\"{o}dinger

\begin{equation}
i\hbar \partial _{t}\left\vert \Psi (t)\right\rangle =H\left( t\right)
\left\vert \Psi (t)\right\rangle ,  \label{ch3.1}
\end{equation}%
associ\'{e}e \`{a} un Hamiltonien $H(t)$ qui d\'{e}pend du temps \`{a}
travers des param\`{e}tres classiques \{$R_{i}$ $(t)$, $i=1,2,\ldots $\} qui
peuvent repr\'{e}senter l'interaction du syst\`{e}me consid\'{e}r\'{e} avec
des champs externes d\'{e}pendant du temps (champ \'{e}lectrique, champ magn%
\'{e}tique, \ldots .) ou, tout simplement, des param\`{e}tres internes d\'{e}%
pendant du temps (la masse, la charge, la profondeur d'un puits de
potentiel, \ldots ) \cite{Messiah,teu}. En g\'{e}n\'{e}rale, le but est de
trouver la solution $\left\vert \Psi (t)\right\rangle $ de l'\'{e}quation de
Schr\"{o}dinger (\ref{ch3.1}) \`{a} partir de la donn\'{e}e de la condition
initiale $\left\vert \Psi (t_{0})\right\rangle $. Il est possible de
reformuler la dynamique des syst\`{e}mes quantiques \`{a} l'aide de l'op\'{e}%
rateur d'\'{e}volution $U(t,t_{0})$ (Voir \'{e}quation (\ref{ch2.2})).

Pour r\'{e}soudre l'\'{e}quation de Schr\"{o}dinger (\ref{ch3.1}), Il est n%
\'{e}cessaire d'user d'approximations telle que: la th\'{e}orie des
perturbations d\'{e}pendante du temps , l'approximation soudaine ,
l'approximation adiabatique ou \ aussi la m\'{e}thode exacte mais formelle
de la th\'{e}orie des invariants de Lewis et Riesenfeld \cite{lewis,lewis2}.

Les premiers travaux sur l'approximation adiabatique en m\'{e}canique
quantique sont dus \`{a} M. Born et V. Fock \cite{fock} qui constituent une
extension des travaux d'Ehrenfest \cite{ehr} en m\'{e}canique classique et
l'ancienne th\'{e}orie des Quanta \cite{Messiah,teu,grif}. Mais \`{a} partir
des ann\'{e}es cinquante, il y a eu une \'{e}tude intense du sujet \cite{gal}%
. Des applications pratiques ont \'{e}t\'{e} trouv\'{e}es dans la physique
des plasmas, la technologie de fusion, les acc\'{e}l\'{e}rateurs des
particules charg\'{e}es, et m\^{e}me dans l'astronomie galactique \cite{avr}.

\section{R\'{e}gimes soudain et adiabatique}

On proc\`{e}de au changement de variable dans l'\'{e}quation de Schr\"{o}%
dinger consistant \`{a} passer en temps r\'{e}duit $s=\frac{t}{T}\in \left[
0,1\right] $ o\`{u} $T$ est la dur\'{e}e de l'interaction consid\'{e}r\'{e}e.%
\begin{equation}
i\hbar \frac{dU_{T}(s,0)}{ds}=TH(s)U_{T}(s,0).
\end{equation}%
On s'int\'{e}resse aux deux r\'{e}gimes dynamiques extr\^{e}mes, le r\'{e}%
gime soudain o\`{u} le param\`{e}tre $T\ll $ \ (l'interaction est tr\`{e}s
rapide), et le r\'{e}gime adiabatique o\`{u} $T\gg $ (l'interaction est tr%
\`{e}s lente).

\subsection{Approximation soudaine}

Dans le r\'{e}gime soudain on a

\begin{equation}
U_{T}(s,0)=1-\frac{i}{\hbar }T\int_{0}^{s}H(s^{\prime })U_{T}(s^{\prime
},0)ds^{\prime }=1+O(T).
\end{equation}%
En premi\`{e}re approximation, pour une interaction tr\`{e}s rapide, l'op%
\'{e}rateur d'\'{e}volution est r\'{e}duit \`{a} l'identit\'{e}. Le syst\`{e}%
me n'ayant pas le temps de s'adapter \`{a} la modification, il reste sur son 
\'{e}tat originel. Supposons par exemple que $H(t)=H_{0}$ pour $t$ $\langle $
$0$ et $H(t)=H_{1}$ pour $t$ $\rangle $ $0$ avec $T$ au voisinage de $0$. Si
l'\'{e}tat initial du syst\`{e}me \'{e}tait un \'{e}tat propre $\psi _{0}$
de $H_{0}$, apr\`{e}s l'interaction $\left\vert \psi _{0}\right\rangle $ est
toujours l'\'{e}tat du syst\`{e}me. Il n'est plus n\'{e}anmoins \'{e}tat
propre de l'Hamiltonien qui est devenu $H_{1}$ et un ph\'{e}nom\`{e}ne
d'oscillations de Rabi doit appara\^{\i}tre.

L'erreur commise dans l'approximation soudaine est

\begin{equation}
e=\left\langle \psi _{0}\right\vert U_{T}(1,0)^{+}\left( 1-\left\vert \psi
_{0}\right\rangle \left\langle \psi _{0}\right\vert U_{T}(1,0)\left\vert
\psi _{0}\right\rangle \right) ,
\end{equation}%
avec $\left\vert \psi _{0}\right\rangle $ \'{e}tat initial norm\'{e}. En
utilisant la d\'{e}composition de $U_{T}(1,0)$ en s\'{e}rie de Dyson, il
vient

\begin{eqnarray}
e &=&\frac{T^{2}}{\hbar ^{2}}\int_{0}^{1}\int_{0}^{1}\left\langle \psi
_{0}\right\vert H(s_{1})H(s_{2})\left\vert \psi _{0}\right\rangle
ds_{1}ds_{2} \\
&&-\frac{T^{2}}{\hbar ^{2}}\int_{0}^{1}\int_{0}^{1}\left\langle \psi
_{0}\right\vert H(s_{1})\left\vert \psi _{0}\right\rangle \left\langle \psi
_{0}\right\vert H(s_{2})\left\vert \psi _{0}\right\rangle
ds_{1}ds_{2}+O(T^{3}).
\end{eqnarray}%
On note $\triangle \bar{H}^{2}=$ $\left\langle \bar{H}^{2}\right\rangle
-\left\langle \bar{H}\right\rangle ^{2}$ la variance dans l'\'{e}tat $\psi
_{0}$ de la moyenne temporelle de l'Hamiltonien $\bar{H}=\int_{0}^{1}H(s)ds$%
, on a alors

\begin{equation}
e=\frac{T^{2}}{\hbar ^{2}}\triangle \bar{H}+O(T^{3}).
\end{equation}%
On voit donc que la condition pour que l'approximation soudaine soit valide
est

\begin{equation}
T\ll \frac{\hbar }{\triangle \bar{H}}.
\end{equation}

\subsection{Approximation adiabatique}

L'approche utilis\'{e}e sera la m\^{e}me que celle propos\'{e}e pas M.V.
Berry \cite{berry} qui introduit l'hypoth\`{e}se adiabatique d\`{e}s le d%
\'{e}part. En \'{e}tudiant l'\'{e}volution des syst\`{e}mes physiques
quantiques r\'{e}gies par un Hamiltonien d\'{e}pendant de param\`{e}tres qui
varient tr\`{e}s lentement en fonction du temps, Berry \cite{berry} a d\'{e}%
couvert un effet g\'{e}om\'{e}trique relatif \`{a} la phase des \'{e}tats
stationnaires. Cet effet apporte un compl\'{e}ment important \`{a} l'\'{e}%
nonc\'{e} du th\'{e}or\`{e}me adiabatique \cite{Messiah}. qui stipule qu'un
syst\`{e}me initialement dans un \'{e}tat stationnaire non d\'{e}g\'{e}n\'{e}%
r\'{e}, rep\'{e}r\'{e} par un ensemble donn\'{e} de nombres quantiques,
restera dans un \'{e}tat sp\'{e}cifi\'{e} par les m\^{e}mes nombres
quantiques lors d'une \'{e}volution adiabatique. Pour fixer les notations
soit $H(s)\equiv H(\vec{R}(s))$ un Hamiltonien admettant des vecteurs
propres non d\'{e}g\'{e}n\'{e}r\'{e}s $\left\vert \phi _{i}(s)\right\rangle $
et de valeurs propres $\lambda _{i}(s)$ 
\begin{equation}
H(s)\left\vert \phi _{i}(s)\right\rangle =\lambda _{i}(s)\left\vert \phi
_{i}(s)\right\rangle .
\end{equation}%
La formulation usuelle du th\'{e}or\`{e}me adiabatique exprime que l'\'{e}%
tat initial $\left\vert \psi (0)\right\rangle =\left\vert \phi
_{i}(0)\right\rangle $ \'{e}tat propre de l'Hamiltonien \`{a} l'instant z%
\'{e}ro $H(0)$, \'{e}volue en un \'{e}tat $\left\vert \psi (t)\right\rangle $
qui, \`{a} tout instant, reste \'{e}tat propre de l'Hamiltonien $H(s)$. la
remarque de Berry est que la phase $\varphi _{i}$ de cet \'{e}tat, d\'{e}%
finit par rapport aux \'{e}tats de r\'{e}f\'{e}rence $\left\vert \phi
_{i}(s)\right\rangle $ par 
\begin{equation}
\left\vert \psi (t)\right\rangle =\exp \left( i\varphi _{i}(t)\right)
\left\vert \phi _{i}(s)\right\rangle ,
\end{equation}%
est enti\`{e}rement d\'{e}termin\'{e}e si on impose \`{a}\ $\left\vert \psi
(t)\right\rangle $ de satisfaire l'\'{e}quation de Schr\"{o}ndinger 
\begin{equation}
\frac{i\hbar }{T}\frac{\partial }{\partial s}\left\vert \psi
(s)\right\rangle =H(s)\left\vert \psi (s)\right\rangle .
\end{equation}%
\textquotedblleft en moyenne\textquotedblright , c'est-\`{a}-dire en
projetant cette \'{e}galit\'{e} sur l'\'{e}tat \ $\left\vert \psi
(s)\right\rangle $ lui-m\^{e}me 
\begin{equation}
\frac{1}{T}\left\langle \psi (s)\right\vert i\hbar \frac{\partial }{\partial
s}\left\vert \psi (s)\right\rangle =\left\langle \psi (s)\right\vert
H(s)\left\vert \psi (s)\right\rangle .
\end{equation}%
On obtient alors une expression explicite pour $\varphi _{i}$ 
\begin{equation*}
\varphi _{i}(t)=[\delta _{i}(t)+\gamma _{i}(t)]
\end{equation*}%
qui contient deux termes. Le premier , appel\'{e} phase dynamique, 
\begin{equation}
\delta _{i}(t)=-\hbar ^{-1}\int_{0}^{s}\lambda _{i}(s^{\prime })ds^{\prime };%
\text{ \ \ \ \ \ \ }\lambda _{i}(s)=\langle \phi _{i}(s)\mid H(s)\mid \phi
_{i}(s)\rangle
\end{equation}%
est `'attendu\textquotedblright\ car il est pr\'{e}sent m\^{e}me si les param%
\`{e}tres ne d\'{e}pendent pas du temps. Le second est la phase de Berry donn%
\'{e}e par

\begin{equation}
\;\gamma _{i}(t)=\int_{\vec{R}(0)}^{\vec{R}(t)}\overrightarrow{A}_{i}(\vec{R}%
(s))d\vec{R};\ \ \ \ \ \ \ \overrightarrow{A}_{i}(\vec{R}(s))=\langle \phi
_{i}(\vec{R}(s))\mid i\nabla _{\vec{R}}\mid \phi _{i}(\vec{R}(s))\rangle .
\end{equation}%
(on v\'{e}rifie que $\overrightarrow{A}_{i}$ et $\gamma _{i}$ sont r\'{e}%
els.) Cette phase est aussi appel\'{e}e phase g\'{e}om\'{e}trique. Son caract%
\`{e}re g\'{e}om\'{e}trique est justifi\'{e} par le fait que, lorsque les
param\`{e}tres effectuent (adiabatiquement) un cycle $\mathcal{C}$, $\gamma
_{i}$ ne d\'{e}pend que du chemin suivi dans l'espace des param\`{e}tres. En
effet, m\^{e}me si on change la base des vecteurs propres de r\'{e}f\'{e}%
rence par une `'transformation de jauge\textquotedblright\ $\mid \phi _{i}(%
\vec{R}(s))\rangle \longrightarrow e^{i\varphi _{i}(\vec{R})}\mid \phi _{i}(%
\vec{R}(s))\rangle $\ , \ ce qui modifie le `'potentiel
vecteur\textquotedblright\ par un terme de gradient $\vec{A}%
_{i}\longrightarrow \vec{A}_{i}-\nabla _{\vec{R}}\varphi _{i}(\vec{R})$ (et
donc modifie $\gamma _{i}(t)$ ), la phase de Berry pour un cycle $\;\gamma
_{i}(t)=\oint_{c}\vec{A}_{i}(\vec{R})d\vec{R}$ reste, elle, inchang\'{e}e.

L'appellation `'potentiel vecteur\textquotedblright\ pour $\vec{A}_{i}$\
n'est pas innocente. Pour le voir, il faut remplacer l'int\'{e}grale de
ligne par une int\'{e}grale de surface au moyen du th\'{e}or\`{e}me de
Stokes: 
\begin{align}
\gamma_{i}(C) 
&= \oint_{c} \vec{A}_{i}(\vec{R}) \, d\vec{R}
   = -\operatorname{Im} \iint_{s} d\vec{S} \cdot \left( \vec{\nabla} \times \vec{A}_{i} \right) \notag \\
&= -\operatorname{Im} \iint_{s} d\vec{S} \cdot 
   \langle \vec{\nabla}\phi_{i}(\vec{R}(s)) \mid \times \mid \vec{\nabla}\phi_{i}(\vec{R}(s)) \rangle
\end{align}

o\`{u} $\times $ est le produit vectoriel habituel. \ La relation de
fermeture donne%
\begin{equation*}
\gamma _{i}(C)=-\operatorname{Im}{Im}\iint_{s}d\vec{S}.\sum_{j\neq i}\langle \phi _{i}(%
\vec{R}(s))\mid \times \mid \phi _{j}(\vec{R}(s))\rangle \times \langle \phi
_{j}(\vec{R}(s))\mid \vec{\nabla}\mid \phi _{i}(\vec{R}(s))\rangle
\end{equation*}%
La r\'{e}duction de la somme \`{a} $j\neq i$ est due au fait que l'\'{e}tat
interm\'{e}diaire \ $j=i$ n'apporte aucune contribution puisque la
connection $\langle \phi _{i}(\vec{R}(s))\mid \vec{\nabla}_{\vec{R}}\mid
\phi _{i}(\vec{R}(s))\rangle $ est purement imaginaire. Comme

\begin{equation}
\langle \phi _{j}(\vec{R}(s)\mid \vec{\nabla}_{\vec{R}}\mid \phi _{i}(\vec{R}%
(s))\rangle =\frac{\langle \phi _{j}(\vec{R}(s)\mid \vec{\nabla}_{\vec{R}}H(%
\vec{R})\mid \phi _{i}(\vec{R}(s))\rangle }{\lambda _{i}(s)-\lambda _{j}(s)}%
,\;\;\;\;\;j\neq i
\end{equation}%
la phase de Berry peut s'\'{e}crire sous forme d'une expression analogue 
\`{a} celle du flux magn\'{e}tique en \'{e}lectromagn\'{e}tisme

\begin{equation}
\gamma _{i}(C)=\iint d\vec{S}.\vec{V}_{i}
\end{equation}%
o\`{u} le champ $\vec{V}_{n}$ est l'analogue du champ magn\'{e}tique et donc
de la courbure dans l'espace des param\`{e}tres :

\begin{equation}
\vec{V}_{i}=-\operatorname{Im}{Im}\sum_{m\neq n}\frac{\langle \phi _{i}(\vec{R}(s))\mid 
\vec{\nabla}_{\vec{R}}H(\vec{R})\mid \phi _{j}(\vec{R}(s))\rangle \times
\langle \phi _{j}(\vec{R}(s)\mid \vec{\nabla}_{\vec{R}}H(\vec{R})\mid \phi
_{i}(\vec{R}(s))\rangle }{(\lambda _{i}(s)-\lambda _{j}(s))^{2}},
\end{equation}%
Signalons enfin qu'un tel potentiel avait aussi \'{e}t\'{e} introduit par
Mead et Truhlar \cite{mead} pour d\'{e}crire, dans le cadre de
l'approximation de Born-Oppenheimer, la r\'{e}troaction sur les mouvements
des noyaux de la phase induite sur les \'{e}tats \'{e}lectroniques par ce m%
\^{e}me mouvement. On le rencontre chaque fois que dans un probl\`{e}me
quantique il est possible de s\'{e}parer des variables lentes et des
variables rapides.

On terminera cette analyse par remarquer que la condition de validit\'{e} de
l'approximation adiabatique est

\begin{equation*}
T\gg \hbar \sup_{s\in \left[ 0,1\right] }\left\vert \frac{\langle \phi _{i}(%
\vec{R}(s))\mid \frac{d}{ds}\mid \phi _{i}(\vec{R}(s))\rangle }{\lambda
_{i}(s)-\lambda _{j}(s)}\right\vert ,\forall i\neq j.
\end{equation*}%
Ce qui suppose que la dur\'{e}e de l'interaction $T$ soit beaucoup plus
grande que les dur\'{e}es de transition entre les \'{e}tats propres $\frac{%
\hbar }{\left\vert \lambda _{i}(s)-\lambda _{j}(s)\right\vert }$ et que les
couplages non-adiabatiques $\langle \phi _{i}(\vec{R}(s))\mid \frac{d}{ds}%
\mid \phi _{i}(\vec{R}(s))\rangle $ soient petits. Physiquement, l'hypoth%
\`{e}se adiabatique signifie que le taux (la vitesse) de transition entre 
\'{e}tats propres est petit par rapport \`{a} la fr\'{e}quence de Bohr $%
\frac{\left( \lambda _{i}(s)-\lambda _{j}(s)\right) }{2\pi \hbar }$ pour la
transition $i\rightarrow j$. Autrement dit, les transitions entre \'{e}tats
propres diff\'{e}rents sont donc infiniment lentes et n'ont en fait pas
lieu. Si le syst\`{e}me se trouve initialement dans le n\`{e}me \'{e}tat
propre, il y restera toujours.

\section{G\'{e}n\'{e}ralisation \`{a} la th\'{e}orie des invariants}

\subsection{Th\'{e}orie quantique des invariants}

Certaines grandeurs physiques scalaires et vectorielles se conservent au
cours du mouvement. Elles servent alors \`{a} caract\'{e}riser le mouvement.
On les appelle invariants ou constantes du mouvement. La th\'{e}orie des
invariants (ou de Lewis et Riesenfeld) \cite{lewis,lewis2} constitue une m%
\'{e}thode de r\'{e}solution l'\'{e}quation de Schr\"{o}dinger d\'{e}%
pendante du temps, elle permet d'obtenir \ la solution de l'\'{e}quation de
Schr\"{o}dinger (\ref{ch3.1}) en fonction des \'{e}tats propres de l'op\'{e}%
rateur invariant Hermitique $\hat{I}(t)\equiv $ $I(t)$ qui v\'{e}rifie l'%
\'{e}quation de Von-Neuman

\begin{equation}
\frac{dI(t)}{dt}=\frac{\partial I(t)}{\partial t}+\frac{1}{i\hbar }\left[
I(t),H(t)\right] =0.\text{ \ }  \label{ch3.2}
\end{equation}%
Remarquons que l'action de l'op\'{e}rateur invariant $I(t)$ sur un vecteur d'%
\'{e}tat $\left\vert \Psi (t)\right\rangle $ solution de l'\'{e}quation de
Schr\"{o}dinger (\ref{ch3.1}) est aussi solution de l'\'{e}quation Schr\"{o}%
dinger 
\begin{equation}
i\hbar \frac{\partial }{\partial t}\left( I(t)\left\vert \Psi
(t)\right\rangle \right) =H(t)\left( I(t)\left\vert \Psi (t)\right\rangle
\right) .  \label{ch3.3}
\end{equation}%
Supposons que l'op\'{e}rateur invariant $I(t)$ admet un ensemble d'\'{e}tats
propres $\left\vert \phi _{\lambda ,k}\right\rangle $%
\begin{equation}
I(t)\left\vert \phi _{\lambda ,k}(t)\right\rangle =\lambda \left\vert \phi
_{\lambda ,k}(t)\right\rangle ,  \label{ch3.4}
\end{equation}%
avec des valeurs propres $\lambda $, o\`{u} $k$ repr\'{e}sente tous les
autres nombres quantiques n\'{e}cessaire sp\'{e}cifiant les \'{e}tats
propres de $I(t)$ (car cet op\'{e}rateur peut avoir un spectre d\'{e}g\'{e}n%
\'{e}r\'{e}). Ces fonctions propres sont suppos\'{e}es orthonorm\'{e}es 
\begin{equation}
\left\langle \phi _{\lambda ^{\prime },k^{\prime }}(t)\right\vert \left.
\phi _{\lambda ,k}(t)\right\rangle =\delta _{\lambda ^{\prime },\lambda
}\delta _{k^{\prime },k}.  \label{ch3.5}
\end{equation}%
En vertu de l\textbf{'}hermiticit\'{e} de $I(t)$, les valeurs propres $%
\lambda $\ sont r\'{e}elles et ind\'{e}pendantes du temps. En effet, la d%
\'{e}riv\'{e}e par rapport au temps de l'\'{e}quation $\left( \ref{ch3.4}%
\right) $ donne 
\begin{equation}
\frac{\partial I}{\partial t}\left\vert \phi _{\lambda ,k}(t)\right\rangle +I%
\frac{\partial }{\partial t}\left\vert \phi _{\lambda ,k}(t)\right\rangle =%
\frac{\partial \lambda }{\partial t}\left\vert \phi _{\lambda
,k}(t)\right\rangle +\lambda \frac{\partial }{\partial t}\left\vert \phi
_{\lambda ,k}(t)\right\rangle ,  \label{ch3.6}
\end{equation}%
multiplions ensuite \`{a} gauche par $\left\langle \phi _{\lambda ,k^{\prime
}}\right\vert $, on aura 
\begin{equation}
\frac{\partial \lambda }{\partial t}=\left\langle \phi _{\lambda
,k}(t)\right\vert \frac{\partial I}{\partial t}\left\vert \phi _{\lambda
,k^{\prime }}(t)\right\rangle .  \label{ch3.7}
\end{equation}%
La valeur moyenne de $\left( \ref{ch3.2}\right) $ dans les \'{e}tats $%
\left\vert \phi _{\lambda ,k}(t)\right\rangle $ s'\'{e}crit%
\begin{equation}
i\hbar \left\langle \phi _{\lambda ^{\prime },k^{\prime }}(t)\right\vert 
\frac{\partial I}{\partial t}\left\vert \phi _{\lambda ,k}(t)\right\rangle
+\left( \lambda ^{\prime }-\lambda \right) \left\langle \phi _{\lambda
^{\prime },k^{\prime }}(t)\right\vert H\left\vert \phi _{\lambda
,k}(t)\right\rangle =0,  \label{ch3.8}
\end{equation}%
qui implique que pour $\lambda ^{\prime }=\lambda $ 
\begin{equation}
\left\langle \varphi _{\lambda ,k}(t)\right\vert \frac{\partial I}{\partial t%
}\left\vert \varphi _{\lambda ,k^{\prime }}(t)\right\rangle =0,
\label{ch3.9}
\end{equation}%
d'o\`{u} on d\'{e}duit que les valeurs propres sont constantes (ind\'{e}%
pendantes du temps).

Le fait que les valeurs propres $\lambda $ sont constantes permet de faire
le lien entre les \'{e}tats propres de $I(t)$ et les solutions de l'\'{e}%
quation de Schr\"{o}dinger, ainsi l'\'{e}quation $\left( \ref{ch3.6}\right) $
multipli\'{e} \`{a} gauche par $\left\langle \phi _{\lambda ^{\prime
},k^{\prime }}(t)\right\vert $ donne 
\begin{equation}
\left( \lambda -\lambda ^{\prime }\right) \left\langle \phi _{\lambda
^{\prime },k^{\prime }}(t)\right\vert \frac{\partial }{\partial t}\left\vert
\phi _{\lambda ,k}(t)\right\rangle =\left\langle \phi _{\lambda ^{\prime
},k^{\prime }}(t)\right\vert \frac{\partial I}{\partial t}\left\vert \phi
_{\lambda ,k}(t)\right\rangle ,  \label{ch3.10}
\end{equation}%
qui, pour $\lambda ^{\prime }\neq \lambda ,$ permet d'\'{e}crire l'\'{e}%
quation $\left( \ref{ch3.8}\right) $ sous la forme suivante

\begin{equation}
i\hbar \left\langle \phi _{\lambda ^{\prime },k^{\prime }}\right\vert \frac{%
\partial }{\partial t}\left\vert \phi _{\lambda ,k}\right\rangle
=\left\langle \phi _{\lambda ^{\prime },k^{\prime }}\right\vert H\left\vert
\phi _{\lambda ,k}\right\rangle .  \label{ch3.11}
\end{equation}

On aurait pu d\'{e}duire imm\'{e}diatement que les fonctions propres $%
\left\vert \phi _{\lambda ,k}(t)\right\rangle $ sont des solutions de l'\'{e}%
quation de Schr\"{o}dinger si $\lambda =\lambda ^{\prime }$. Cela pourrait 
\^{e}tre le cas si on utilisait le fait que les phases des \'{e}tats
stationnaires ne sont pas fix\'{e}es. En effet, on peut donc tr\`{e}s bien
multiplier $\left\vert \varphi _{\lambda ,k}(t)\right\rangle $ par un
facteur de phase d\'{e}pendant du temps: \ 
\begin{equation}
\left\vert \phi _{\lambda ,k}(t)\right\rangle _{\alpha }\equiv \exp \left[
i\alpha _{\lambda k}(t)\right] \left\vert \phi _{\lambda ,k}(t)\right\rangle
,  \label{ch3.12}
\end{equation}%
o\`{u} $\alpha _{\lambda k}(t)$ est une fonction r\'{e}elle du temps
arbitrairement choisie. Ces $\left\vert \phi _{\lambda ,k}(t)\right\rangle
_{\alpha }$ sont aussi des \'{e}tats propres orthonorm\'{e}s de $I(t)$ associ%
\'{e}s aux valeurs propres $\lambda .$ Si on choisit bien les phases $\alpha
_{\lambda k}(t),$ l'\'{e}quation $\left( \ref{ch3.11}\right) $ sera v\'{e}%
rifi\'{e}e pour $\lambda =\lambda ^{\prime }$ et donc l'objectif sera
atteint. Il faut juste avoir les phase $\alpha _{\lambda k}(t)$ tel que 
\begin{equation}
\hbar \delta _{kk^{\prime }}\frac{d\alpha _{\lambda k}}{dt}=\left\langle
\phi _{\lambda ,k^{\prime }}(t)\right\vert \left( i\hbar \frac{\partial }{%
\partial t}-H\right) \left\vert \phi _{\lambda ,k}(t)\right\rangle .
\label{ch3.13}
\end{equation}%
Ce choix montre que l'\'{e}quation $\left( \ref{ch3.11}\right) $ pour $%
\left\vert \varphi _{\lambda ,k}\right\rangle _{\alpha }$ est v\'{e}rifi\'{e}%
e pour $\lambda =\lambda ^{\prime }\ $et les \'{e}l\'{e}ment non diagonaux $%
\left\langle \phi _{\lambda ,k^{\prime }}(t)\right\vert \left( i\hbar \frac{%
\partial }{\partial t}-H\right) \left\vert \phi _{\lambda
,k}(t)\right\rangle $ sont identiquement nuls. Pour $k=k^{\prime },$ la
phase $\alpha _{\lambda k}(t)$ v\'{e}rifie l'\'{e}quation:%
\begin{equation}
\hbar \frac{d\alpha _{\lambda k}}{dt}=\left\langle \phi _{\lambda
,k}(t)\right\vert \left( i\hbar \frac{\partial }{\partial t}-H\right)
\left\vert \phi _{\lambda ,k}(t)\right\rangle .  \label{ch3.14}
\end{equation}%
La solution de l'\'{e}quation de Schr\"{o}dinger s'\'{e}crit comme une
combinaison lin\'{e}aire des \'{e}tats propres%

\begin{equation}
\left\vert \Psi (t)\right\rangle =\underset{\lambda k}{\sum }C_{\lambda
k}(0)\exp \left[ i\alpha _{\lambda k}(t)\right] \left\vert \phi _{\lambda
,k}(t)\right\rangle .  \label{cha3.15}
\end{equation}

\subsection{Evolutions p\'{e}riodiques et th\'{e}orie de Floquet}

Les \'{e}volutions r\'{e}gies par des Hamiltoniens p\'{e}riodiques

\begin{equation}
H\left( t+T\right) =H\left( t\right) .  \label{ch3.2.1}
\end{equation}%
rel\`{e}vent, elles aussi, de l'approche des invariants (ou de Lewis et
Riesenfeld). En effet la d\'{e}composition de Floquet de l'op\'{e}rateur d'%
\'{e}volution \cite{floq,shir,ara,gri,hol} 
\begin{equation}
U\left( t\right) =Z\left( t\right) e^{iMt}\text{ },\text{ \ \ \ }Z\left(
t+T\right) =Z\left( t\right) ,\text{ \ \ \ \ }Z\left( T\right) =Z\left(
0\right) =1,  \label{ch3.2.2}
\end{equation}%
montre que les \'{e}tats $\ $%
\begin{equation}
\left\vert \phi _{j}(t)\right\rangle =Z\left( t\right) \left\vert \phi
_{j}(0)\right\rangle ,
\end{equation}%
sont des \'{e}tats propres de l'op\'{e}rateur invariant 
\begin{equation}
I(t)=Z\left( t\right) MZ^{+}\left( t\right) ,  \label{ch3.2.3}
\end{equation}
\ 
\begin{equation}
I(t)\left\vert \phi _{\lambda ,k}(t)\right\rangle =\lambda \left\vert \phi
_{\lambda ,k}(t)\right\rangle .
\end{equation}%
et o\`{u} $\left\vert \phi _{j}(0)\right\rangle $, \`{a} l'instant $t=0,$
sont des \'{e}tats propres de l'op\'{e}rateur Hermitique $M.$ A partir de l'%
\'{e}quation (\ref{ch2.2}), on d\'{e}duit l'Hamiltonien du syst\`{e}me se d%
\'{e}compose sous la forme suivante:%
\begin{equation}
H=i\text{ }\hbar \dot{Z}\text{ }Z^{+}-\hbar Z\left( t\right) \text{ }M\text{ 
}Z^{+}\left( t\right) .  \label{ch3.2.4}
\end{equation}

\chapter{Les syst\`{e}mes non Hermitiques d\'{e}pendants du temps}

\section{Hamiltonien pseudo-Hermitique}

L'\'{e}volution temporelle des syst\`{e}mes Hamiltoniens est une question
centrale et fondamentale en m\'{e}canique quantique, en particulier en ce
qui concerne les applications physiques. Les principes cl\'{e}s sont tr\`{e}%
s bien compris depuis longtemps pour les syst\`{e}mes Hamiltoniens
Hermitiques et peuvent \^{e}tre trouv\'{e}s dans presque n'importe quel
livre standard sur la m\'{e}canique quantique. Cependant, la situation est
assez diff\'{e}rente pour la classe de syst\`{e}mes non Hermitiques qui poss%
\`{e}dent des spectres de valeurs propres r\'{e}els ou au moins
partiellement r\'{e}els. Pour les syst\`{e}mes ind\'{e}pendants du temps,
les principes directeurs sont maintenant \'{e}tablis et de nombreuses exp%
\'{e}riences existent pour confirmer les principales conclusions, par ex. 
\cite{mussl,makr,guo}. Pour des revues r\'{e}centes sur le sujet, voir par
exemple \cite{mos11,bender5,znojil5}.

En revanche, les syst\`{e}mes non-Hermitiques d\'{e}pendant du temps sont
beaucoup moins \'{e}tudi\'{e}s et il semble que jusqu'\`{a} pr\'{e}sent
aucun consensus n'ait \'{e}t\'{e} atteint sur un certain nombre de questions
centrales. Sachant que le traitement des syst\`{e}mes Hamiltoniens non
Hermitiques d\'{e}pendant du temps avec des op\'{e}rateurs m\'{e}triques ind%
\'{e}pendants du temps \cite{fring3,fring4}, est largement accept\'{e}, la g%
\'{e}n\'{e}ralisation aux op\'{e}rateurs m\'{e}triques d\'{e}pendant du
temps a soulev\'{e} beaucoup de questions \cite%
{mos8,mos9,mos10,znojil2,znojil3,znojil4,gong,maam,fring1,fring2}. Le d\'{e}%
bat de 2007 \cite{mos8,mos9,mos10,znojil2,znojil1,znojil3,znojil4} et les r%
\'{e}sultats qui en d\'{e}coulent r\'{e}v\`{e}lent que l'unitarit\'{e} de l'%
\'{e}volution temporelle peut \^{e}tre garantie mais le Hamiltonien (le g%
\'{e}n\'{e}rateur de l'\'{e}volution temporelle de Schr\"{o}dinger) est en g%
\'{e}n\'{e}ral inobservable. Les r\'{e}cents r\'{e}sultats \cite%
{fring1,fring2} font valoir qu'il est incompatible de maintenir une \'{e}%
volution temporelle unitaire pour les Hamiltoniens non-Hermitiques d\'{e}%
pendant du temps lorsque l'op\'{e}rateur m\'{e}trique d\'{e}pend
explicitement du temps. Nous rappelons bri\`{e}vement les diff\'{e}rents
points de vue concernant ce d\'{e}bat.

Mostafazadeh \cite{mos8} a affirm\'{e} qu'\`{a} l'aide d'un op\'{e}rateur m%
\'{e}trique d\'{e}pendant du temps, on ne peut pas assurer l'unitarit\'{e}
de l'\'{e}volution temporelle en m\^{e}me temps que l'observabilit\'{e} de
l'Hamiltonien. Ce point de vue, a \'{e}t\'{e} adopt\'{e} par Fring et al 
\cite{fring1,fring2}. Tandis que, certains auteurs ont recours \`{a} une 
\'{e}volution temporelle non-unitaire \cite{znojil4,gong,maam} et insistent
sur la relation de quasi-Hermiticit\'{e} entre un \guillemotleft Hamiltonien%
\guillemotright\ Hermitique et un \guillemotleft Hamiltonien\guillemotright\ %
non Hermitique d\'{e}pendant du temps.

Nous allons \'{e}voquer bri\`{e}vement, lorsqu'il y a d\'{e}pendance en
temps de l'op\'{e}rateur m\'{e}trique, les diff\'{e}rents points de vue qui 
\'{e}mergent concernant les syst\`{e}mes quantiques non Hermitiques d\'{e}%
pendant explicitement du temps. L'un est celui de Ali Mostafazadeh qui
affirme que : l'ind\'{e}pendance du temps de l'op\'{e}rateur m\'{e}trique
est une condition n\'{e}cessaire pour assurer la pseudo Hermiticit\'{e} de
l'Hamiltonien $H(t)$, l'autre celui de Miloslav Znojil qui stipule que l'ind%
\'{e}pendance du temps n'est pas une condition n\'{e}cessaire pour garantir
la pseudo-Hermiticit\'{e} de $H(t)$ et enfin celui de Fring et Moussa qui
rejoint le point de vue de Mostafazadeh et d\'{e}finit une nouvelle relation
de quasi-Hermiticit\'{e} lorsque la m\'{e}trique est d\'{e}pendante du temps

\subsection{Point de vue d'Ali Mostafazadeh}

Le point de vue de Mostafazadeh \cite{mos8,mos9,mos10} affirme que : l'ind%
\'{e}pendance du temps de l'op\'{e}rateur m\'{e}trique est une condition n%
\'{e}cessaire pour assurer la pseudo Hermiticit\'{e} de l'Hamiltonien $H(t)$%
. Soit $U^{H}\left( t\right) $ l'op\'{e}rateur d'\'{e}volution associ\'{e} 
\`{a} l'Hamiltonien non-Hermitique $H\left( t\right) $

\begin{equation}
i\hbar \frac{\partial }{\partial t}U^{H}\left( t\right) =U^{H}\left(
t\right) H\left( t\right) ,U(0)=I,  \label{ch5.1.1}
\end{equation}%
et $\psi \left( t\right) $ et $\phi \left( t\right) $ des vecteurs d'\'{e}%
tat \'{e}voluant sous l'action de $U^{H}\left( t\right) $

\begin{equation}
\psi \left( t\right) =U^{H}\left( t\right) \psi \left( 0\right) ,\ \ \phi
\left( t\right) =U^{H}\left( t\right) \phi \left( 0\right) .
\end{equation}%
Le pseudo produit scalaire $\left\langle .,.\right\rangle _{\eta \left(
t\right) }$ (\ref{ch4.2.13}) valable aussi pour\ $\eta \left( t\right) $
ainsi que l'unitarit\'{e} de l'\'{e}volution conf\`{e}re au produit scalaire 
$\left\langle \psi \left( t\right) ,\phi \left( t\right) \right\rangle
_{\eta \left( t\right) }$ son ind\'{e}pendance par-rapport au temps

\begin{eqnarray}
\left\langle \psi \left( t\right) ,\phi \left( t\right) \right\rangle _{\eta
\left( t\right) } &=&\left\langle \psi \left( t\right) \right\vert \eta
\left( t\right) \phi \left( t\right) \rangle  \notag \\
&=&\left\langle \psi \left( t\right) \left\vert U^{+H}\left( t\right) \eta
\left( t\right) U^{H}\left( t\right) \right\vert \phi \left( 0\right)
\right\rangle  \notag \\
&=&\left\langle \psi \left( 0\right) \left\vert \eta \left( 0\right)
\right\vert \phi \left( 0\right) \right\rangle ,
\end{eqnarray}%
d'o\`{u}, on d\'{e}duit que

\begin{equation}
U^{+H}\left( t\right) \eta \left( t\right) U^{H}\left( t\right) =\eta \left(
0\right) \Rightarrow \eta \left( t\right) =\left[ U^{+H}\left( t\right) %
\right] ^{-1}\eta \left( 0\right) U^{H}\left( t\right) ^{-1}
\end{equation}%
ce qui nous permet d'obtenir

\begin{equation}
\eta \left( t\right) ^{-1}=U^{H}\left( t\right) \eta \left( 0\right)
^{-1}U\left( t\right) ^{+H},  \label{ch5.1.2}
\end{equation}%
en utilisant (\ref{ch5.1.1}), la diff\'{e}renciation de (\ref{ch5.1.2}) donne

\begin{equation}
H^{+}\left( t\right) =\eta \left( t\right) H\left( t\right) \eta \left(
t\right) ^{-1}-i\hbar \eta \left( t\right) \frac{\partial }{\partial t}\eta
\left( t\right) ^{-1}.  \label{ch5.1.3}
\end{equation}%
L'\'{e}quation (\ref{ch5.1.3}) montre que $H\left( t\right) $ est $\eta -$%
pseudo-Hermitique au sens usuel ,i.e; $H^{+}\left( t\right) =\eta \left(
t\right) H\left( t\right) \eta \left( t\right) ^{-1}$ si et seulement si $%
\eta $ est ind\'{e}pendant du temps.

\subsection{Point de vue de Miloslav Znojil}

L'\'{e}quation de Schr\"{o}dinger d\'{e}pendante du temps associ\'{e}e \`{a}
l'Hamiltonien Hermitique $h\left( t\right) $ est d\'{e}finie par%
\begin{equation}
i\hbar \frac{\partial }{\partial t}\left\vert \varphi (t)\right\rangle
=h\left( t\right) \left\vert \varphi (t)\right\rangle ,\text{ \ \ \ \ \ \ \ }%
\left\vert \varphi (t)\right\rangle =U^{h}(t)\left\vert \varphi
(0)\right\rangle
\end{equation}%
qui en terme d'op\'{e}rateur d'\'{e}volution unitaire $U^{h}(t)$ s'\'{e}crit%
\begin{equation}
i\hbar \frac{\partial }{\partial t}U^{h}\left( t\right) =h\left( t\right)
U^{h}\left( t\right) \text{\ \ \ }
\end{equation}%
o\`{u} $h\left( t\right) $ et l'\'{e}quivalent non-Hermitique $H(t)$ sont
reli\'{e}s par%
\begin{equation}
h(t)=\rho (t)H(t)\rho ^{-1}(t).  \label{ch5.1.4}
\end{equation}%
La solution formelle de l'\'{e}quation de Schr\"{o}dinger ci dessus s'\'{e}%
crit alors 
\begin{equation}
\left\vert \varphi (t)\right\rangle =U^{h}\left( t\right) \left\vert \varphi
(0)\right\rangle ,  \label{ch5.1.5}
\end{equation}%
et par cons\'{e}quent elle satisfait la relation de la constance de la norme
reste constante \`{a} tout instant%
\begin{equation}
\langle \varphi (t)\left\vert \varphi (t)\right\rangle =\langle \varphi
(0)\left\vert \varphi (0)\right\rangle .
\end{equation}

M. Znojil \cite{znojil2,znojil1,znojil3,znojil4} stipule que l'ind\'{e}%
pendance du temps de la m\'{e}trique $\eta $ n'est pas une condition n\'{e}%
cessaire pour garantir la pseudo-Hermiticit\'{e} de $H(t)$, ce qui l'a
conduit \`{a} d\'{e}finir un g\'{e}n\'{e}rateur d'\'{e}volution $H_{gen}(t)$
non observable diff\'{e}rent de $H(t)$, dans ce cas fait la distinction
entre deux \'{e}volutions formelles d\'{e}finies par 
\begin{equation}
\left\vert \phi \left( t\right) \right\rangle =U_{D}\left( t\right)
\left\vert \phi \left( 0\right) \right\rangle ,\text{ \ \ \ \ \ \ }%
U_{D}\left( t\right) =\text{\ }\rho ^{-1}\left( t\right) U^{h}\left(
t\right) \rho \left( 0\right) ,
\end{equation}%
\begin{equation}
\left\langle \langle \phi \left( t\right) \right\vert =\left\langle \langle
\phi \left( 0\right) \right\vert U_{G}\left( t\right) ,\text{ \ \ \ \ \ \ \ }%
U_{G}\left( t\right) =\text{\ }\rho ^{-1}\left( 0\right) U^{+h}\left(
t\right) \rho \left( t\right) ,
\end{equation}%
o\`{u} les op\'{e}rateurs $U_{D}\left( t\right) $ et $U_{G}\left( t\right) $
agissent sur le ket $\left\vert \phi \right\rangle =\rho ^{-1}\left(
t\right) \left\vert \varphi (t)\right\rangle $ et \ le bra $\left\langle
\langle \phi \right\vert =\left\langle \varphi (t)\right\vert \rho \left(
t\right) $\ respectivement.\ Cette convention refl\`{e}te bien le fait
qu'apparait deux mani\`{e}res diff\'{e}rentes de repr\'{e}senter la fonction
d'onde (\ref{ch5.1.5}). Les \'{e}quations diff\'{e}rentielles relatives aux
deux op\'{e}rateurs d'\'{e}volution $U_{D}\left( t\right) $ et $U_{G}\left(
t\right) $ sont donc,

\begin{equation}
i\hbar \partial _{t}U_{D}\left( t\right) =-i\hbar \rho ^{-1}\left( t\right) 
\left[ \partial _{t}\rho \left( t\right) \right] U_{D}\left( t\right)
+H\left( t\right) U_{D}\left( t\right) ,
\end{equation}%
et

\begin{equation}
i\hbar \partial _{t}U_{G}^{+}\left( t\right) =H^{+}\left( t\right)
U_{G}^{+}\left( t\right) +\left[ i\hbar \partial _{t}\rho ^{+}\left(
t\right) \right] \left[ \rho ^{-1}\left( t\right) \right] ^{+}U_{G}^{+}%
\left( t\right) .
\end{equation}%
Par cons\'{e}quent les \'{e}tats $\left\vert \phi \right\rangle $ et $%
\left\vert \phi \right\rangle \rangle $ $\ $satisfont s\'{e}par\'{e}ment les
\ \'{e}quations de Schr\"{o}dinger suivantes 
\begin{equation}
i\hbar \frac{\partial }{\partial t}\left\vert \phi (t)\right\rangle
=H_{(gen)}\left( t\right) \left\vert \phi (t)\right\rangle ,\text{ \ \ \ \ \
\ \ }
\end{equation}%
\ 
\begin{equation}
i\hbar \frac{\partial }{\partial t}\left\vert \phi (t)\right\rangle \rangle
=H_{(gen)}^{+}\left( t\right) \left\vert \phi (t)\right\rangle \rangle ,%
\text{\ \ \ \ \ }
\end{equation}%
o\`{u}

\begin{eqnarray}
H_{gen}\left( t\right) &=&H\left( t\right) -i\hbar \rho ^{-1}\left( t\right)
\partial _{t}\rho \left( t\right) ,  \notag \\
H_{gen}^{+}\left( t\right) &=&H^{+}\left( t\right) +i\hbar \partial _{t}\rho
^{+}\left( t\right) \left( \rho ^{-1}\right) ^{+}\left( t\right) .
\end{eqnarray}%
\ \ \ Un calcul \'{e}l\'{e}mentaire montre que lorsqu'on effectue la diff%
\'{e}renciation de la norme $\left\langle \langle \phi \right\vert \left.
\phi \right\rangle $ par rapport au temps donne

\begin{equation}
\frac{\partial }{\partial t}\left\langle \langle \phi \right\vert \left.
\phi \right\rangle =0
\end{equation}%
qui, de ce point de vue, montre aussi que l'\'{e}volution par rapport au
temps est unitaire. Nous constatons que ces deux points de vue sont oppos%
\'{e}es lorsqu'on traite les syst\`{e}mes quantiques quasi-Hermitiques d\'{e}%
pendants du temps.

\subsection{Point de vue de Fring et Moussa}

Fring et al \cite{fring1,fring2} soutiennent que les relations de
quasi-Hermiticit\'{e} (\ref{ch4.2.2}) et (\ref{ch4.2.10}) ne sont plus
valable dans le cas d'une m\'{e}trique $\eta \left( t\right) $ d\'{e}%
pendante du temps\ et par ainsi ils approuvent le point de vue de
Mostafazadah. Ils partent des \'{e}quations de Schr\"{o}dinger 
\begin{equation}
i\hbar \frac{\partial }{\partial t}\left\vert \psi (t)\right\rangle =h\left(
t\right) \left\vert \psi (t)\right\rangle ,\text{ \ \ \ \ \ \ \ }i\hbar 
\frac{\partial }{\partial t}\left\vert \phi (t)\right\rangle =H\left(
t\right) \left\vert \phi (t)\right\rangle ,  \label{ch5.1.6}
\end{equation}%
o\`{u} $h\left( t\right) =h^{+}\left( t\right) $ et $H\left( t\right) \neq
H^{+}\left( t\right) $. Ils insistent sur le fait que les op\'{e}rateurs ne
peuvent \^{e}tre appel\'{e}s Hamiltoniens que s'ils g\'{e}n\`{e}rent l'\'{e}%
volution temporelle du syst\`{e}me consid\'{e}r\'{e}. Ils supposent ensuite
que les deux solutions $\left\vert \psi (t)\right\rangle $ et $\left\vert
\phi (t)\right\rangle $ sont reli\'{e}es par un op\'{e}rateur inversible d%
\'{e}pendant du temps $\rho \left( t\right) $ 
\begin{equation}
\left\vert \psi (t)\right\rangle =\rho \left( t\right) \left\vert \phi
(t)\right\rangle  \label{ch5.1.7}
\end{equation}%
Il s'ensuit imm\'{e}diatement par substitution directe de (\ref{ch5.1.7})
dans (\ref{ch5.1.6}) que les deux Hamiltoniens $h\left( t\right) $ et $%
H\left( t\right) $ sont reli\'{e}s par\qquad 
\begin{equation}
h\left( t\right) =\rho \left( t\right) H\left( t\right) \rho ^{-1}\left(
t\right) -i\hbar \rho ^{-1}\left( t\right) \partial _{t}\rho \left( t\right)
,  \label{ch5.1.8}
\end{equation}%
on voit que $h\left( t\right) $et $H\left( t\right) $ ne sont plus reli\'{e}%
s par une transformation de similarit\'{e} comme dans le sc\'{e}nario compl%
\`{e}tement ind\'{e}pendant du temps ou dans le sc\'{e}nario d\'{e}pendant
du temps avec une m\'{e}trique ind\'{e}pendante du temps. Ils d\'{e}%
finissent l'\'{e}quation (\ref{ch5.1.8}) \ comme \'{e}tant la relation de
Dyson d\'{e}pendante du temps g\'{e}n\'{e}ralisant ainsi le sc\'{e}nario ind%
\'{e}pendant du temps. La conjugaison Hermitique de l'\'{e}quation (\ref%
{ch5.1.8}) et l'utilisation l'Hermiticit\'{e} de $h(t)$ permettra d'obtenir
une relation entre $H(t)$ et son conjugu\'{e} Hermitique

\begin{equation}
H^{+}\left( t\right) \eta \left( t\right) -\eta \left( t\right) H\left(
t\right) =i\hbar \partial _{t}\eta \left( t\right) ,  \label{ch5.1.9}
\end{equation}%
interpr\'{e}tant $\eta \left( t\right) $=$\rho ^{+}\left( t\right) \rho
\left( t\right) $ en tant qu'op\'{e}rateur m\'{e}trique cette relation (\ref%
{ch5.1.9}) remplace la relation quasi-Hermiticit\'{e} standard bien connue
dans le contexte quantique non-Hermitique ind\'{e}pendant du temps.

\section{Hamiltonien pseudo $\mathcal{PT}$-sym\'{e}trique}

Comme il a \'{e}t\'{e} d\'{e}ja signal\'{e} au chapitre 4, il est difficile
de r\'{e}soudre l'\'{e}quation de Schr\"{o}dinger d\'{e}pendante du temps (%
\ref{ch3.1}) de fa\c{c}on exacte dans le cas Hermitique. Dans le cas
non-Hermitique d\'{e}pendant du temps,\ et par analogie au cas Hermitique,
on peut faire appel \`{a} l'hypoth\`{e}se adiabatique introduite dans \cite%
{gong} qui se base sur le point de vue de Znojil \ et par cons\'{e}quent le g%
\'{e}n\'{e}rateur de l'\'{e}volution n'est pas observable, ce point ne fera
pas l'objet de ce cours. Il nous reste \`{a} \'{e}tudier la th\'{e}orie de
Floquet pour les syst\`{e}mes p\'{e}riodiques \cite{maam, sarra} et la th%
\'{e}orie des pseudo invariants \cite{khan,kalt,wali}.

Avant d'aborder la th\'{e}orie de Floquet pour les syst\`{e}mes p\'{e}%
riodiques non-Hermitiques, il est utile de signaler le nouveau concept de la
pseudo-$\mathcal{PT}$-sym\'{e}trie introduit par Luo et al \cite{luo} dans l'%
\'{e}tude des syst\`{e}mes optiques \`{a} modulation p\'{e}riodique avec
gain et perte \'{e}quilibr\'{e} o\`{u} l'op\'{e}rateur renversement du temps 
$\mathcal{T}$ \ agit sur le param\`{e}tre temps $t$ du Hamiltonien.\ \ Ils
proc\`{e}dent tout d'abord par moyenner les termes de haute fr\'{e}quence du
Hamiltonien $\mathcal{PT-}$sym\'{e}trique ou non $\mathcal{PT-}$sym\'{e}%
trique initial pour ainsi obtenir un Hamiltonien effectif non modul\'{e} et
diagonalis\'{e} dont des valeurs propres ou quasi-\'{e}nergies sont ind\'{e}%
pendantes du temps. Deux remarques s'imposent:

i) cette m\'{e}thode n'est pas r\'{e}versible c'est \`{a} dire qu'on ne
pourra pas revenir \`{a} notre syst\`{e}me initial. ii) le renversement du
temps en m\'{e}canique quantique est intimement li\'{e} \`{a} la conjugaison
complexe. Puisque le temps $t$ est un param\`{e}tre en m\'{e}canique
quantique, l'op\'{e}rateur $\mathcal{T}$ ne peut pas agir sur $t$ dans le
Hamiltonien.

Dans la suite de ce paragraphe nous utiliserons la th\'{e}orie de Floquet 
\cite{floq} pour ainsi obtenir l'Hamiltonien de Floquet ind\'{e}pendant du
temps o\`{u} la $\mathcal{PT-}$sym\'{e}trie d\'{e}finie au chapitre 4 est
facilement applicable c'est \`{a} dire que l'op\'{e}rateur renversement du
temps $\mathcal{T}$ concerne uniquement la conjugaison complexe.

\subsection{La th\'{e}orie de Floquet pour les syst\`{e}mes pseudo $\mathcal{%
PT-}$sym\'{e}trique}

L'\'{e}quation de Schr\"{o}dinger r\'{e}git par l'Hamiltonien p\'{e}riodique
non-Hermitique $H(t)=H(t+\tau )$ est donn\'{e}e par:

\begin{equation}
i\hbar \frac{\partial }{\partial t}\left\vert \Phi (t)\right\rangle
=H(t)\left\vert \Phi (t)\right\rangle ,  \label{Ae}
\end{equation}%
o\`{u} $\tau =2\pi /\omega $ est la p\'{e}riode de l'\'{e}volution. L'op\'{e}%
rateur d'\'{e}volution non unitaire $U(t)$ dans l'intervalle $[0,t]$ \
satisfait l'\'{e}quation diff\'{e}rentielle

\begin{equation}
\text{\ }i\hbar \frac{\partial }{\partial t}U(t)=H(t)U(t),  \label{A}
\end{equation}%
tel que $U(0)=1$. La d\'{e}composition de Floquet de l'op\'{e}rateur d'\'{e}%
volution\ \cite{maam,chou,moor} est

\begin{equation}
U(t)=Z(t)e^{-iMt},  \label{C}
\end{equation}%
o\`{u} $Z(t)$ est un op\'{e}rateur p\'{e}riodique non unitaire tel que $%
Z(0)=Z(\tau )=1$, $M$ est un op\'{e}rateur non-Hermitique ind\'{e}pendant du
temps. La transformation non-unitaire

\begin{equation}
\left\vert \Phi (t)\right\rangle =Z(t)\left\vert \psi (t)\right\rangle
=U(t)e^{iMt}\left\vert \psi (t)\right\rangle .
\end{equation}%
conduit \`{a} une \'{e}quation de Schr\"{o}dinger gouvern\'{e} par
l'Hamiltonien non-Hermitique $M$ ind\'{e}pendant du temps

\begin{equation}
i\hbar \frac{\partial }{\partial t}\left\vert \psi (t)\right\rangle
=M\left\vert \psi (t)\right\rangle ,  \label{EQ}
\end{equation}%
qui nous permettra d'\'{e}tudier la $\mathcal{PT-}$sym\'{e}trie o\`{u} l'op%
\'{e}rateur renversement du temps $\mathcal{T}$ est cette fois ci li\'{e} 
\`{a} la conjugaison complexe, contrairement \`{a} ce qui a \'{e}t\'{e} adopt%
\'{e} par Luo et al \cite{luo}.

L'Hamiltonien $M$ est dit $\mathcal{PT}$-Sym\'{e}trique s'il est invariant
sous la transformation $\mathcal{PT}$, c'est-\`{a}-dire qu'il commute n\'{e}%
cessairement avec l'op\'{e}rateur $\mathcal{PT}$

\begin{equation}
M=\mathcal{PT}M\mathcal{PT}.
\end{equation}
La $\mathcal{PT}$-sym\'{e}trie est dite non bris\'{e}e si toutes les
fonctions propres de l'Hamiltonien $\mathcal{PT}$-sym\'{e}trique $M$ sont en
m\^{e}me temps des fonctions propres de l'op\'{e}rateur $\mathcal{PT}$. \
Par contre, s'il existe des fonctions propres de l'Hamiltonien $\mathcal{PT}$%
-sym\'{e}trique $M$ qui ne sont pas des fonctions propres de l'op\'{e}rateur 
$\mathcal{PT}$, elle est dite bris\'{e}e.

Jusqu'\`{a} pr\'{e}sent, nous avons exig\'{e} que la sym\'{e}trie ne soit
pas bris\'{e}e pour construire une th\'{e}orie quantique \`{a} partir des
Hamiltoniens $\mathcal{PT}$-sym\'{e}triques. Il faut cependant noter qu'il
existe des Hamiltoniens non $\mathcal{PT}$-sym\'{e}triques dont les spectres
sont r\'{e}els, cette sym\'{e}trie induite est appel\'{e}e \guillemotleft %
pseudo-$\mathcal{PT-}$ sym\'{e}trie\guillemotright . L'op\'{e}rateur $M$
joue un r\^{o}le tr\`{e}s important dans l'\'{e}tude de la notion de la $%
\mathcal{PT}$ (ou pseudo $\mathcal{PT}$)-sym\'{e}trie d'o\`{u} la d\'{e}%
composition de Floquet (\ref{C}) est fortement li\'{e}e \`{a} la stabilit%
\'{e} dynamique du syst\`{e}me et d\'{e}pend de la nature $\mathcal{PT}$-sym%
\'{e}trique de l'op\'{e}rateur $M$ (bris\'{e}e ou non bris\'{e}e). L'action
de l'op\'{e}rateur de Floquet $U(\tau )=e^{-iM\tau }$ au bout d'une p\'{e}%
riode $\tau $ sur les \'{e}tats propres $\left\vert \phi
_{n}(0)\right\rangle $ de $M$ donne 
\begin{equation}
e^{-iM\tau }\left\vert \phi _{n}(0)\right\rangle =\left\vert \phi _{n}(\tau
)\right\rangle =e^{-i\epsilon _{n}\tau }\left\vert \phi _{n}(0)\right\rangle
,
\end{equation}%
o\`{u} $\epsilon _{n}$ sont les quasi-\'{e}nergies. L'\'{e}volution \`{a}
tout instant est donn\'{e}e par

\begin{equation}
e^{-iMt}\left\vert \phi _{n}(0)\right\rangle =\left\vert \phi
_{n}(t)\right\rangle =e^{-i\epsilon _{n}t}\left\vert \phi
_{n}(0)\right\rangle .
\end{equation}%
D\'{e}finissons les fonctions $\tau $-p\'{e}riodique de Floquet, c'est-\`{a}%
-dire $\left\vert \phi _{n}(t)\right\rangle =\left\vert \phi _{n}(t+\tau
)\right\rangle ,$ comme suit

\begin{equation}
\left\vert \phi _{n}(t)\right\rangle =Z(t)\left\vert \phi
_{n}(0)\right\rangle .
\end{equation}%
Les \'{e}tats \'{e}volu\'{e}s de Floquet sont reli\'{e}s aux fonctions de
Floquet \`{a} l'aide d'un facteur de phase

\begin{equation}
\left\vert \Phi _{n}(t)\right\rangle =e^{-i\epsilon _{n}t}\left\vert \phi
_{n}(t)\right\rangle .  \label{D}
\end{equation}

Notons que ces \'{e}tats $\left\vert \Phi _{n}(t)\right\rangle $, sont des
solutions de l'\'{e}quation de Schr\"{o}dinger d\'{e}pendante du temps (\ref%
{Ae}). En substituant (\ref{D}) dans l'\'{e}quation (\ref{Ae}), nous
obtenons l'\'{e}quation aux valeurs propres

\begin{equation}
K(t)\left\vert \phi _{n}(t)\right\rangle =\epsilon _{n}\left\vert \phi
_{n}(t)\right\rangle ,
\end{equation}%
de l'op\'{e}rateur non Hermitique de Floquet

\begin{equation}
K(t)=H(t)-i\hbar \dfrac{\partial }{\partial t}I,
\end{equation}%
ce qui conf\`{e}re aux valeurs propres $\epsilon _{n}$ le nom de quasi-\'{e}%
nergie et aux \'{e}tats $\left\vert \phi _{n}(t)\right\rangle $ le nom \'{e}%
tats de quasi-\'{e}nergie \cite{zel}. Ainsi, l'analyse des Hamiltoniens p%
\'{e}riodiques non-Hermitiques $H(t)$ peut \^{e}tre r\'{e}duite \`{a} l'\'{e}%
tude de l'op\'{e}rateur de Floquet $K(t)=H(t)-i\hbar (\partial /(\partial
t))I$ ind\'{e}pendant du temps.

En r\'{e}sum\'{e}, en utilisant la d\'{e}composition de Floquet de l'op\'{e}%
rateur d'\'{e}volution non-unitaire associ\'{e} aux syst\`{e}mes p\'{e}%
riodiques non -Hermitiques, nous avons pr\'{e}sent\'{e}\ une analyse
rigoureuse de la dynamique r\'{e}git par des Hamiltoniens p\'{e}riodiques
non-Hermitiques et introduit le concept de la pseudo-$\mathcal{PT}$-sym\'{e}%
trie. Nous avons montr\'{e} que la stabilit\'{e} de la dynamique se produit
lorsque la $\mathcal{PT}$ sym\'{e}trie de l'op\'{e}rateur de Floquet $%
U(T)=e^{iMT}$ ou plus pr\'{e}cisement de l'operateur $M$ est non-bris\'{e}e
ce qui correspond aux quasi-\'{e}nergies r\'{e}elles $\epsilon _{n}$.
Lorsque la $\mathcal{PT-}$sym\'{e}trie de $M$ est bris\'{e}e, c'est \`{a}
dire les quasi-\'{e}nergies sont complexe conjugu\'{e}es par pair, la
dynamique est instable. Nous insistons sur le fait que la stabilit\'{e} de
la dynamique d\'{e}pend de la brisure de la $\mathcal{PT}$ -sym\'{e}trie de
l'op\'{e}rateur $M$, et non pas de celle de l'Hamiltonien $H(t)$.

\subsection{ Oscillateur harmonique dans un champ lin\'{e}aire imaginaire p%
\'{e}riodique}

Dans ce paragraphe nous appliquerons la th\'{e}orie de Floquet d\'{e}crite
ci dessus \`{a} un oscillateur harmonique interagissant avec un champ lin%
\'{e}aire imaginaire p\'{e}riodique d\'{e}pendant du temps \cite{sarra}.
Nous allons d\'{e}river dans le cas non-Hermitique les valeurs propres et
les fonctions propres de Floquet et \'{e}tudier la $\mathcal{PT}\emph{-}$sym%
\'{e}trie ou pseudo $\mathcal{PT}\emph{-}$sym\'{e}trie de l'op\'{e}rateur de
Floquet.

L'Hamiltonien consid\'{e}r\'{e} est 
\begin{equation}
H(t)=\dfrac{1}{2m}p^{2}+\dfrac{m\omega _{0}^{2}}{2}x^{2}-if(t)x,  \label{H}
\end{equation}%
o\`{u} $f(t)=$ $\lambda \cos (\omega t+\phi )$. \ Introduisons la
transformation non unitaire $D(t)$ d\'{e}pendante du temps%
\begin{equation}
D(t)=e^{i(x\mathcal{P}_{c}(t)-p\mathcal{X}_{c}(t))}=e^{-\tfrac{i\hbar }{2}%
\mathcal{X}_{c}(t)\mathcal{P}_{c}(t)}e^{ix\mathcal{P}_{c}(t)}e^{-ip\mathcal{X%
}_{c}(t)},
\end{equation}%
et son inverse 
\begin{equation}
D^{-1}(t)=e^{+\tfrac{i\hbar }{2}\mathcal{X}_{c}(t)\mathcal{P}_{c}(t)}e^{ip%
\mathcal{X}_{c}(t)}e^{-ix\mathcal{P}_{c}(t)}.
\end{equation}%
o\`{u} les param\`{e}tres 
\begin{eqnarray}
\mathcal{X}_{c}(t) &=&x_{c}(t)-x_{c}(0),  \notag \\
\mathcal{P}_{c}(t) &=&p_{c}(t)-p_{c}(0)
\end{eqnarray}%
sont reli\'{e}s aux solutions classiques \ $x_{c}(t)$ et $p_{c}(t)$ associ%
\'{e}es \`{a} l'\'{e}quivalent classique de l'Hamiltonien $H(t)$ (\ref{H}) 
\begin{eqnarray}
\dot{x}_{c} &=&\frac{1}{m}p_{c}  \notag \\
\dot{p}_{c} &=&-m\omega _{0}^{2}x_{c}+if(t)  \notag \\
\overset{\cdot \cdot }{x_{c}}+m\omega _{0}^{2}x_{c} &=&if(t).
\end{eqnarray}
Les solutions classiques seront de la forme

\begin{equation}
x_{c}(t)=\frac{\lambda \cos (\omega t+\phi )}{m\left( \omega _{0}^{2}-\omega
^{2}\right) },\ \ \ \dot{x}_{c}=-i\frac{\lambda \omega \sin (\omega t+\phi )%
}{m\left( \omega _{0}^{2}-\omega ^{2}\right) }.
\end{equation}%
Il est facile de montrer que l'action de l'op\'{e}rateur $D(t)$ sur les op%
\'{e}rateurs $x$ et $p$ est donn\'{e}e par :

\begin{eqnarray}
D^{-1}(t)xD(t) &=&x+\mathcal{X}_{c}(t)  \notag \\
D^{-1}(t)pD(t) &=&p+\mathcal{P}_{c}(t).  \label{x_c}
\end{eqnarray}%
et que son action sur une fonction d'onde $G(x)$ est\textit{\ }d\'{e}termin%
\'{e}e\textit{\ }par 
\begin{equation}
D(t)G(x)=\exp \left[ -\tfrac{i}{2\hbar }\mathcal{X}_{c}(t)\mathcal{P}_{c}(t)%
\right] \exp \left[ \frac{i}{\hbar }x\mathcal{P}_{c}(t)\right] G(x-\mathcal{X%
}_{c}(t)).
\end{equation}%
Ainsi l'\'{e}quation de l'\'{e}quation de Schr\"{o}dinger associ\'{e}e \`{a} 
$H(t)$ (\ref{H}) se transforme en une \'{e}quation d'\'{e}volution dont le
nouveau Hamiltonien est donn\'{e} par

\begin{equation}
h(t)=D^{-1}(t)H(t)D(t)-i\hbar D^{-1}(t)\frac{\partial }{\partial t}D(t),
\end{equation}%
qui s'\'{e}crit dans ce cas : \ 
\begin{eqnarray}
h(t) &=&h_{d}^{OQL}+h_{f}^{OC}(t)+x(m\omega _{0}^{2}x_{c}-if(t)+\dot{p}%
_{c})-x_{c}(0)(m\omega _{0}^{2}x_{c}-if(t)+\dot{p}_{c})  \notag \\
&&+\frac{1}{2}\left( \dot{p}_{c}x_{c}-p_{c}\dot{x}_{c}\right) +\frac{1}{2}%
\left( \dot{p}_{c}x_{c}(0)-p_{c}(0)\dot{x}_{c}\right) ,  \label{11a}
\end{eqnarray}%
ou simplement%
\begin{equation}
h(t)=h_{d}^{OQL}+L(t),  \label{hL}
\end{equation}%
o\`{u} 
\begin{equation}
h_{d}^{OQL}=\frac{1}{2m}\left( p-p_{c}(0)\right) ^{2}+\frac{m\omega _{0}^{2}%
}{2}(x-x_{c}(0))^{2},
\end{equation}%
et%
\begin{eqnarray}
L(t) &=&h_{f}^{OC}(t)+\frac{1}{2}\left( \dot{p}_{c}x_{c}-p_{c}\dot{x}%
_{c}\right) +\frac{1}{2}\left( \dot{p}_{c}x_{c}(0)-p_{c}(0)\dot{x}_{c}\right)
\label{20} \\
&=&\frac{1}{4}\frac{\lambda ^{2}}{m\left( \omega _{0}^{2}-\omega ^{2}\right) 
}+\frac{1}{4}\frac{\lambda ^{2}}{m\left( \omega _{0}^{2}-\omega ^{2}\right) }%
\cos 2(\omega t+\phi )+\frac{1}{2}\frac{\lambda ^{2}\omega ^{2}}{m\left(
\omega _{0}^{2}-\omega ^{2}\right) ^{2}}\cos \omega t  \notag
\end{eqnarray}%
repr\'{e}sentent respectivement l'Hamiltonien ind\'{e}pendant du temps de
l'oscillateur harmonique d\'{e}plac\'{e} et le Lagrangien classique.
L'Hamiltonien classique $h_{f}^{OC}(t)$

\begin{equation}
h_{f}^{OC}(t)=\frac{1}{2m}\left( p_{c}\right) ^{2}+\frac{m\omega _{0}^{2}}{2}%
(x_{c})^{2}-if(t)x_{c},
\end{equation}%
d\'{e}crit l'oscillateur harmonique forc\'{e} d\'{e}pendant du temps.

Le Lagrangien (\ref{20}) peut \^{e}tre \'{e}limin\'{e} de l'Hamiltonien (\ref%
{11a}) par la transformation%
\begin{equation}
\left\vert \Phi (t)\right\rangle =\exp \left[ -\frac{i}{\hbar }%
\int\limits_{0}^{t}d\acute{t}L(\acute{t})\right] D(t)e^{-\tfrac{i}{\hbar }%
h_{d}^{OQL}t}\left\vert \Psi (0)\right\rangle .
\end{equation}%
\ En \'{e}valuant l'int\'{e}grale 
\begin{equation*}
\int\limits_{0}^{t}d\acute{t}L(\acute{t})=\frac{1}{4}\frac{\lambda ^{2}t}{%
m\left( \omega _{0}^{2}-\omega ^{2}\right) }+\frac{\lambda ^{2}}{8m\omega
\left( \omega _{0}^{2}-\omega ^{2}\right) }\left( \sin 2(\omega t+\phi
)-\sin 2\phi \right) +\frac{\lambda ^{2}\omega }{2m\left( \omega
_{0}^{2}-\omega ^{2}\right) ^{2}}\sin \omega t
\end{equation*}%
on d\'{e}duit l'expression de l'op\'{e}rateur d'\'{e}volution non-unitaire
associ\'{e} \`{a} l'Hamiltonien p\'{e}riodique (\ref{H}) non-Hermitique 
\begin{eqnarray}
U(t) &=&\exp \left[ -\tfrac{i}{\hbar }\left( \frac{\lambda ^{2}}{8m\omega
\left( \omega _{0}^{2}-\omega ^{2}\right) }\left( \sin 2(\omega t+\phi
)-\sin 2\phi \right) +\frac{\lambda ^{2}\omega }{2m\left( \omega
_{0}^{2}-\omega ^{2}\right) ^{2}}\sin \omega t\right) \right]  \notag \\
&&D(t)\exp \left[ -\tfrac{i}{\hbar }\left( h_{d}^{OQL}+\frac{1}{4}\frac{%
\lambda ^{2}}{m\left( \omega _{0}^{2}-\omega ^{2}\right) }\right) t\right]
\label{U}
\end{eqnarray}

L'\'{e}volution au cours du temps peut \^{e}tre obtenue par l'action de l'op%
\'{e}rateur de l'\'{e}volution (\ref{U}) sur l'\'{e}tat initial $\left\vert
\Phi (0)\right\rangle =$ $\left\vert \Psi (0)\right\rangle $. Comme a \'{e}t%
\'{e} d\'{e}ja not\'{e}, l'op\'{e}rateur d'\'{e}volution $U(t)$ (\ref{U})
est le produit de deux op\'{e}rateurs

\begin{equation}
U(t)=Z(t)e^{-\tfrac{i}{\hbar }Mt}
\end{equation}%
o\`{u} $Z(t)$ est p\'{e}riodique 
\begin{equation}
Z(t)=\exp \left[ -\tfrac{i}{\hbar }\left( \frac{\lambda ^{2}}{8m\omega
\left( \omega _{0}^{2}-\omega ^{2}\right) }\left( \sin 2(\omega t+\phi
)-\sin 2\phi \right) +\frac{\lambda ^{2}\omega }{2m\left( \omega
_{0}^{2}-\omega ^{2}\right) ^{2}}\sin \omega t\right) \right] D(t)
\end{equation}%
et $M$ est donn\'{e} par%
\begin{equation}
M=h_{d}^{OQL}+\frac{1}{4}\frac{\lambda ^{2}}{m\left( \omega _{0}^{2}-\omega
^{2}\right) }  \label{F}
\end{equation}%
dont les valeurs propres appel\'{e}es quasi-\'{e}nergies sont: 
\begin{equation}
\mathcal{E}_{n}=\hbar \omega _{0}(n+1/2)+\frac{1}{4}\frac{\lambda ^{2}}{%
m\left( \omega _{0}^{2}-\omega ^{2}\right) }  \label{ch65}
\end{equation}
Au point de r\'{e}sonance $\omega =\omega _{0}$ le spectre de quasi-\'{e}%
nergie devient infini par cons\'{e}quent l'\'{e}quation (\ref{ch65}) n'est
plus d'actualit\'{e}.

Finalement, la fonction d'onde solution de l'\'{e}quation de Schr\"{o}dinger
associ\'{e} \`{a} l'Hamiltonien (\ref{H}) est proportionnel aux fonctions
Hermite $\mathcal{H}_{n}$

\begin{eqnarray}
\Phi _{n}(x,t) &=&\frac{1}{\sqrt{2^{n}n!}}(\frac{m\omega _{0}}{\pi }%
)^{1/4}\exp \left[ -i\frac{\omega }{m}\left( \frac{\lambda \sin (\omega
t+\phi )}{2(\omega _{0}^{2}-\omega ^{2})}\right) ^{2}\right] \exp \left[
i\left( (n+1/2)\omega _{0}+\frac{1}{4}\frac{\lambda ^{2}}{m\left( \omega
_{0}^{2}-\omega ^{2}\right) }\right) t\right]  \notag \\
&&\exp \left[ -i\left( \frac{\lambda ^{2}}{8m\omega \left( \omega
_{0}^{2}-\omega ^{2}\right) }\left( \sin 2(\omega t+\phi )-\sin 2\phi
\right) +\frac{\lambda ^{2}\omega }{2m\left( \omega _{0}^{2}-\omega
^{2}\right) ^{2}}\sin \omega t\right) \right]  \notag \\
&&\exp \left[ -ip_{c}(x-x_{c})\right] \exp \left[ -\frac{m\omega _{0}}{2}%
(x-x_{c})^{2}\right] \mathcal{H}_{n}(\sqrt{\omega _{0}m}(x-x_{c}))  \label{w}
\end{eqnarray}%
et correspond aux modes de Floquet p\'{e}riodiques.

\qquad \qquad

\section{Th\'{e}orie des op\'{e}rateurs pseudo-invariants pour les syst\`{e}%
mes non-Hermitiques}

La th\'{e}orie des invariants de Lewis et Riesenfeld est une technique tr%
\`{e}s utile pour d\'{e}terminer les solutions des syst\`{e}mes quantiques d%
\'{e}crits par des Hamiltoniens Hermitiques d\'{e}pendants explicitement du
temps. La solution de l'\'{e}quation de Schr\"{o}dinger s'exprime en
fonction des \'{e}tats propres \cite{lewis2} de l'invariant multipli\'{e}
par une phase, d'o\`{u} le probl\`{e}me se r\'{e}duit \`{a} trouver la forme
explicite de l'op\'{e}rateur invariant et les phases associ\'{e}es \`{a} l'%
\'{e}volution.

Lorsque la m\'{e}trique est d\'{e}pendante du temps, la notion de
quasi-Hermiticit\'{e} analogue au cas stationnaire pourrait \^{e}tre utilis%
\'{e}e dans la th\'{e}orie des op\'{e}rateurs invariants pseudo-Hermitique,
il suffit de remplacer l'Hamiltonien stationnaire par l'invariant
pseudo-Hermitique dans les \'{e}quations (\ref{ch4.2.2})-(\ref{ch4.2.10}).

A cet effet, nous d\'{e}velopperons la th\'{e}orie des invariants
pseudo-Hermitique associ\'{e} \`{a} l'Hamiltonien non-Hermitique d\'{e}%
pendant du temps \cite{khan,kalt,wali}, puis nous pr\'{e}senterons la
solution de l'\'{e}quation de Schr\"{o}dinger d\'{e}pendante du temps en
fonction des \'{e}tats propres de l'op\'{e}rateur invariant
pseudo-Hermitique $I^{PH}(t)$ avec des phases r\'{e}elles. Nous illustrons
cette th\'{e}orie en \'{e}tudiant l'Hamiltonien de Swanson non Hermitique d%
\'{e}pendant du temps \cite{kalt,wali}.

\subsection{D\'{e}rivation de l'op\'{e}rateur pseudo-invariant}

Ce paragraphe utilise la m\^{e}me d\'{e}marche que celle du chapitre 4
relatif au cas Hermitique. Soit $H(t)$ un Hamiltonien non-Hermitique d\'{e}%
pendant du temps dont l'\'{e}quation de Schr\"{o}dinger associ\'{e}e est 
\begin{equation}
i\hbar \partial _{t}\left\vert \Phi _{n}^{H}(t)\right\rangle =H(t)\left\vert
\Phi _{n}^{H}(t)\right\rangle .  \label{schr}
\end{equation}%
Nous supposons par ailleurs l'existence d'un invariant pseudo-Hermitique $%
I^{PH}$ $(t)$ explicitement d\'{e}pendant du temps v\'{e}rifiant l'\'{e}%
quation de Von\textbf{-}Neumann%
\begin{equation}
\frac{\partial I^{PH}(t)}{\partial t}=\frac{i}{\hbar }\left[ I^{PH}\left(
t\right) ,H(t)\right] .  \label{LewisPH}
\end{equation}%
En compl\`{e}te analogie avec le sc\'{e}nario ind\'{e}pendant du temps, la
relation temporelle de la quasi-Hermiticit\'{e} pour l'op\'{e}rateur
invariant s'\'{e}crit%
\begin{equation}
I^{PH\dag }\left( t\right) =\eta (t)I^{PH}\left( t\right) \eta ^{-1}(t)\text{
}\Leftrightarrow I^{h}(t)=\rho (t)I^{PH}(t)\rho ^{-1}(t)=I^{h\dag }(t),
\label{quas}
\end{equation}%
ainsi l'op\'{e}rateur m\'{e}trique d\'{e}pendant du temps $\eta (t)=\rho
^{+}(t)\rho (t)$ relie $I^{PH}(t)$ \`{a} son conjugu\'{e} Hermitique $%
I^{PH\dag }\ (t)$ et \`{a} l'op\'{e}rateur invariant Hermitique $I^{h}(t)$
au moyen de la transformation de similarit\'{e} $\rho (t)$. Remarquons
aussi, que l'action de l'op\'{e}rateur invariant sur une solution de l'\'{e}%
quation de Schr\"{o}dinger est aussi une solution de l'\'{e}quation de Schr%
\"{o}dinger, c'est-\`{a}-dire%
\begin{equation}
H(t)\left( I^{PH}\left( t\right) \left\vert \Phi ^{H}(t)\right\rangle
\right) =i\hbar \partial _{t}\left( I^{PH}\left( t\right) \left\vert \Phi
^{H}(t)\right\rangle \right) .
\end{equation}%
Supposons que l'op\'{e}rateur invariant pseudo-Hermitique $I^{PH}(t)$ admet
un ensemble d'\'{e}tats propres $\left\vert \phi _{n}^{H}\left( t\right)
\right\rangle $%
\begin{equation}
\text{ \ }I^{PH}\left( t\right) \left\vert \phi _{n}^{H}(t)\right\rangle
=\lambda _{n}\left\vert \phi _{n}^{H}(t)\right\rangle ,  \label{IPH}
\end{equation}%
en introduisant la m\'{e}trique d\'{e}pendante du temps $\eta (t)$, le
pseudo-produit scalaire entre \'{e}tats propres s'\'{e}crit 
\begin{equation}
\left\langle \phi _{m}^{H}(t)\right\vert \eta (t)\left\vert \phi
_{n}^{H}(t)\right\rangle =\delta _{m,n}.  \label{psD}
\end{equation}%
Les valeurs propres $\lambda _{n}$ sont \'{e}galement ind\'{e}pendantes du
temps pour le voir il suffit de diff\'{e}rencier l'\'{e}quation (\ref{IPH})
par rapport au temps%
\begin{equation}
\text{ }\frac{\partial I^{PH}}{\partial t}\text{\ }\left\vert \phi
_{n}^{H}(t)\right\rangle +\text{\ }I^{PH}\text{ }\frac{\partial \left\vert
\phi _{n}^{H}(t)\right\rangle }{\partial t}\text{\ }=\frac{\partial \lambda
_{n}}{\partial t}\left\vert \phi _{n}^{H}(t)\right\rangle +\lambda _{n}\frac{%
\partial \left\vert \phi _{n}^{H}(t)\right\rangle }{\partial t},  \label{dif}
\end{equation}%
multiplions l'\'{e}quation (\ref{dif}) par $\left\langle \phi _{n}^{H}\left(
t\right) \right\vert \eta (t)$ et en utilisant l'\'{e}quation (\ref{LewisPH}%
), nous obtenons 
\begin{equation}
\frac{\partial \lambda _{n}}{\partial t}=\left\langle \phi
_{n}^{H}(t)\right\vert \eta (t)\frac{\partial I^{PH}}{\partial t}\text{\ }%
\left\vert \phi _{n}^{H}(t)\right\rangle =0,  \label{VP}
\end{equation}%
qui montre que les valeurs propres $\lambda _{n}$ sont des constantes. La r%
\'{e}alit\'{e} des valeurs propres $\lambda _{n}$ est garantie, puisque
l'invariant Hermitique $I^{h}(t)$ non-Hermitique $I^{PH}(t)$ li\'{e}s par
une transformation de similarit\'{e} (\ref{quas}) admettent les m\^{e}mes
valeurs propres.

Afin d'\'{e}tudier la relation entre les \'{e}tats propres de $I^{PH}(t)$ et
les solutions de l'\'{e}quation de Schr\"{o}dinger (\ref{schr}), nous
projetons l'\'{e}quation (\ref{dif}) sur $\left\langle \phi
_{m}^{H}(t)\right\vert \eta (t)$ et en utilisant l'\'{e}quation (\ref{VP}),
nous obtenons%
\begin{equation}
i\hbar \left\langle \phi _{m}^{H}(t)\right\vert \eta (t)\frac{\partial }{%
\partial t}\ \left\vert \phi _{n}^{H}(t)\right\rangle =\left\langle \phi
_{m}^{H}(t)\right\vert \eta (t)H(t)\ \left\vert \phi
_{n}^{H}(t)\right\rangle ,\ \ \ (m\neq n).
\end{equation}%
Pour $m=n$\ on peut v\'{e}rifier que $\left\vert \phi
_{n}^{H}(t)\right\rangle $ est une solution de l'\'{e}quation de Schr\"{o}%
dinger. Cela pourrait \^{e}tre le cas si on utilisera le fait que les phases
des \'{e}tats stationnaires ne sont pas fix\'{e}es. En effet, on peut donc tr%
\`{e}s bien multiplier $\left\vert \phi _{n}^{H}(t)\right\rangle $ par un
facteur de phase d\'{e}pendant du temps, les nouveaux \'{e}tats propres $%
\left\vert \Phi _{n}^{H}(t)\right\rangle $ de $I^{PH}(t)$ sont%
\begin{equation}
\left\vert \Phi _{n}^{H}(t)\right\rangle =e^{i\gamma _{n}(t)}\left\vert \phi
_{n}^{H}(t)\right\rangle ,  \label{sol}
\end{equation}%
et doivent satisfaire l'\'{e}quation de Schr\"{o}dinger. C'est-\`{a}-dire, $%
\left\vert \Phi _{n}^{H}(t)\right\rangle $ est une solution particuli\`{e}re 
\`{a} l'\'{e}quation de Schr\"{o}dinger. Cette condition permet d'obtenir l'%
\'{e}quation diff\'{e}rentielle de premier ordre satisfaite par la phase $%
\gamma _{n}(t)$%
\begin{equation}
\frac{d\gamma _{n}(t)}{dt}=\left\langle \phi _{n}^{H}(t)\right\vert \eta (t)%
\left[ i\hbar \frac{\partial }{\partial t}-H(t)\right] \text{\ }\left\vert
\phi _{n}^{H}(t)\right\rangle .  \label{Phase}
\end{equation}%
Le premier terme de l'\'{e}quation (\ref{Phase}) correspond \`{a} la phase g%
\'{e}om\'{e}trique non-adiabatique, le second terme repr\'{e}sente la phase
dynamique. Il est facile de montrer en utilisant (\ref{psD}) que la phase $%
\gamma _{n}(t)$ est r\'{e}elle.

La solution g\'{e}n\'{e}rale de l'\'{e}quation de Schr\"{o}dinger associ\'{e}%
e \`{a} un Hamiltonien non-Hermitique $H(t)$ d\'{e}pendant du temps est donn%
\'{e}e par%
\begin{equation}
\left\vert \Phi ^{H}(t)\right\rangle =\sum_{n}C_{n}e^{i\gamma
_{n}(t)}\left\vert \phi _{n}^{H}(t)\right\rangle ,  \label{SolGen}
\end{equation}%
o\`{u} les coefficients $C_{n}=\left\langle \phi _{n}^{H}(0)\right\vert \eta
(0)\left\vert \Phi ^{H}(0)\right\rangle $ sont ind\'{e}pendants du temps.

\subsection{Application: Hamiltonien de Swanson g\'{e}n\'{e}ralis\'{e} non
Hermitique d\'{e}pendant du temps}

Nous appliquons la th\'{e}orie des invariants pseudo-Hermitiques sur un
exemple tr\`{e}s \'{e}tudi\'{e} dans la litt\'{e}rature \cite%
{jones,Swanson,bagchi,musumbu,quesne} : l'oscillateur de Swanson g\'{e}n\'{e}%
ralis\'{e}. En effet, consid\'{e}rons l'Hamiltonien de Swanson
non-Hermitique avec des coefficients d\'{e}pendant du temps \cite{fring2}%
\begin{equation}
H(t)=\omega (t)\left( a^{+}a+\frac{1}{2}\right) +\alpha (t)a^{2}+\beta
(t)a^{+2},  \label{HH}
\end{equation}%
o\`{u} $(\omega (t)$, $\alpha (t)$, $\beta (t))$ sont des param\`{e}tres
complexes d\'{e}pendant du temps et les op\'{e}rateurs $\left(
a,a^{+}\right) $ sont respectivement les op\'{e}rateurs d'annihilation et de
cr\'{e}ation v\'{e}rifiant la relation de commutation $\left[ a,a^{+}\right]
=1.$

Supposons que l'op\'{e}rateur invariant pseudo-Hermitique $I^{PH}(t)$ s'\'{e}%
crit sous la forme%
\begin{equation}
I^{PH}(t)=\delta _{1}(t)\left( a^{+}a+\frac{1}{2}\right) +\delta
_{2}(t)a^{2}+\delta _{3}(t)a^{+2},  \label{IH}
\end{equation}%
o\`{u} les param\`{e}tres $(\delta _{1}(t),\delta _{2}(t),\delta _{3}(t))$
sont des r\'{e}els et d\'{e}pendants du temps. L'invariant (\ref{IH}) est 
\'{e}videmment non-Hermitique lorsque $\delta _{2}(t)\neq \delta _{3}(t)$.

Reprenons la m\^{e}me d\'{e}marche que celle des r\'{e}f\'{e}rences \cite%
{kalt,wali} et introduisons l'op\'{e}rateur $\rho (t)$ tel que: 
\begin{align}
\rho \left( t\right) & =\exp \left[ \epsilon \left( t\right) \left( a^{+}a+%
\frac{1}{2}\right) +\mu \left( t\right) a^{2}+\mu ^{\ast }\left( t\right)
a^{+2}\right] ,  \notag \\
& =\exp \left[ \vartheta _{+}\left( t\right) K_{+}\right] \exp \left[ \ln
\vartheta _{0}\left( t\right) K_{0}\right] \exp \left[ \vartheta _{-}\left(
t\right) K_{-}\right] ,
\end{align}%
o\`{u} $K_{0}=\frac{1}{2}\left( a^{+}a+\frac{1}{2}\right) ,$ $K_{-}=\frac{1}{%
2}a^{2},$ et $K_{+}=\frac{1}{2}a^{+2}$ sont les g\'{e}n\'{e}rateurs de l'alg%
\`{e}bre de Lie $su(1,1)$%
\begin{equation}
\left\{ 
\begin{array}{c}
\left[ K_{0},K_{+}\right] =K_{+} \\ 
\left[ K_{0},K_{-}\right] =-K_{-} \\ 
\left[ K_{+},K_{-}\right] =-2K_{0}%
\end{array}%
\right. ,  \label{su(1,1)}
\end{equation}%
et les param\`{e}tres ($\vartheta _{+}\left( t\right) ,\vartheta _{0}\left(
t\right) ,\vartheta _{-}\left( t\right) $) sont reli\'{e}s \`{a} ($\epsilon
\left( t\right) ,\mu \left( t\right) $) \ \`{a} l'aide des relations
suivantes

\begin{align}
\vartheta _{+}\left( t\right) & =\frac{2\mu ^{\ast }\sinh \theta }{\theta
\cosh \theta -\epsilon \sinh \theta }=-\Phi (t)e^{-i\varphi (t)},  \notag \\
\vartheta _{0}\left( t\right) & =\left( \cosh \theta -\frac{\epsilon }{%
\theta }\sinh \theta \right) ^{-2}=\Phi ^{2}(t)-\chi (t), \\
\vartheta _{-}\left( t\right) & =\frac{2\mu \sinh \theta }{\theta \cosh
\theta -\epsilon \sinh \theta }=-\Phi (t)e^{i\varphi (t)},  \notag \\
\chi (t)& =-\frac{\cosh \theta +\frac{\epsilon }{\theta }\sinh \theta }{%
\cosh \theta -\frac{\epsilon }{\theta }\sinh \theta }\text{ \ \ \ \ \ \ ,\ \
\ }\theta =\sqrt{\epsilon ^{2}-4\left\vert \mu \right\vert ^{2}},  \notag
\end{align}%
$\mu ^{\ast }\left( t\right) $ indique le complexe conjugu\'{e} $\mu \left(
t\right) $. En utilisant les relations suivantes

\begin{align}
\rho \left( t\right) K_{+}\rho ^{-1}\left( t\right) & =\frac{1}{\vartheta
_{0}}\left[ -2\vartheta _{-}\chi K_{0}+\vartheta _{-}^{2}K_{-}+\chi ^{2}K_{+}%
\right] ,  \notag \\
\rho \left( t\right) K_{0}\rho ^{-1}\left( t\right) & =\frac{1}{\vartheta
_{0}}\left[ -\left( \vartheta _{-}\vartheta _{+}+\chi \right)
K_{0}+\vartheta _{-}K_{-}+\chi \vartheta _{+}K_{+}\right] , \\
\rho \left( t\right) K_{-}\rho ^{-1}\left( t\right) & =\frac{1}{\vartheta
_{0}}\left[ -2\vartheta _{+}K_{0}+K_{-}+\vartheta _{+}^{2}K_{+}\right] , 
\notag
\end{align}%
l'op\'{e}rateur invariant quasi-Hermitique $I^{h}\left( t\right) =\rho
\left( t\right) I^{PH}(t)\rho ^{-1}\left( t\right) $ \ se simplifie en: 
\begin{align}
I^{h}(t)& =\frac{2}{\vartheta _{0}}\left[ \left[ -\delta _{1}\left(
\vartheta _{-}\vartheta _{+}+\chi \right) -2\left( \delta _{2}\vartheta
_{+}+\delta _{3}\chi \vartheta _{-}\right) \right] K_{0}\right.  \notag \\
& \left. +\left( \delta _{1}\vartheta _{-}+\delta _{2}+\delta _{3}\vartheta
_{-}^{2}\right) K_{-}+\left( \delta _{1}\chi \vartheta _{+}+\delta
_{2}\vartheta _{+}^{2}+\delta _{3}\chi ^{2}\right) K_{+}\right] .
\label{invh}
\end{align}%
Pour que $I^{h}\left( t\right) $ soit Hermitique ($I^{h}\left( t\right) $ $=$
$I^{\dag h}\left( t\right) $), nous imposons que le coefficient de $K_{0}$
soit r\'{e}el et que les coefficients de $K_{-}$ et $K_{+}$ soient des
conjugu\'{e}s complexes l'un de l'autre, d'o\`{u}%
\begin{align}
\delta _{2}& =\delta _{3}\chi ,  \notag \\
\delta _{1}& =-\frac{\delta _{3}\left( \vartheta _{-}^{2}+\chi \right) }{%
\vartheta _{-}}=-\frac{\delta _{3}\left( \vartheta _{+}^{2}+\chi \right) }{%
\vartheta _{+}}.  \label{delta}
\end{align}%
Ces derni\`{e}res \'{e}quations (\ref{delta}) impliquent que que $\vartheta
_{+}=\vartheta _{-}\equiv -\Phi (t)$ et impose la r\'{e}alit\'{e} du param%
\`{e}tre $\mu (t)$, c'est-\`{a}-dire $\mu (t)=\mu ^{\ast }(t)$. Finalement,
l'op\'{e}rateur invariant $I^{h}(t)$ est r\'{e}duit \`{a}

\begin{equation}
I^{h}\left( t\right) =-\frac{2}{\vartheta _{0}}\left[ \delta _{1}\left( \Phi
^{2}+\chi \right) -4\delta _{3}\chi \Phi \right] K_{0}.  \label{invh1}
\end{equation}%
Soit $\left\vert \psi _{n}^{h}\right\rangle $ un \'{e}tat propre de $K_{0}$
avec la valeur propre $k_{n}$ i.e.%
\begin{equation}
K_{0}\left\vert \psi _{n}^{h}\right\rangle =k_{n}\left\vert \psi
_{n}^{h}\right\rangle .  \label{vp}
\end{equation}%
Les \'{e}tats propres de $I^{h}(t)$ (\ref{invh1}) sont \'{e}videmment donn%
\'{e}s par%
\begin{equation}
I^{h}\left( t\right) \left\vert \psi _{n}^{h}(t)\right\rangle =-\frac{2}{%
\vartheta _{0}}\left[ \delta _{1}\left( \Phi ^{2}+\chi \right) -4\delta
_{3}\chi \Phi \right] k_{n}\left\vert \psi _{n}^{h}\right\rangle \text{. \ }
\label{eqvpdt}
\end{equation}%
Sachant que les valeurs propres de l'invariant $I^{h}(t)$ sont r\'{e}elles
et constantes, alors le pr\'{e}-facteur de (\ref{eqvpdt}) doit \^{e}tre
constant. Ainsi, on choisit le facteur $-[\delta _{1}(\Phi ^{2}+\chi
)-4\delta _{3}\chi \Phi ]/\vartheta _{0}$ \'{e}gal \`{a} $1$. Il s'ensuit
que les \'{e}tats propres $\left\vert \phi _{n}^{H}(t)\right\rangle $ de $%
I^{PH}(t)$ sont d\'{e}duit de ceux $\left\vert \psi _{n}^{h}\right\rangle $
de $I^{h}(t)$ \`{a} travers la transformation de similarit\'{e} $\left\vert
\phi _{n}^{H}(t)\right\rangle $ $=\rho ^{-1}(t)\left\vert \psi
_{n}^{h}\right\rangle $.

La condition de quasi-Hermiticit\'{e} (\ref{quas}) nous permet d'obtenir%
\begin{equation}
I^{PH}(t)=-\frac{2}{\vartheta _{0}}\left[ \left( \Phi ^{2}+\chi \right)
K_{0}+\chi \Phi K_{-}+\Phi K_{+}\right] .  \label{PH1}
\end{equation}%
Par cons\'{e}quent, la condition d'invariance (\ref{LewisPH}) donne les \'{e}%
quations diff\'{e}rentielles que doivent v\'{e}rifier les coefficients $%
\delta _{1}(t)$, $\delta _{2}(t)$, $\delta _{3}(t)$ 
\begin{eqnarray}
\dot{\delta}_{1}(t) &=&2i\left[ \beta (t)\delta _{2}-\alpha (t)\delta _{3}%
\right] \\
\dot{\delta}_{2}(t) &=&i\left[ \omega \left( t\right) \delta _{2}-\alpha
\left( t\right) \delta _{1}\right] \\
\dot{\delta}_{3}(t) &=&i\left[ \beta \left( t\right) \delta _{1}-\omega
\left( t\right) \delta _{3}\right] ,
\end{eqnarray}%
la s\'{e}paration des parties r\'{e}elles et imaginaires de ces deux derni%
\`{e}res \'{e}quations diff\'{e}rentielles conduisent aux contraintes
suivantes%
\begin{equation}
\dot{\vartheta}_{0}=\frac{\vartheta _{0}}{\Phi }\left[ -2\Phi \left\vert
\omega \right\vert \sin \varphi _{\omega }+\left\vert \alpha \right\vert
\sin \varphi _{\alpha }+\left( 2\Phi ^{2}+\chi \right) \left\vert \beta
\right\vert \sin \varphi _{\beta }\right] ,  \label{cont1}
\end{equation}%
\begin{equation}
\dot{\Phi}=-\Phi \left\vert \omega \right\vert \sin \varphi _{\omega
}+\left\vert \alpha \right\vert \sin \varphi _{\alpha }+\Phi ^{2}\left\vert
\beta \right\vert \sin \varphi _{\beta },  \label{cont2}
\end{equation}%
\begin{equation}
\ 
\begin{array}{c}
\chi \left\vert \beta \right\vert \cos \varphi _{\beta }=\left\vert \alpha
\right\vert \cos \varphi _{\alpha } \\ 
\left( \Phi ^{2}+\chi \right) \left\vert \alpha \right\vert \cos \varphi
_{\alpha }=\chi \Phi \left\vert \omega \right\vert \cos \varphi _{\omega }
\\ 
\Phi \left\vert \omega \right\vert \cos \varphi _{\omega }=\left( \Phi
^{2}+\chi \right) \left\vert \beta \right\vert \cos \varphi _{\beta }%
\end{array}%
,  \label{rel}
\end{equation}%
o\`{u} $\varphi _{\omega }$, $\varphi _{\alpha }$ et $\varphi _{\beta }$
sont respectivement les angles polaires de $\omega $, $\alpha $ et $\beta $.

Pour obtenir la solution de l'\`{e}quation de Schr\"{o}dinger (\ref{sol});
calculons tout d'abord la phase 
\begin{align}
\frac{d\gamma _{n}(t)}{dt}& =\left\langle \phi _{n}^{H}(t)\right\vert \eta
(t)\left[ i\frac{\partial }{\partial t}-H(t)\right] \text{\ }\left\vert \phi
_{n}^{H}(t)\right\rangle  \notag \\
& =\left\langle \psi _{n}^{h}\right\vert \left[ i\rho \dot{\rho}^{-1}-\rho
H\rho ^{-1}\right] \text{\ }\left\vert \psi _{n}^{h}\right\rangle .
\label{Phase1}
\end{align}%
En remarquant que%
\begin{equation}
i\rho \dot{\rho}^{-1}-\rho H\rho ^{-1}=2W\left( t\right) K_{0}+2U\left(
t\right) K_{-}+2V\left( t\right) K_{+},
\end{equation}%
o\`{u} les fonctions de coefficient $W\left( t\right) $, $U\left( t\right) $
et $V\left( t\right) $ sont donn\'{e}es par%
\begin{align}
W\left( t\right) & =\frac{1}{\vartheta _{0}}\left[ \omega \left( \Phi
^{2}+\chi \right) -2\Phi \left( \alpha +\beta \chi \right) -\frac{i}{2}%
\left( \dot{\vartheta}_{0}-2\Phi \dot{\Phi}\right) \right] , \\
U\left( t\right) & =\frac{1}{\vartheta _{0}}\left[ \omega \Phi -\alpha
-\beta \Phi ^{2}+\frac{i}{2}\dot{\Phi}\right] , \\
V\left( t\right) & =\frac{1}{\vartheta _{0}}\left[ \omega \chi \Phi -\alpha
\Phi ^{2}-\beta \chi ^{2}+\frac{i}{2}\left( \vartheta _{0}\dot{\Phi}+\Phi
^{2}\dot{\Phi}-\Phi \dot{\vartheta}_{0}\right) \right] .
\end{align}%
et en utilisant des \'{e}quations (\ref{rel}), ces coefficients se
simplifient \'{e}norm\'{e}ment%
\begin{align}
W\left( t\right) & =\frac{1}{\vartheta _{0}}\left[ \left\vert \omega
\right\vert \left( \Phi ^{2}+\chi \right) \cos \varphi _{\omega }-4\Phi
\left\vert \alpha \right\vert \cos \varphi _{\alpha }\right.  \notag \\
& \left. -i\frac{\vartheta _{0}}{2\Phi }\left[ -\Phi \left\vert \omega
\right\vert \sin \varphi _{\omega }+\left\vert \alpha \right\vert \sin
\varphi _{\alpha }+\chi \left\vert \beta \right\vert \sin \varphi _{\beta }%
\right] \right] , \\
U\left( t\right) & =0, \\
V\left( t\right) & =0.
\end{align}%
Sachant que la phase $\gamma _{n}(t)$ (\ref{Phase1}) est r\'{e}elle, il faut
imposer que la fr\'{e}quence $W(t)$ doit \^{e}tre r\'{e}elle, condition qui
conduit \`{a}%

\begin{equation}
\gamma_{n}(t) = k_{n} 
\int_{0}^{t} \frac{2}{\vartheta_{0}} 
\left[
  \lvert \omega \rvert \left( \Phi^{2} + \chi \right) \cos \varphi_{\omega}
  - 4 \Phi \lvert \alpha \rvert \cos \varphi_{\alpha}
\right] 
\, dt^{\prime}.
\end{equation}
Par cons\'{e}quent, la solution g\'{e}n\'{e}rale de l'\'{e}quation de Schr%
\"{o}dinger (\ref{schr}) s'\'{e}crit%
\begin{equation}
\left\vert \Phi ^{H}(t)\right\rangle =\sum_{n}C_{n}(0)\exp \left(
ik_{n}\int\limits_{0}^{t}\frac{2}{\vartheta _{0}}\left[ \left\vert \omega
\right\vert \left( \Phi ^{2}+\chi \right) \cos \varphi _{\omega }-4\Phi
\left\vert \alpha \right\vert \cos \varphi _{\alpha }\right] dt^{\prime
}\right) \left\vert \phi _{n}^{H}(t)\right\rangle .
\end{equation}%
Dans le cas particulier o\`{u} les coefficients $(\omega (t)$, $\alpha (t)$, 
$\beta (t))$ de l'Hamiltonien (\ref{HH}) sont r\'{e}els, c'est \`{a} dire $%
\varphi _{\omega }=\varphi _{\alpha }=\varphi _{\beta }=0$ , les \'{e}%
quations (\ref{cont1}- \ref{rel}) se simplifient et s'\'{e}crivent 
\begin{equation}
\dot{\vartheta}_{0}=0\text{ \ \ \ \ \ }\dot{\Phi}=0,\text{ \ }\vartheta
_{0},\Phi \text{ sont constants,}
\end{equation}%
\begin{equation}
\ 
\begin{array}{c}
\chi \left\vert \beta \right\vert =\left\vert \alpha \right\vert \\ 
\left( \Phi ^{2}+\chi \right) \left\vert \alpha \right\vert =\chi \Phi
\left\vert \omega \right\vert \\ 
\Phi \left\vert \omega \right\vert =\left( \Phi ^{2}+\chi \right) \left\vert
\beta \right\vert%
\end{array}%
.
\end{equation}%
Ainsi, on retrouve le cas d'une m\'{e}trique $\eta $ ind\'{e}pendante du
temps, situation d\'{e}ja \'{e}tudi\'{e}e par C. Figueira de Morisson Faria
et Fring \cite{fring3,fring4}.

En conclusion, nous avons pr\'{e}sent\'{e} des m\'{e}thodes appropri\'{e}es
pour r\'{e}soudre les syst\`{e}mes quantiques non-Hermitiques d\'{e}pendants
du temps. Notons que le d\'{e}bat \cite%
{mos8,mos9,mos10,znojil2,znojil3,znojil1} soulev\'{e} en 2007 pour d\'{e}%
crire le cadre g\'{e}n\'{e}ral de l'\'{e}volution temporelle unitaire pour
les Hamiltoniens non Hermitiens d\'{e}pendant du temps dure jusqu'\`{a} nos
jours \cite{znojil4,gong,maam,fring1,fring2,znojil9,bila,maam1,znojil8}. La
th\'{e}orie des op\'{e}rateurs pseudo-invariants s'attelle \`{a} jeter les
bases d'une approche alternative pour r\'{e}soudre l'\'{e}volution
temporelle des syst\`{e}mes quantiques non-Hermitiques en utilisant la
relation de quasi-Hermiticit\'{e} (\ref{quas}) des op\'{e}rateurs invariants.

\thispagestyle{empty} 
\markboth{}{Références} \addcontentsline{toc}{chapter}{Références}.

\end{document}